\documentclass[aps,prx,twocolumn,footinbib,floatfix]{revtex4-2}
\usepackage{graphicx}
\usepackage{indentfirst}
\usepackage{braket}
\usepackage{float}
\usepackage{amsmath}
\usepackage{physics}
\usepackage{amssymb}
\usepackage{CJK}
\usepackage{esint}
\usepackage{color}
\usepackage[T1]{fontenc}
\usepackage{subfigure}
\usepackage{amsfonts}
\usepackage{footmisc}
\usepackage{scrextend}
\usepackage{multirow}
\usepackage[outdir=./]{epstopdf}

\usepackage[english]{babel}
\usepackage{url}
\usepackage{comment}
\usepackage{array}
\usepackage{subfiles}
\usepackage{tikz}
\usetikzlibrary{shapes,arrows}
\usepackage{mathrsfs}
\usepackage[mathscr]{eucal}

\usepackage[hyperfootnotes=false]{hyperref}
\usepackage{cleveref}
\definecolor{darkblue}{rgb}{0,0,0.5}
\hypersetup{
colorlinks=true,
linkcolor=black,
filecolor=blue,
citecolor=darkblue,  
urlcolor=black,
}

\newtheorem{theorem}{Theorem}

\newenvironment{proof}[1][Proof]{\noindent\textbf{#1.} }{\ \rule{0.5em}{0.5em}}

\def\be{\begin{equation}}
\def\ee{\end{equation}}
\def\ba{\begin{eqnarray}}
\def\ea{\end{eqnarray}}
\def\bal{\begin{equation}\begin{aligned}}
\def\eal{\end{aligned}\end{equation}}

\def\bp{\begin{pmatrix}}
\def\ep{\end{pmatrix}}

\usepackage{bm}

\newcommand{\calA}{{\cal A}}
\newcommand{\calB}{{\cal B}}

\newcommand{\calI}{{\cal I}}
\newcommand{\calL}{{\cal L}}

\newcommand{\calN}{{\cal N}}

\newcommand{\calJ}{{\cal J}}
\newcommand{\bbJ}{{\mathbb J}}

\newcommand{\calU}{{\cal U}}

\newcommand{\hrho}{\hat{\rho}}
\newcommand{\1}{^{(1)}}

\newcommand{\state}[1]{\ketbra{#1}{#1}}
\newcommand{\bn}{{\bm n}}

\usepackage{pict2e,picture,graphicx}
\makeatletter
\DeclareRobustCommand{\Arrow}[1][]{%
\check@mathfonts
\if\relax\detokenize{#1}\relax
\settowidth{\dimen@}{$\m@th\rightarrow$}%
\else
\setlength{\dimen@}{#1}%
\fi
\sbox\z@{\usefont{U}{lasy}{m}{n}\symbol{41}}%
\begin{picture}(\dimen@,\ht\z@)
\roundcap
\put(\dimexpr\dimen@-.7\wd\z@,0){\usebox\z@}
\put(0,\fontdimen22\textfont2){\line(1,0){\dimen@}}
\end{picture}%
}
\makeatother

\newcommand{\QZ}[1]{{{\textcolor{black}{#1}}}}
\newcommand{\hw}[1]{{{\textcolor{black}{#1}}}}

\usepackage[normalem]{ulem}
\usepackage{xr}

\begin{document}

\title{Ultimate precision limit of noise sensing and dark matter search}
\author{Haowei Shi$^1$}
\email{hwshi@usc.edu} 

\author{Quntao Zhuang$^{1,2}$}
\email{qzhuang@usc.edu}

\affiliation{
$^1$Ming Hsieh Department of Electrical and Computer Engineering, University of Southern California, Los
Angeles, California 90089, USA
\\
$^2$Department of Physics and Astronomy, University of Southern California, Los
Angeles, California 90089, USA
}

\begin{abstract}
The nature of dark matter is unknown and calls for a systematical search. For axion dark matter,
such a search relies on finding feeble random noise arising from the weak coupling between dark matter and microwave haloscopes. We model such process as a quantum channel and derive the fundamental precision limit of noise sensing. An entanglement-assisted strategy based on two-mode squeezed vacuum is thereby demonstrated optimal, while the optimality of a single-mode squeezed vacuum is found limited to the lossless case. We propose a `nulling' measurement (squeezing and photon counting) to achieve the optimal performances. In terms of the scan rate, even with 20-decibel of strength, single-mode squeezing still underperforms the vacuum limit which is achieved by photon counting on vacuum input; while the two-mode squeezed vacuum provides large and close-to-optimum advantage over the vacuum limit, \hw{thus more exotic quantum resources are no longer required}. Our results highlight the necessity of entanglement assistance and microwave photon counting in dark matter search.

\end{abstract}

\maketitle

\tableofcontents

\section{Introduction}
A fundamental question that puzzles us today is the nature of the hypothetical dark matter (DM) that makes up a large portion of the entire Universe's energy density, as inferred from multiple astrophysical and cosmological observations and simulations~\cite{aghanim2020planck,massey2010dark,sofue2001rotation}. Due to its weak interaction with ordinary matter, DM is extremely challenging to search for. Moreover, as the frequency of DM is unknown, a search requires a scan over a huge frequency range from Terahertz to hertz, involving different systems ranging from opto-mechanical~\cite{carney2020PRD,dal2020resonators,dal2021VDM,carney2021DM,carney2021white_paper,monteiro2020PRL,moore2021levitated,afek2022trapped_sensors,yin2022} and microwave~\cite{Sikivie:1983ip,girvin2016axdm,malnou2019,dixit2021,backes2021}, which can easily take hundred of years with the state-of-the-art technology~\cite{berlin2022searches,backes2021}. As much attention has been on utilizing quantum metrology, empowered by quantum resources such as squeezing~\cite{girvin2016axdm,malnou2019,backes2021} and entanglement~\cite{brady2022entangled}, to boost the DM search, it is crucial to understand the ultimate precision limits of DM search allowed by quantum physics.


Axion dark matter search relies on microwave haloscopes---microwave cavities in presence of magnetic field~\cite{Sikivie:1983ip,girvin2016axdm,malnou2019,dixit2021,backes2021} that allows axion particles to convert to microwave photons.
Such a search process can be modelled as a quantum sensing problem over a covariant bosonic quantum channel~\cite{holevo2007one,weedbrook2012gaussian}, whose additive noise level reveals the existence of DM. The ultimate precision limit of DM search can therefore be understood from the ultimate precision limit of additive noise sensing. However, while the ultimate limits of phase sensing~\cite{escher2011general}, displacement sensing~\cite{zhuang2018DQSCV,xia2020demonstration}, loss sensing~\cite{monras2007optimal,nair2018quantum} and amplifier gain sensing~\cite{nair2022optimal} have been explored, little is known about the limit of noise sensing~\cite{pirandola2017ultimate} in bosonic quantum channels when there is energy constraint.

In this paper, we derive the ultimate precision limit of energy-constrained\QZ{~\cite{nair2018quantum,nair2022optimal}} noise sensing in covariant bosonic quantum channels, and therefore reveal the DM search performance limit allowed by quantum physics. Via quantum Fisher information (QFI) calculations, we show that entangled source in the form of two-mode squeezed vacuum (TMSV) is optimal for noise sensing in the parameter region of interest. \hw{On the other hand,} (single-mode) squeezed vacuum source is only optimal in the lossless case, and even underperforms the vacuum limit when loss is large.

Next, we consider measurement protocols. Although it has been shown~\cite{zheng2016accelerating,malnou2019,backes2021} that squeezed vacuum input improves the performance of homodyne measurement, our analyses show that protocols with homodyne detection are sub-optimal in general. Instead, a `nulling' measurement strategy based on squeezing and photon counting is optimal and beats the homodyne detection by orders of magnitude. For vacuum input, the nulling strategy simplifies to direct photon counting. In particular, when implemented ideally in the lossless case, it takes $15$dB of squeezing for a homodyne-based strategy to overcome the (photon counting) vacuum limit.

Finally, we interpret our results in DM search in the setting of microwave haloscopes. We show that the total Fisher information is proportional to the previously well-accepted figure of merit---the scan rate \QZ{(information acquisition rate across all frequencies~\cite{malnou2019,backes2021,brady2022entangled})} in homodyne detection. Then, we recover squeezing's advantage over vacuum homodyne. However, we show that these strategies are below the vacuum-limit scan-rate (vacuum+photon counting), in the practical range of squeezing ($<20$dB). \QZ{In contrast,} an entanglement-assisted strategy based on TMSV enables optimal advantage over the vacuum limit at an arbitrary squeezing level, achievable with our nulling receiver based on photon counting. This provides a guideline for the next-generation haloscopes---developing good microwave photon counting measurement is of high priority; while more exotic quantum sources such as Gottesman–Kitaev–Preskill state~\cite{gottesman2001} are not necessary \QZ{since a simple TMSV state is optimal}. Our results can also apply to a sensor-network~\cite{brady2022entangled,brady2022entanglement}, all quantum advantages hold with an additional scaling advantage due to coherent signal processing.

\begin{figure}[t]
    \centering
    \includegraphics[width=1\linewidth]{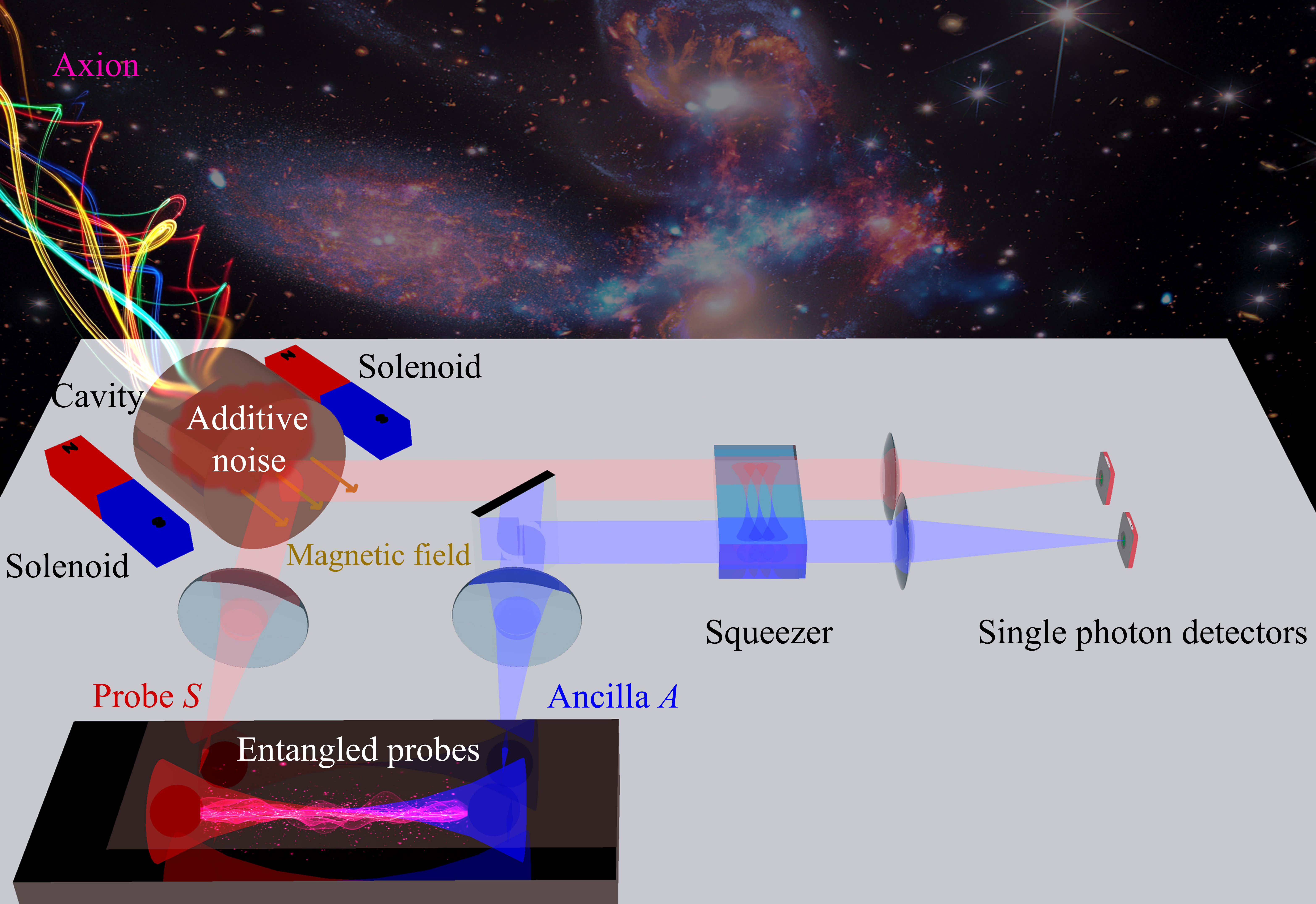}
    \caption{
    Conceptual plot of the entanglement-enhanced microwave haloscopes for axion dark matter search. A pair of entangled probes, including the signal (red beam) and the ancilla (blue beam), are generated at the bottom-right box. Note that the blue beam becomes inaccessible when entanglement assistance is forbidden. The signal probe is then shined on the input port of a microwave cavity, which is coupled with the axion (top-right beam) via a strong magnetic field. At this moment, the unknown parameter of axion is encoded on the signal probe, while the ancilla is shared with the receiver intact. The receiver applies a nonlinear processing to the returned probes jointly, e.g. a two-mode squeezer for the entanglement-enhanced case. Finally, the receiver collects the photon counts and estimate the unknown parameter of axion based on the readout. Background image credit: \copyright James Webb Space Telescope.
}
    \label{fig:DM_search}
\end{figure}



\section{Results}
\subsection{ DM search as additive noise sensing}
We consider dark matter search for the axion DM model (see Fig.~\ref{fig:DM_search}), while we note that our results may also apply to other DM hypotheses. In a search for axion dark matter, an important experimental set-up involves a cavity in presence of electric and magnetic field (microwave haloscopes), where axion DM can couple to cavity modes. Due to the large number density, axion DM is assumed to behave as classical waves~\cite{brady2022entangled}---the mean field at position $\vec{x}$ has the form
\be 
\expval{ a}(m_{\rm a},\Vec{k_0},t)\propto \cos\left(\omega_0t+\Vec{k_0}\cdot \vec{x} +\phi_{\rm a}\right)\,,
\ee 
where the center frequency $\omega_0$ is determined by the axion DM mass $m_{\rm a}$, $\Vec{k_0}$ is the wave factor and $\phi_{\rm a}$ is a phase factor. As the potential DM induced cavity signal is weak, to determine the presence or absence of DM, one considers a long observation time, during which $\phi_{\rm a}$ is completely random in $[0,2\pi)$. Due to the randomness of the axion field, the input-output relation of the cavity at each frequency $\omega$ can be effectively modeled as
\be 
\hat{a}_{\rm out}=\chi_{mm} \hat{a}_{\rm in}+\chi_{ma}\mu_{\rm a}+\sqrt{1-\chi_{mm}^2}\hat{a}_{\rm B},
\label{eq:a_in_a_out}
\ee 
where 
\QZ{\begin{align}
&\chi_{mm}^2(\omega)\simeq \frac{(\gamma_m-\gamma_\ell)^2/4+\omega^2}{(\gamma_m + \gamma_\ell)^2/4+\omega^2}\,,
\label{chi_mm}
\\
&\chi_{ma}^2(\omega)\simeq \frac{\gamma_m \gamma_a}{(\gamma_m+\gamma_\ell)^2/4+\omega^2}\,,
\label{chi_ma}
\end{align}
are the susceptibilities determined by the cavity coupling rates ($\gamma_m$, $\gamma_\ell$ and $\gamma_a$ for measurement port, loss and axion)}, $\hat{a}_{\rm B}$ describes the thermal background with mean photon number determined by the cavity temperature. \QZ{In general, $\gamma_a\ll \gamma_m,\gamma_\ell$.} For simplicity, we have omitted the noise independent phase factors, as it does not affect our analyses. Most importantly, the DM induced signal contributes to the additive noise---$\mu_{\rm a}$ is a complex Gaussian random number with variance equaling $n_{\rm a}$, the number of axion particles in the cavity. The search for DM is therefore a parameter estimation task of the additional additive noise \QZ{$\chi_{ma}^2 n_{\rm a}$ from axion DM.}

The input-output relation in Eq.~\eqref{eq:a_in_a_out} is a special case of a phase-covariant bosonic Gaussian channel (BGC)~\cite{holevo2007one,weedbrook2012gaussian} $\calN_{\kappa,{n_{\rm B}}}$ with transmissivity $\kappa$ and dark count noise of mean photon number ${n_{\rm B}}$, which maps a vacuum input state to a thermal state with mean photon number ${n_{\rm B}}$ and an input mean field $\alpha$ to output mean field $\sqrt{\kappa}\alpha$. The transmissivity $\kappa$ ranges from 0 to $\infty$: for $0\le \kappa< 1$, $\calN_{\kappa,{n_{\rm B}}}$ is a thermal-loss channel, which corresponds to the ones in dark matter search Eq.~\eqref{eq:a_in_a_out}; for $\kappa=1$, $\calN_{\kappa,{n_{\rm B}}}$ is an additive white Gaussian noise (AWGN) channel; for $\kappa>1$, $\calN_{\kappa,{n_{\rm B}}}$ is a thermal amplifier channel. For the channel to be physical, the dark photon count must be larger than the intrinsic amplification noise: ${n_{\rm B}}\ge \max\{\kappa -1,0\}$. \QZ{Note that our definition of noise $n_{\rm B}$ is different from some other conventions for the purpose of simplifying our notations, see Appendix~\ref{app:methods}.}

\QZ{For the case of Eq.~\eqref{eq:a_in_a_out}, the quantum channel is the thermal loss case, with transmissivity $\kappa(\omega)=\chi_{mm}^2(\omega)$ and the noise coming from the thermal background and DM axion
\be 
{n_{\rm B}(\omega)}=\left[1-\chi_{mm}^2\left(\omega\right)\right]n_{T}+\chi_{ma}^2\left(\omega\right) n_{\rm a},
\label{nB_NS}
\ee 
where the thermal photon number of the background environment mode,
$
n_{T}\equiv 1/[\exp\left(\hbar \Omega/k_{\rm B}T\right)-1],
$
is approximately taken at the center frequency $\Omega$.
}

\QZ{Before we proceed with our analyses, we provide some realistic parameter settings. At practical operating condition~\cite{backes2021,brady2022entangled} of temperature $T=35$mK and frequency $f\simeq 7$GHz,  the environment thermal photon number $n_T\sim 10^{-4}$ from the Bose-Einstein distribution. According to theoretical predictions, axion density $n_a/V\sim 10^{15}\frac{\lambda}{\rm km} {\rm cm}^{-3}$~\cite{brady2022entangled} is noticeably large in the microwave region, where $\lambda$ is the De Broglie wavelength of DM. At the same time, the coupling between axion and cavity $\chi_{ma}^2\left(\omega\right)$ is extremely weak, such that the added noise $\chi_{ma}^2\left(\omega\right) n_a\ll1$ is infinitesimal and therefore hard to verify or nullify.}


In the above we have modeled a single sensor case. However, as we will discuss at the end of the paper, identical sensor arrays can be reduced to the above single sensor case~\cite{brady2022entangled}, and therefore our results below can be adopted to sensor-networks.

\subsection{General noise sensing strategies}

\begin{figure}[tbp]
    \centering
    \includegraphics[width=0.4\textwidth]{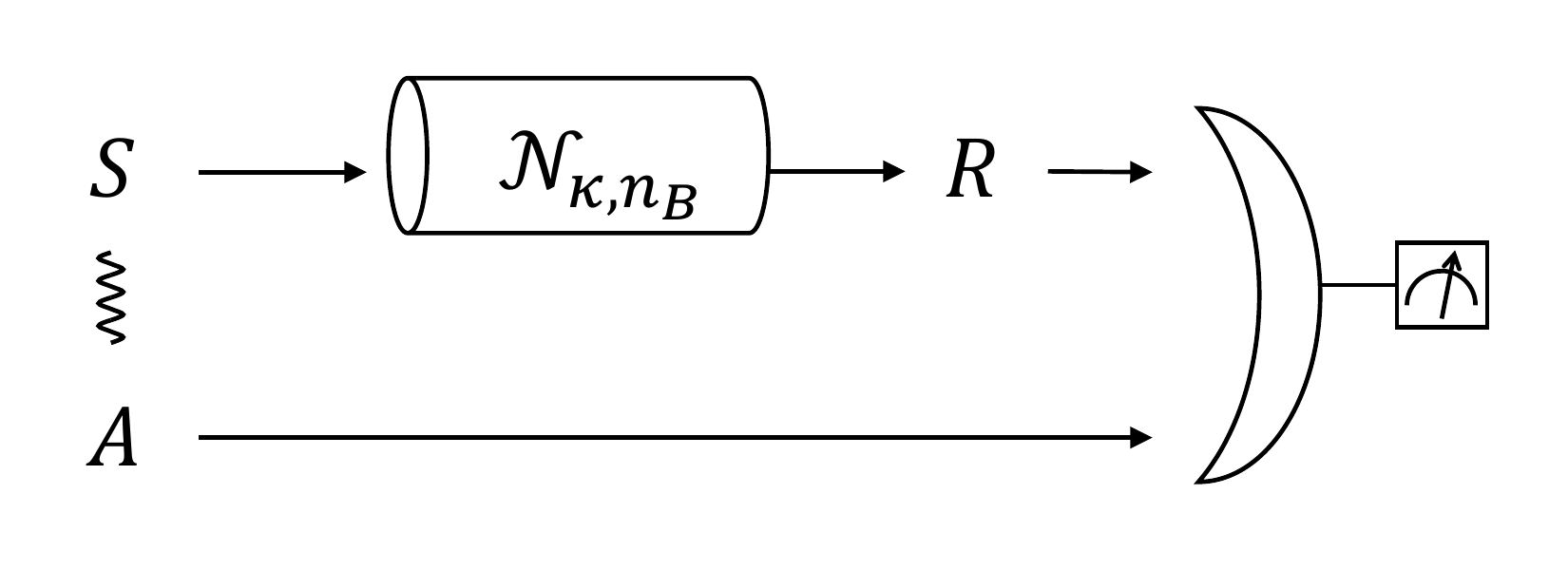}
    \caption{Schematic of the entanglement-assisted strategy. \hw{A probe system $S$ is allowed to be entangled with an ancilla $A$. The probe $S$ is input to the channel $\calN_{\kappa,n_{\rm B}}^{\otimes M}(\hrho)$. The output system $R$ is jointly measured with the ancilla $A$.}}
    \label{fig:EAschematic}
\end{figure}

We consider the estimation of additive noise $n_{\rm B}$ in BGCs, assuming that the transmissivity is known from prior calibration. To measure the additive noise $n_{\rm B}$, one can in general input a probe system $S$ in state $\hrho$ and measure the output system $R$ of the channel $\calN_{\kappa,n_{\rm B}}^{\otimes M}(\hrho)$. Here, to achieve a good performance, we have considered probing the channel $M$ times with input state $\hat{\rho}$ potentially entangled across $M$ probings and a joint measurement on the entire output. To model a physically meaningful setting \QZ{operating with finite energy}, the total mean photon number of the input state is constrained to be $M N_{\rm S}$ over $M$ modes.

A general strategy can also rely on entanglement to boost the sensing performance~\cite{tan2008quantum,shi2020entanglement,zhuang2022}. As shown in Fig.~\ref{fig:EAschematic}, this is implemented by allowing an ancilla $A$ entangled with the probe, such that the joint state of system $AS$ is pure. In general, one can write out the joint input-ancilla pure state as
\be 
\ket{\psi_0}_{AS}=\sum_{\bn\ge \bm 0} \sqrt{p_\bn}\ket{\chi_\bn}_A\ket \bn_{S},
\label{AS_state}
\ee
where $\ket{\bn}=\otimes_{\ell=1}^M \ket{n_\ell}$ is the number state basis of the $M$-mode signal $S$, $p_{\bm n}$'s are probabilities and normalized to unity and \QZ{ancilla states} $\{\ket{\chi_\bn}\}$ are normalized but not necessarily orthogonal~\cite{nair2022optimal,nair2018quantum}. We adopt the vector notation $\bn=\{n_1,\cdots,n_M\}$. \QZ{For such a state, the energy constraint is specified as $\sum_{j=1}^M \expval{\hat a^\dagger_{S,j} \hat a_{S,j}}=\sum_{\bm n}\left(p_{\bm n}\sum_j n_j\right)\le M N_{\rm S}$, where $\hat a_{S,j}$ is the $j$-th signal mode.}
In an entangled strategy, one can perform measurement on the output $R$ and the ancilla system in the quantum state 
\be 
\hrho(n_{\rm B})=\calN_{\kappa,n_{\rm B}}^{\otimes M}\otimes \calI (\state{\psi_0}),
\label{rho_nb_definition}
\ee 
where identity channel $\calI$ models perfect ancilla storage. 

In both strategies, we quantify the performance via the root-mean-square (rms) error $\delta {n_{\rm B}}$. In the following, we provide the ultimate limits on $\delta {n_{\rm B}}$, when one is allowed to optimize any entangled input-ancilla state $\ket{\psi_0}_{AS}$ (subject to the energy constraint) and any measurement strategy. Then we consider practical protocols consisting of a source and a measurement to achieve the limit in the parameter region of interest. 

\subsection{Ultimate limit on noise sensing}

Given a fixed state $\hat{\rho}(n_{\rm B})$ dependent on parameter $n_{\rm B}$, the rms error in estimating $n_{\rm B}$ when allowing an arbitrary measurement is lower bounded by the asymptotically tight quantum Cram\'{e}r-Rao bound~\cite{Helstrom_1967,Yuen_1973,holevo2011probabilistic} 
\begin{equation}
    \delta {n_{\rm B}}^2 \geq \frac{1}{\calJ\left[\hat \rho\left({n_{\rm B}}\right)\right]}.
\end{equation}
The quantity $\calJ\left[\hat \rho\left({n_{\rm B}}\right)\right]$ is the QFI~\cite{braunstein1994statistical} defined via
\begin{align}
\calJ\left[\hat \rho\left({n_{\rm B}}\right)\right]&=\lim_{\epsilon\to0}\frac{8}{\epsilon^2}\left\{1-F\left[\hrho({n_{\rm B}}),\hrho({n_{\rm B}}+\epsilon) \right]\right\}
\\
&=-4\partial^2_{{n_{\rm B}}^\prime} F\left[\hrho({n_{\rm B}}),\hrho({n_{\rm B}}^\prime) \right] \bigg|_{{n_{\rm B}}^\prime={n_{\rm B}}},
\label{eq:QFI_fidelity}
\end{align} 
where the fidelity between two quantum states $\hrho_0,\hrho_1$ is defined as $F(\hrho_0,\hrho_1)\equiv \Tr \sqrt{\sqrt{\hrho_0}\hrho_1\sqrt{\hrho_0}}$.

As QFI $\calJ\left[\hat \rho\left({n_{\rm B}}\right)\right]$ depends on the input-ancilla state via Eq.~\eqref{rho_nb_definition}, in order to understand the ultimate limit of noise sensing precision, we need to maximize $\calJ\left[\hat \rho\left({n_{\rm B}}\right)\right]$ over all $2M$-mode general quantum states $\ket{{\psi}_0}_{AS}$, subject to the total photon number constraint of $M N_{\rm S}$ on the input system $A$. This is in general a challenging task, as the states can be arbitrary and entangled across $2M$ modes; however, we are able to obtain the following upper bound on $\calJ\left[\hat \rho\left({n_{\rm B}}\right)\right]$, via making use of the fidelity interpretation of QFI in Eq.~\eqref{eq:QFI_fidelity}. We detail the full proof based on unitary extension (UE) of channels in Appendix~\ref{app:proof}.

\begin{theorem}
\label{thereom:JUB}
The quantum Fisher information per mode for energy constrained additive noise sensing of a phase-covariant Bosonic Gaussian channel $\calN_{\kappa,n_{\rm B}}$ has the following upper bound
\be 
\calJ_{\rm UB, {UE}}=
\frac{1}{{n_{\rm B}}({n_{\rm B}}+1)}+\frac{\kappa  N_{\rm S}
   \left(2 {n_{\rm B}}-\kappa +1\right)}{{n_{\rm B}} \left({n_{\rm B}}+1\right)^2
   \left({n_{\rm B}}-\kappa +1\right)}\,,
   \label{eq:QFI_UB}
\ee
where $N_{\rm S}$ is the input mean photon number per mode.
Furthermore, the upper bound is additive: $\calJ\left[\hat \rho\left({n_{\rm B}}\right)\right]\le M \calJ_{\rm UB,UE}$ for any $2M$-mode input-ancilla state subject to mean photon number constraint $M N_{\rm S}$.
\end{theorem}

The additivity of the above upper bound can also be proven in a more general setting, where multiple channels are dependent on a global noise parameter $\theta$. Consider a compound channel $\otimes_{\ell=1}^K \calN_{\kappa_\ell, n_{{\rm B},\ell}(\theta)}$, the noise of each sub-channel $n_{{\rm B},\ell}(\theta)$ is a general smooth function of $\theta$. Suppose one utilizes $N_{{\rm S},\ell}$ mean photon number on each sub-channel, the total Fisher information about $\theta$ is upper bounded by
\be 
\calJ^{\rm UB, {UE}}_\theta=\sum_{\ell=1}^K \left[\partial_\theta n_{{\rm B},\ell}\right]^2 \calJ_{\rm UB,{UE}}(N_{{\rm S},\ell},\kappa_\ell,n_{{\rm B},\ell}),
\label{J_UB_additivity}
\ee 
where $\calJ_{\rm UB}(N_{\rm S},\kappa,n_{\rm B})$ makes the functional dependence explicit in Eq.~\eqref{eq:QFI_UB}. The detailed proof is presented in Appendix~\ref{app:proof}. This additivity property is non-trivial as in general the inputs to different channels can be entangled.

Before proceeding to applying the upper bound, we would like to compare with some known results. In Ref.~\cite{pirandola2017ultimate}, there is an upper bound from teleportation (TP)-stretching that holds for the energy unconstrained problem of noise estimation. In our notations, it provides an upper bound 
\be 
\calJ_{\rm UB,TP}=\frac{1}{n_{\rm B}(n_{\rm B}+1-\kappa)},
\label{eq:QFT_UB_TP}
\ee 
which holds true for arbitrary values of $N_{\rm S}$. In our analyses, we will obtain the best upper bound 
\be 
\calJ_{\rm UB}=\min[\calJ_{\rm UB, {UE}},\calJ_{\rm UB,TP}],
\label{eq:UB_all}
\ee 
combining both Eq.~\eqref{eq:QFI_UB} and Eq.~\eqref{eq:QFT_UB_TP}. As shown in Fig.~\ref{fig:source_opt_UB}, in the practical region of squeezing (region below the cyan dashed line), $\calJ_{\rm UB}=\calJ_{\rm UB, {UE}}$ is adopted; while in the large squeezing region (above the cyan dashed line), $\calJ_{\rm UB}=\calJ_{\rm UB,TP}$ is adopted. \QZ{It is noteworthy that the teleportation-based bound $\calJ_{\rm UB,TP}$ does not depend on the photon-number constraint $N_{\rm S}$. This is due to its derivation allows infinite energy---it is the QFI achieved by the Choi state of the channel, which is the channel output when the TMSV input becomes infinitely squeezed~\cite{pirandola2017ultimate}. Naturally, it is a loose bound for finite $N_{\rm S}$. In contrast, our unitary extension bound $\calJ_{\rm UB, {UE}}$ is much tighter for small $N_{\rm S}$, but the assumption of receiver being able to access the environment of the unitary extension makes $\calJ_{\rm UB, {UE}}$ loose at the limit of large $N_{\rm S}$: When $N_{\rm S}$ increases, eventually the environment contains too much information about the noise level; therefore, such assumed access to the environment increases the QFI drastically and makes the resulting QFI upper bound loose.}

\subsection{Performance of Gaussian sources} 

With the ultimate limit in hand, we now consider QFI enabled by different type of input-ancilla states $\ket{\psi_0}$. We consider $M$ identical probes, each with mean photon number $N_{\rm S}$, \QZ{i.e. $\expval{\hat a_S^\dagger \hat a_S}\le N_{\rm S}$}. Due to the additive nature of QFI for multi-copies, we will just consider the Fisher information for a single probe. All sources considered here are Gaussian~\cite{weedbrook2012gaussian} and the QFI can be evaluated analytically, as we detail in Appendix~\ref{app:Gaussian}. 

We begin with the $N_{\rm S}=0$ case of vacuum input. The vacuum-limit (VL) QFI can be evaluated as
\be 
\calJ_{\rm VL}=
\frac{1}{{n_{\rm B}}({n_{\rm B}}+1)}= \calJ_{\rm UB}|_{N_{\rm S}=0},
\label{eq:QFI_vl}
\ee
which also coincides with $ \calJ_{\rm UB}$ at zero input photon. In this case the performance is limited by vacuum noise fluctuations, thus, we name the corresponding QFI as the vacuum limit. As we will address later, this vacuum limit is much better than the performance of vacuum input with homodyne detection, the latter often considered as benchmark in previous works~\cite{malnou2019,backes2021}.

Now we consider exotic quantum resources to overcome the vacuum limit. We begin with the squeezed vacuum state, in absence of any entangled ancilla. Squeezed vacuum states have been considered in DM search~\cite{malnou2019}, however, the QFI enabled by it remains unclear. A squeezed vacuum state is prepared by applying single-mode squeezing $\hat S(r)=\exp\left[-r\left(\hat a_S^2-\hat a_S^{\dagger 2}\right)/2\right]$ to a vacuum mode, where $\hat a_S$ is the annihilation operator of the initial mode and $r$ is the squeezing parameter. The resulting mode has mean photon number $N_{\rm S}=\sinh^2 r$ and quadrature variances $G\equiv\exp(2r)$ and $1/G\equiv\exp(-2r)$ for position and momentum, where we have chosen vacuum variance as unity. Here $G$ is often referred to in decibels (dB) as the squeezing strength. A single-mode squeezed vacuum (SV) yields the following QFI
\begin{widetext}
\be 
\calJ_{\rm SV}=
\frac{({n_{\rm B}}+1)^2 + ({n_{\rm B}}+2 \kappa  N_{\rm S})^2+
2 \kappa  N_{\rm S} \left(\kappa +1\right)}{\left[\kappa  N_{\rm S} \left(2 {n_{\rm B}}-\kappa +1\right)+{n_{\rm B}} \left({n_{\rm B}}+1\right)\right] \left[2 {n_{\rm B}} \left({n_{\rm B}}+2
   \kappa  N_{\rm S}+1\right)-2 (\kappa -1) \kappa  N_{\rm S}+1\right]}.
   \label{eq:QFI_SQZ}
\ee
\end{widetext}
First, as a sanity check, with zero mean photon number $N_{\rm S}=0$, the QFI result $\calJ_{\rm SV}=
1/{{n_{\rm B}}({n_{\rm B}}+1)}=\calJ_{\rm VL}$ agrees with the vacuum limit.
\begin{figure}[b]
    \centering
    \includegraphics[width=\linewidth]{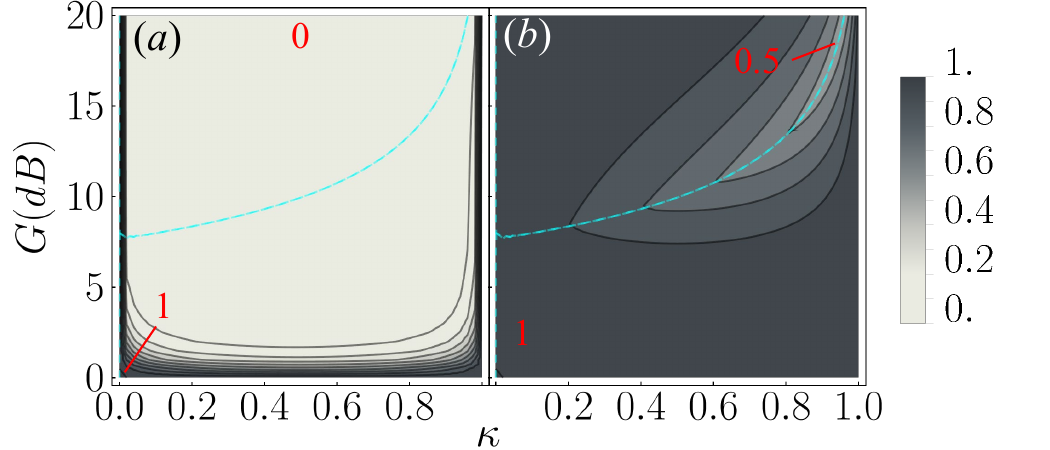}
    \caption{Quantum Fisher information of squeezed vacuum and TMSV sources, normalized by the ultimate limit $\calJ_{\rm UB}$.
    \QZ{(a) $\calJ_{\rm SV}/\calJ_{\rm UB}$;
    (b) $\calJ_{\rm TMSV}/\calJ_{\rm UB}$.}
    Cyan dashed: $\calJ_{\rm UB, {UE}}=\calJ_{\rm UB, {TP}}$\QZ{; $\calJ_{\rm UB, {UE}}>\calJ_{\rm UB, {TP}}$ for larger $G$, namely one adopts ${{{{\mathcal{J}}}}}_{{{{\rm{UB}}}}}=\calJ_{\rm UB, {TP}}$ above the cyan line, ${{{{\mathcal{J}}}}}_{{{{\rm{UB}}}}}=\calJ_{\rm UB, {UE}}$ otherwise}. 
    The values of maximum and minimum in each subplot are highlighted in red. The range between the adjacent two contours is 0.1. $n_{\rm B}=10^{-3}$. As a reminder, $G=e^{2r}=\exp{2\sinh^{-1}\sqrt{N_{\rm S}}}$. 
    \label{fig:source_opt_UB}
    }
\end{figure}
\begin{figure}[b]
    \centering
    \includegraphics[width=\linewidth]{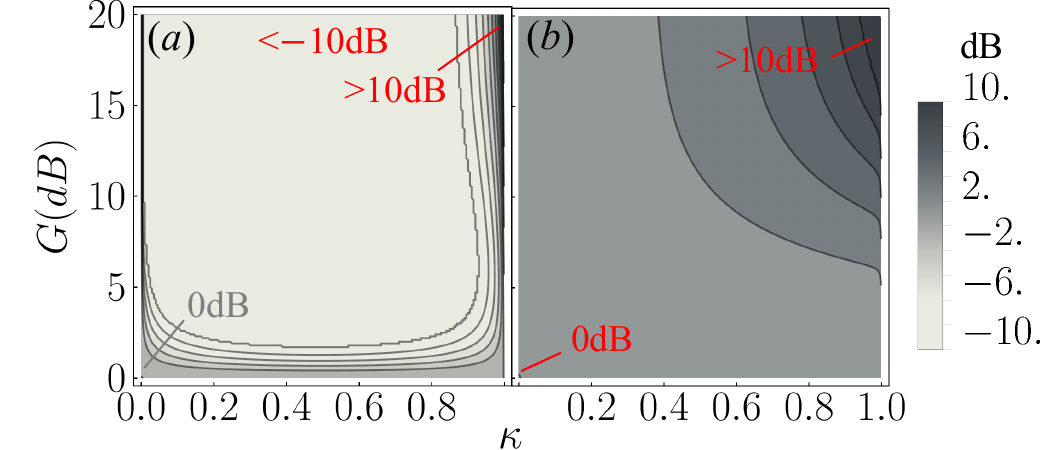}
    \caption{Quantum Fisher information of squeezed vacuum and TMSV sources, normalized by the standard quantum limit $\calJ_{\rm VL}$.
    \QZ{
    (a) $\calJ_{\rm SV}/\calJ_{\rm VL}$;
    (b) $\calJ_{\rm TMSV}/\calJ_{\rm VL}$.
    }
    The values of maximum and minimum in each subplot are highlighted in red. The range between the adjacent two contours is \QZ{2 decibel}. 
    $n_{\rm B}=10^{-3}$. Note that in larger $n_{\rm B}$ region neither source has much of advantage compared to vacuum. $G=e^{2r}=\exp{2\sinh^{-1}\sqrt{N_{\rm S}}}$. 
    \label{fig:source_opt_VL}
    }
\end{figure}
While when the input mean photon number $N_{\rm S}\gg1$ is large, $\calJ_{\rm SV}\simeq 2/(1-\kappa+2n_{\rm B})^2$ converges to a finite value; From the above, we see that single-mode squeezing can even worsen the performance when $n_{\rm B}\ll1$ in the lossy case of $\kappa<1$. While at the lossless case of $\kappa=1$, the squeezed state QFI $\calJ_{\rm SV}=\calJ_{\rm UB}$ achieves the upper bound when ${n_{\rm B}}\ll \min[1,1/N_{\rm S}]$. The above performance from asymptotic analyses can be verified in an example in Fig.~\ref{fig:source_opt_UB} (a) and Fig.~\ref{fig:source_opt_VL}(a), where we plot the relative ratio to the upper bound and the vacuum limit. Overall, we see close-to-optimal performance of single-mode squeezed vacuum only when $\kappa \sim 1$ is close to the lossless limit. Note that the optimality at $\kappa\sim 0$ is trivial and input-independent.

To further improve the performance, we consider entanglement-assisted strategies, where one stores an ancilla $A$ entangled with the input signal $S$ and jointly measure the signal and the ancilla for noise estimation. In this work, we consider entanglement in the form of TMSV, which are readily available in both optical domain and microwave domain. A TMSV state can be prepared by applying the two mode squeezing $\hat S_2(r)=\exp\left[-r\left(\hat a_S\hat a_A-\hat a_S^\dagger \hat a_A^\dagger\right)\right]$ to two vacuum modes. After the two-mode squeezing, the signal mode has mean photon number $N_{\rm S}\equiv \expval{\hat a_S^\dagger \hat a_S}=\sinh^2 r$. Similar to the single-mode squeezing case, we define the squeezing strength $G=\exp(2r)$, as such a TMSV state becomes two independent squeezed states of strength $G$ via passing through a balanced beamsplitter. The QFI for noise estimation enabled by TMSV can be evaluated as
\be 
\calJ_{\rm TMSV}=
\frac{(2 {n_{\rm B}}-\kappa+1) N_{\rm S}+{n_{\rm B}}-\kappa+1}{{n_{\rm B}} \left({n_{\rm B}}-\kappa +1\right) \left[ (2 {n_{\rm B}}-\kappa+1) N_{\rm S}+{n_{\rm B}}+1\right]}.
\label{eq:QFI_TMSV}
\ee
Different from single-mode squeezing, we find that TMSV always overcomes the vacuum limit:
\be 
\frac{\calJ_{\rm TMSV}}{\calJ_{\rm VL}}=1+\frac{\kappa  N_{\rm S} \left(2 n_{\rm B}+1-\kappa \right)}{\left(n_{\rm B}+1-\kappa \right) \left[(2 n_{\rm B}+1-\kappa )
   N_{\rm S}+n_{\rm B}+1\right]}\,.
\label{TMSV_VL_ratio}
\ee
In particular, for the ideal lossless scenario of $\kappa=1$, the TMSV source can be proven to be optimal in the weak noise limit, namely $\calJ_{\rm TMSV}/\calJ_{\rm UB}\simeq 1$ when ${n_{\rm B}}\ll \min[1,1/N_{\rm S}]$. Furthermore, the TMSV source achieves the teleportation bound exactly at the limit of large squeezing, $N_{\rm S}\to \infty$.
We verify the above conclusions numerically. In Fig.~\ref{fig:source_opt_UB}(b), we indeed see the ratio $\calJ_{\rm TMSV}/\calJ_{\rm UB}$ is close to unity in most of the parameter space, not limiting to $\kappa=1$. In Fig.~\ref{fig:source_opt_VL}(b), we see that the TMSV source yields an appreciable advantage over the vacuum limit, which survives in the entire parameter region and is largest in the high squeezing and high transmissivity region, as expected from Eq.~\eqref{TMSV_VL_ratio}.

\subsection{Measurement protocols on Gaussian sources}

Now we consider the measurement to achieve the QFI for the various types of input quantum states.

\begin{figure}[t]
    \centering
    \includegraphics[width=0.4\textwidth]{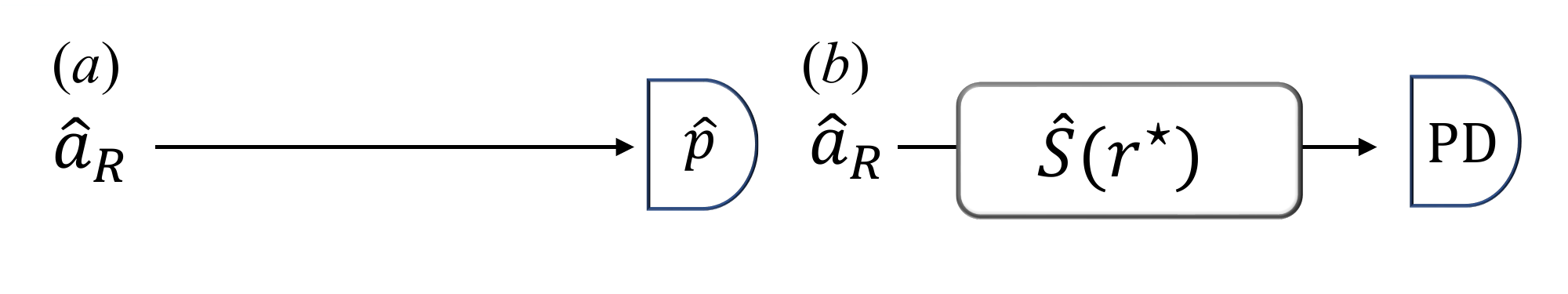}
    \caption{Single-mode squeezing-assisted measurement strategies. (a) Homodyne measurement; (b) nulling receiver based on single-mode squeezing $\hat S(r^\star)$ and photon detection (PD).}
    \label{fig:SQZmeasurement}
\end{figure}

When the input is vacuum, the output state is a thermal state with mean photon number $n_{\rm B}$, which is a photon-number diagonal state. Consequently, the vacuum limit can be achieved by a photon-counting measurement. 
Note that in this case homodyne detection on vacuum input is strictly sub-optimal, with the Fisher information
\be 
\calI_{\rm Vac-hom}=\frac{2}{(1+2{n_{\rm B}})^2}<\calJ_{\rm VL}.
\label{Fisher_vac_homo}
\ee 
And the performance degradation from the vacuum limit is large when ${n_{\rm B}}$ is small, as $\calI_{\rm Vac-hom}/\calJ_{\rm VL}\sim 2{n_{\rm B}}$ in such a weak noise limit. This vacuum-homodyne performance is often regarded as the `standard quantum limit' in the literature~\cite{dixit2021,malnou2019,backes2021}. The quantum optimum vacuum limit $\calJ_{\rm VL}$ has an infinite-fold advantage over the vacuum homodyne $\calI_{\rm Vac-hom}$ as $n_{\rm B}\to 0$. Via this Fisher information analyses, we make the advantage of photon counting proposed in Ref.~\cite{dixit2021} rigorous.

Now we proceed to consider measurement for single-mode squeezed vacuum input. We propose two strategies shown in Fig.~\ref{fig:SQZmeasurement}. First, let us begin with the homodyne measurement shown in Fig.~\ref{fig:SQZmeasurement}(a). As we detail in Appendix~\ref{app:SVhom}, a simple homodyne detection on the squeezed quadrature (here the momentum quadrature $\hat p$) of a single-mode squeezed vacuum state provides the Fisher information
\be 
\calI_{\rm SV-hom}=\frac{2}{\left(2 n_{\rm B}+2 \kappa  \left(N_{\rm S}-\sqrt{N_{\rm S} \left(N_{\rm S}+1\right)}\right)+1\right)^2}\,.
\label{Fisher_SV_homo}
\ee 
Note that the protocol of squeezing followed up with anti-squeezing in the HAYSTAC experiment~\cite{backes2021} in theory yields the same Fisher information as direct homodyne (see Appendix~\ref{app:SVhom}). In the HAYSTAC experiment, anti-squeezing is applied to make the signal robust against additional detection noises. 

\begin{figure}[t]
    \centering
    \includegraphics[width=0.4\textwidth]{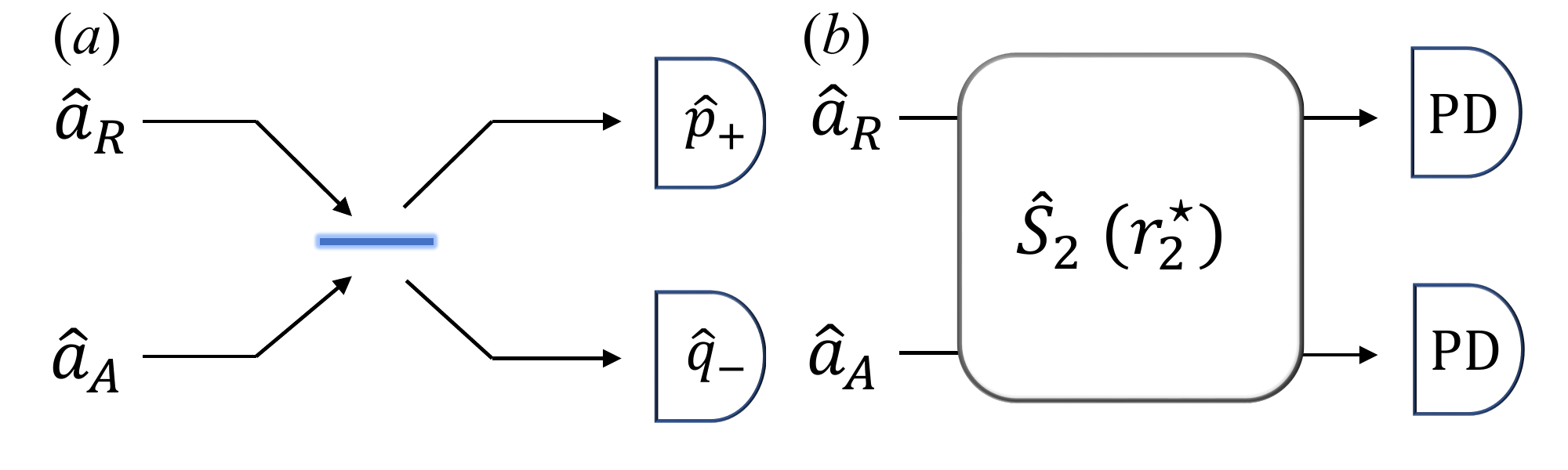}
    \caption{Entanglement-assisted measurement strategies. (a) Bell measurement; (b) nulling receiver based on two-mode squeezing $\hat S_2(r_2^\star)$ and photon detection (PD).}
    \label{fig:EAmeasurement}
\end{figure}

\begin{figure}[t]
    \centering
    \includegraphics[width=0.5\textwidth]{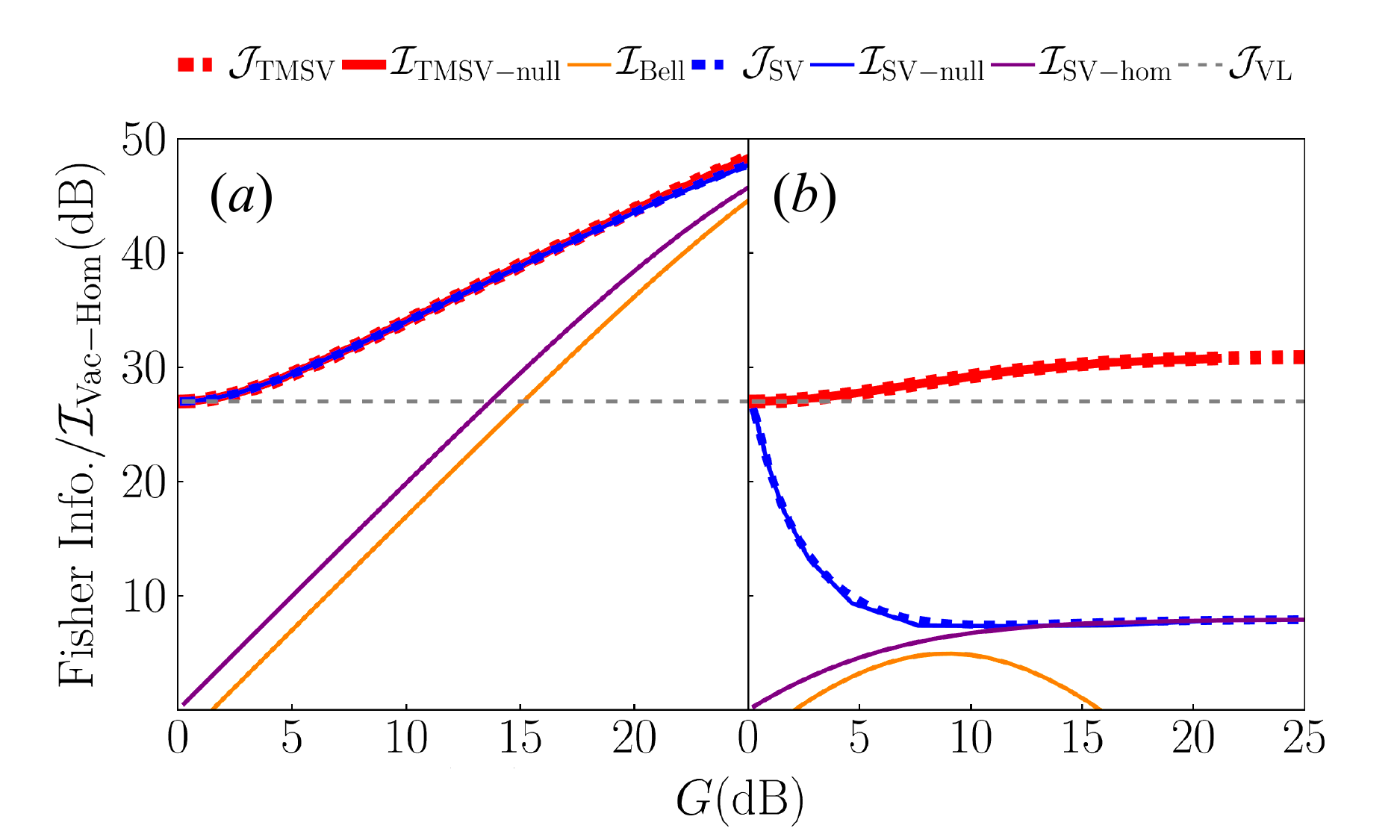}
    \caption{Comparison of the Fisher information of practical measurements (solid) with the quantum limits (dashed), normalized by the standard Fisher information of homodyne on vacuum input. (a)$\kappa=1$. (b)$\kappa=0.6$. Both axes are plotted in the decibel (dB) unit. $n_{\rm B}=10^{-3}$. As a reminder, $G=e^{2r}=\exp{2\sinh^{-1}\sqrt{N_{\rm S}} }$. }
    \label{fig:FI_null}
\end{figure}

Indeed, assuming homodyne detection, squeezed vacuum input provides a better performance, $\calI_{\rm SV-hom}\ge \calI_{\rm Vac-hom}$, with equality achieved at $N_{\rm S}=0$ as expected.
When the squeezing strength $G$ is limited, the bottleneck is the homodyne measurement that fails to achieve the full potential of the squeezed vacuum source.
With unlimited energy budget such that $G\to\infty$, we have $\calI_{\rm SV-hom}=2/(2n_{\rm B}+1-\kappa)^2$ which converges to the squeezed-vacuum quantum limit $\calJ_{\rm SV}$. However, compared with the vacuum limit $\calJ_{\rm VL}$ in Eq.~\eqref{eq:QFI_vl}, $\calI_{\rm SV-hom}$ for squeezing-homodyne is only advantageous when $\kappa$ is very close to unity. Fig.~\ref{fig:FI_null} confirms the gap between the $\calI_{\rm SV-hom}$ (purple) and the QFI of squeezed vacuum source $\calJ_{\rm SV}$ (blue dashed), also the promised advantage of $\calJ_{\rm SV}$ over the vacuum case $\calI_{\rm Vac-hom}$.

To exploit more advantage from single-mode squeezing, as shown in Fig.~\ref{fig:SQZmeasurement}(b) we design a nulling receiver which is proven to be optimum at the ${n_{\rm B}}\to 0$ limit. Specifically, the receiver first aims to null the return mode for squeezed-state sources by performing an anti-squeezing $\hat S(r^\star)$ with $r^\star=-\sinh^{-1}(\sqrt{N_{\rm S}})$. In experiments, such an anti-squeezing can be realized via optical parametric amplification. Indeed, it successfully nulls the return mode to vacuum for an identity channel, while it leaves residue noise for a general BGC. At this moment, the photon count is subject to the probability distribution in the Legendre function, which yields Fisher information in Eq.~\eqref{eq:FI_Sqznull_app}. 
At the identity-channel limit $n_B\to 0, \kappa\to 1$, we find that the nulling receiver achieves the SV limit (see Appendix~\ref{app:measurement})
\bal  
\calI_{\rm SV-null}\simeq \frac{(1+2N_{\rm S})^2}{(1-\kappa)N_{\rm S}} \simeq\calJ_{\rm SV}.
\eal 
As shown in Fig.~\ref{fig:FI_null}, our numerical results of the nulling receiver (blue) achieves the optimal performance allowed by the squeezed vacuum source $\calJ_{\rm SV}$ (blue dashed) in both $\kappa=1$ (subplot a) and $\kappa=0.6<1$ (subplot b) cases. The nulling receiver secures the optimal advantage over the vacuum limit (gray dashed) when $\kappa=1$, whereas the squeezed state source per se fails to beat the vacumm limit when $\kappa=0.6$.

In the entanglement-assisted case, the receiver has access to both the ancilla $A$ and the return $R$. We propose two measurement schemes shown in Fig.~\ref{fig:EAmeasurement}, one is based on Bell measurement [subplot (a)], the other is an extension of the nulling receiver proposed above [subplot (b)].
To begin with, we introduce the Bell measurement, where one performs homodyne detection after passing the return mode $\hat{a}_R$ and the ancilla mode $\hat{a}_A$ through a balanced beamsplitter. The Bell measurement on the TMSV input yields the classical Fisher information
\be 
\calI_{\rm Bell}=\frac{1}{\left({n_{\rm B}}+\kappa  N_{\rm S}-2 \sqrt{\kappa  N_{\rm S} \left(N_{\rm S}+1\right)}+N_{\rm S}+1\right)^2},
\label{I_Bell}
\ee
which is sub-optimal in general. In particular, at the limit $n_{\rm B}\to 0$, one can analytically show that $\calI_{\rm Bell}$ is worse than $\calI_{\rm SV-hom}$, the resolution of single-mode squeezing and homodyne. This is confirmed in Fig.~\ref{fig:FI_null}. In the $\kappa=1$ case of subplot (a), $\calI_{\rm Bell}$ (orange) is constantly 3dB worse than $\calI_{\rm SV-hom}$ (purple); in the $\kappa<1$ case of subplot (b), $\calI_{\rm Bell}$ drops after the squeezing strength $G$ surpassed a threshold, as expected from Eq.~\eqref{I_Bell}.

By contrast, the nulling receiver, now based on two-mode-anti-squeezing, is again near optimum at the ${n_{\rm B}}\to 0$ limit for TMSV input. Specifically, the receiver aims to null the returned signal mode to vacuum for TMSV sources by $\hat S_2(r_2^\star)$ with $r_2^\star=-\sinh^{-1}(\sqrt{\kappa  N_{\rm S}/[(1-\kappa)  N_{\rm S}+1]})$. The return mode is nulled to vacuum over a pure loss channel at the limit $n_{\rm B}\to 0$ (which does not work for amplifier channels or $n_{\rm B}>0$). At this moment, the joint photon count statistics at the signal and ancilla modes can be analytically solved, which yields Fisher information Eq.~\eqref{eq:FI_TMSVnull_app}. At the low-noise limit $n_B\to 0$, we find that the nulling receiver achieves the TMSV limit (see Appendix~\ref{app:measurement})
\bal  
\calI_{\rm TMSV-null}\simeq \frac{1+N_{\rm S}}{[1+N_{\rm S}(1-\kappa)]n_{\rm B}}\simeq\calJ_{\rm TMSV}\,.
\label{TMSV_null}
\eal 
We numerically evaluate it in Fig.~\ref{fig:FI_null}. In a wide range of $G$, the nulling receiver (red solid) is shown to achieve the QFI of the TMSV state source (red dashed). Remarkably, when $\kappa<1$, the EA nulling receiver yields an appreciable advantage over the quantum limit of the single-mode squeezed-state source (blue dashed).

\QZ{In the above, we have considered photon counting on both the signal and ancilla. }
\QZ{It is noteworthy that the optimality still holds if the receiver only measures the signal, 
$  
\calI_{\rm TMSV-null, signal}\simeq 
\calJ_{\rm TMSV} \,;
$
while if one only measures the ancilla, the performance is much worse, $\calI_{\rm TMSV-null, ancilla}\simeq {\kappa ^2 N_{\rm S}}/{[(1-\kappa ) \left(1+(1-\kappa ) N_{\rm S}\right)^3]}$.
}
\QZ{On the other hand, the actual implemented nulling parameter $r_2$ in experiments can deviate from our proposed value $r_2^\star$. In Appendix~\ref{app:nullRx}, we numerically compare the measure-both strategy and the measure-signal strategy---the measure-both strategy is much more robust than the latter against the deviation.}

Overall, in the noise sensing scenario, nulling receivers based on (single-mode or two-mode) squeezing and photon counting performs much better than the quadrature measurements (homodyne for single-mode squeezing and Bell measurement for two-mode squeezing). Remarkably, even the classical vacuum source yields an achievable advantage up to $\sim 30$dB with the assistance of photon number resolvable measurement (see gray dashed lines in Fig.~\ref{fig:FI_null}). This is in contrast to the phase sensing scenario, as in noise sensing the photon number carries the information, while in phase sensing the quadratures carry the information .

{\em Notes added.---} Upon the completion of our manuscript, a related work~\cite{gorecki2022} appeared. There, a different model of displacement statistics is taken and the anti-squeezing and photon counting strategy for single-mode squeezed vacuum source is proposed, while no entanglement assistance is considered. Ref.~\cite{gorecki2022} considers no additional loss or noise and therefore does not directly apply to dark matter haloscopes, where the predominant scenario is lossy for any off-resonance detuning or any detuning under unbalanced coupling, as we will explain in the next section.

\subsection{Implications on axion dark matter search}

Now we focus on the axion DM search with microwave cavity haloscopes and analyze the performance boost in more details. To maximize the initial search efficienty, the typical cavity linewidth is much larger than the predicted bandwidth of axion DM, thus an axion signal can be considered monochromatic~\cite{berlin2022searches}. \QZ{The formal input-output relation can be found in Eq.~\eqref{eq:a_in_a_out}.}
\QZ{In this paper, we will frequently use the normalized coupling rates $\tilde{\gamma}_m\equiv \gamma_m/\gamma_\ell $ and $\tilde{\gamma}_s\equiv \gamma_a/\gamma_\ell$. }In the formalism of bosonic Gaussian channel, here the input probe at detuning $\omega$ is subject to transmissivity $\kappa(\omega)=\chi_{mm}^2(\omega)$. From Eq.~\eqref{eq:a_in_a_out}, the noise $n_{\rm B}$ has contribution from both the environment thermal bath and the DM perturbation. \QZ{As the contribution from DM is $\chi_{ma}^2 n_{\rm a}$,} we can relate the Fisher information $\calJ_{n_{\rm a}}$ about DM density $n_{\rm a}$ and the Fisher information $\calJ_{n_{\rm B}}$ about bosonic channel additive noise $n_{\rm B}$ via the parameter change rule, $\calJ_{n_{\rm a}}=\chi_{ma}^4\calJ_{n_{\rm B}}$.
The classical Fisher information achieved by measurement, denoted as $\calI$, can also be related similarly.

Now we consider the dark matter search process in more detail.
As the center frequency of axion is unknown, one has to search through the whole frequency domain with a uniform prior. Consider a search protocol consisting of $2n+1$ measurements with large $n$, where the cavity resonance frequency is tuned such that the detuning $\omega$ to a fixed frequency covers the range of $[-n\Delta\omega,n\Delta\omega]$ with a small discrete step $\Delta\omega$. Note that Fisher information is additive for independent measurements (c.f. joint quantum measurement), the total Fisher information about axion DM occupation number at the fixed frequency is the summation of the Fisher information at each measurement,
$
\sum_{k=-n}^{n}\calJ_{n_{\rm a}}(k\Delta\omega)\,.
$ 
Take the continuous limit of $\Delta\omega\to 0$, $n\to\infty$, the total Fisher information $\lim_{\Delta\omega\to 0}\frac{1}{\Delta\omega} \sum_{k=-\infty}^{\infty}J_{n_{\rm a}}(k\Delta\omega) \Delta \omega$ is proportional to the continuous-spectrum total Fisher information~\cite{polloreno2022opportunities}
\bal 
  \mathbb{J}\equiv \int_{-\infty}^\infty {\rm d}\omega\, \calJ_{n_{\rm a}}(\omega)\,,
  \label{J_total_define}
\eal 
up to a constant prefactor of $1/\Delta\omega$.
This prefactor shall not lead to divergence in practice because the scanning step is finite. Our approximation of the sum to the integral is valid as long as the susceptibility functions $\chi_{ma}(\omega)$, $\chi_{mm}(\omega)$ are smooth enough relative to the discrete step $\Delta\omega$, which is indeed the case in DM scan~\cite{brady2022entangled}.
The same procedure applies to define the continuous-spectrum total classical Fisher information for a particular measurement protocol as 
\be 
\mathbb I=  \int_{-\infty}^\infty {\rm d}\omega\, \calI_{n_{\rm a}}(\omega).
\label{I_total_define}
\ee 

As we will show later, considering homodyne measurement, the Fisher information $\calI_{n_{\rm a}}$ (see Eq.~\eqref{I_na_vac_homo} for noisy vacuum input and Eq.~\eqref{eq:FisherSpectrum_SVhomo} for noisy squeezed vacuum) is equivalent to the squared signal visibility $\alpha^2(\omega)$~\cite{malnou2019,brady2022entangled}, up to a constant factor. This Fisher information interpretation of the scan rate allows us to obtain insights to the DM scan rate from the total Fisher information $\mathbb I$. At the same time, $\bbJ$ therefore characterizes the quantum limit of the scan rate given a specific input source and its upper bound will bring the ultimate limit of the DM scan rate.

\subsubsection{Upper bound on DM scan rate}
To obtain an upper bound on the performance of DM search, we assume that the input port can be arbitrarily engineered, without being affected by any additional thermal noise that often appears in experiments. 

In general, we will utilize Eq.~\eqref{eq:UB_all} to obtain the upper bound. However, as most of our evaluations are with limited $N_{\rm S}$ of 10 or 20 dB of squeezing, we will focus on Eq.~\eqref{eq:QFI_UB} to obtain analytical solution, while our numerical evaluation utilizes Eq.~\eqref{eq:UB_all}. For convenience, we will denote all results as `UB' without specifying `TP' or `UE'. From Eq.~\eqref{eq:QFI_UB} and the parameter change rule, the Fisher information upper bound about $n_{\rm a}$ can be derived as
\begin{align}
&\calJ_{n_{\rm a}}^{\rm UB}=
\nonumber
\\
&\chi_{ma}^4
\Bigg[\frac{1}{{n_{\rm B}}({n_{\rm B}}+1)}+\frac{\chi_{mm}^2  N_{\rm S}
   \left(2 {n_{\rm B}}-\chi_{mm}^2 +1\right)}{{n_{\rm B}} \left({n_{\rm B}}+1\right)^2
   \left({n_{\rm B}}-\chi_{mm}^2 +1\right)}\Bigg]\,,
   \label{eq:QFI_UB_na}
\end{align}
where we have not made the frequency dependence on $\omega$ explicit and ${n_{\rm B}}$ can be taken as $(1-\chi_{mm}^2)n_{T}$ due to the axion signal being weak. 

Thanks to the additivity property of Eq.~\eqref{J_UB_additivity}, the total Fisher information upper bound can be directly obtained through integration, 
$
\mathbb{J}_{\rm UB}= \int {\rm d}\omega \calJ_{n_{\rm a}}^{\rm UB}(\omega)$. Note that this upper bound is general---it allows arbitrary entanglement across all frequencies. While the closed form solution is lengthy (see Appendix~\ref{app:QFI_UB_total}), at the low temperature limit of $n_T\ll1$, the scan rate upper bound has a simple form
\begin{align}
\mathbb{J}_{\rm UB}&\simeq 2\pi \gamma_l \frac{\tilde{\gamma}_m\tilde{\gamma_a}^2}{n_T}\left[\frac{1}{1+\tilde{\gamma}_m}+\frac{N_{\rm S}\left(1+\tilde{\gamma}_m^2\right)}{\left(1+\tilde{\gamma}_m\right)^3}\right].
\label{J_total_UB_asym}
\end{align}
The maximum of $\mathbb{J}_{\rm UB}$ can be obtained as
\be 
\mathbb{J}^\star_{\rm UB }=2\pi \gamma_l \tilde{\gamma_a}^2 \frac{1+n_T+N_{\rm S}(1+2n_T)}{n_T(1+n_T)},
\label{eq:ScanRateUB_opt}
\ee 
at the over coupling limit of $\tilde{\gamma}_m\to\infty$. \QZ{It reveals an increasing quantum advantage proportional to $N_{\rm S}$ over the vacuum limit ($N_{\rm S}=0$).}

With the upper bound in hand, in the following we consider the performance with different sources and measurements. 
As homodyne measurement using vacuum probes is prevalent in the experiment proposals at the current stage~\cite{malnou2019,backes2021,brady2022entangled}, we take it as a benchmark for classical schemes, which is to be surpassed by the non-classical probes and receivers. We will focus on parameters on par with the experiment reported in Ref.~\cite{backes2021}, where the cavity is cooled to $61$mK and the cavity resonant frequency is at around $10$GHz.

\subsubsection{Ideal input engineering}
\QZ{Here we focus on} the ideal input engineering case where the input port is not affected by thermal noise before the probing. For these cases, all of our previous results can be directly translated to DM search with the parameter change rule of Fisher information---the Fisher information about $n_{\rm a}$ at detuning $\omega$ is
\be 
F_{n_{\rm a}}(\omega)=\chi_{ma}^4(\omega) F_{{n_{\rm B}}}\left[\chi_{mm}^2(\omega), (1-\chi_{mm}^2)n_{T}, N_{\rm S} \right],
\label{eq:QFI_freq_perfect}
\ee
where $F_{n_{\rm B}}[\kappa,n_{\rm B},N_{\rm S}]$ is the corresponding Fisher information about additive noise $n_{\rm B}$ for the channel $\calN_{\kappa,n_{\rm B}}$ when the signal mean photon number is $N_{\rm S}$.

\begin{figure}[t]
    \centering
    \includegraphics[width=0.45\textwidth]{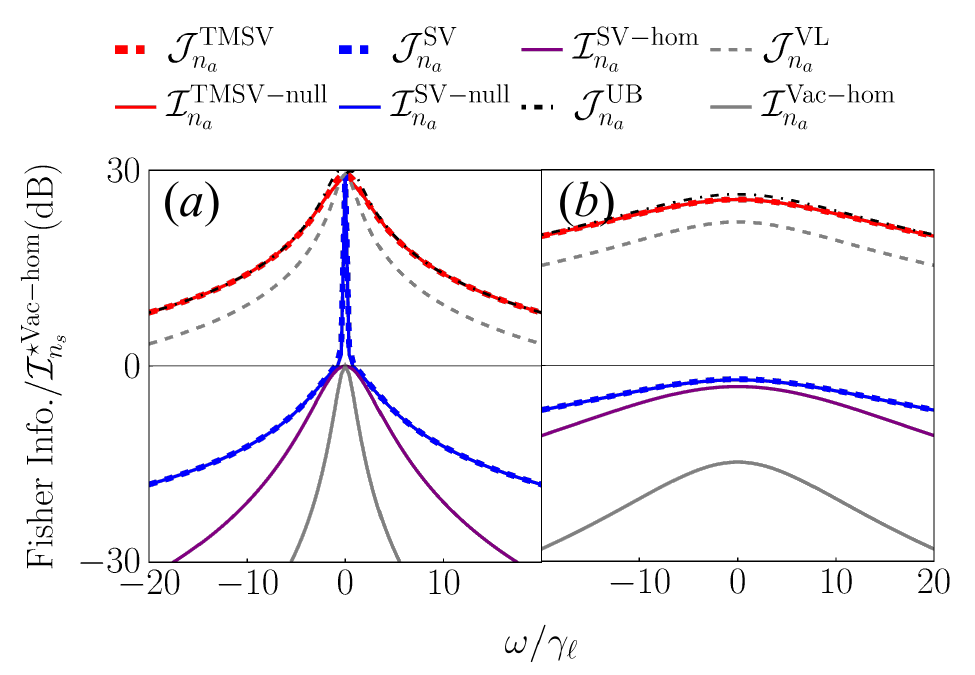}
    \caption{ Frequency spectrum of the Fisher information with respect to the axion occupation number $n_{\rm a}$, normalized by the optimized peak value of vacuum-homodyne Fisher ${\calI^\star}_{n_{\rm a}}^{ \rm Vac-hom} $and plotted in decibel unit. (a) $\gamma_m/\gamma_\ell=1$; (b) $\gamma_m/\gamma_\ell=2G$. Temperature $T=61mK$, cavity resonant frequency $\omega_c=2\pi\cdot 10$GHz, squeezing strength $G=10$dB, $\gamma_a/\gamma_\ell=10^{-12}$.  }
    \label{fig:FisherSpectrum_idealDM}
\end{figure}

Applying the above relation to Eq.~\eqref{eq:QFI_vl} produces the vacuum limit of DM search $\calJ^{\rm VL}_{n_{\rm a}}$ and
Eq.~\eqref{Fisher_vac_homo} produces the vacuum homodyne performance
$
\calI_{n_{\rm a}}^{\rm Vac-hom}(\omega)$, which is the classical benchmark commonly referred to. With single-mode squeezing, applying the relation to Eq.~\eqref{eq:QFI_SQZ} produces the single-mode squeezed vacuum performance limit $\calJ^{\rm SV}_{n_{\rm a}}$ and Eq.~\eqref{Fisher_SV_homo} produces the squeezing-homodyne performance $\calI^{\rm SV-hom}_{n_{\rm a}}$. For entanglement-assisted strategies, applying the relation to Eq.~\eqref{eq:QFI_TMSV} produces the TMSV performance limit $\calJ^{\rm TMSV}_{n_{\rm a}}$. Similar results also apply to the performance of the nulling receivers for both the single-mode squeezing input and TMSV input.

We evaluate the spectrum of the Fisher information for various probes and receivers in Fig.~\ref{fig:FisherSpectrum_idealDM}. Here the Fisher information quantities are normalized by ${\calI^\star}_{n_{\rm a}}^{ \rm Vac-hom}\equiv\calI_{n_{\rm a}}^{\rm Vac-hom}(\omega=0)|_{\gamma_m=\gamma_m^\star}$, the peak Fisher information of the classical benchmark at the critical coupling ratio $\gamma_m^\star= \gamma_\ell$. In general, we see Lorentzian-type of envelops due to the $\chi_{ma}^4(\omega)$ term in Eq.~\eqref{eq:QFI_freq_perfect}. In subplot (a) \QZ{ of the critical coupling case ($\gamma_m/\gamma_\ell=1$), the transmissivity $\kappa(\omega)= 0$ at resonance $\omega=0$, and any input trivially converges to vacuum. When $\omega$ deviates from resonance, generally the Fisher information deviates from the peak value (peak sensitivity). Interestingly, the single-mode squeezed vacuum performance limit $\calJ_{n_{\rm a}}^{\rm SV}$ (blue dashed) is worse than the vacuum limit (gray dashed) for the range of detuning $\omega$ of interest, due to the loss in the probing. This can be intuitively understood by considering the anti-squeezed quadrature contributing greatly to noise while the squeezed quadrature is still almost vacuum at the large loss limit. This is also seen in Fig.~\ref{fig:source_opt_VL}(a), where the ratio to vacuum limit also sharply decay when $\kappa$ increases near $\kappa=0$.}
By contrast, the TMSV state (red dashed) demonstrates a huge advantage over the whole frequency domain. Remarkably, in the presented scenarios $\calJ_{n_{\rm a}}^{\rm TMSV}$ achieves the upper bound (black dot-dash) almost everywhere, which indicates that the TMSV is the optimal input state here. Squeezed-vacuum homodyne performance (purple solid) is better than vacuum homodyne (gray solid) but worse than the vacuum limit (gray dashed) enabled by photon counting. In subplot (b), we consider the over-coupling case and the Fisher information spectrum broadens, while the peak sensitivity at $\omega=0$ decreases as expected. In both subplots, we find the nulling receiver (blue solid and red solid) to be optimal as expected.

\begin{figure}[t]
    \centering
    \includegraphics[width=0.5\textwidth]{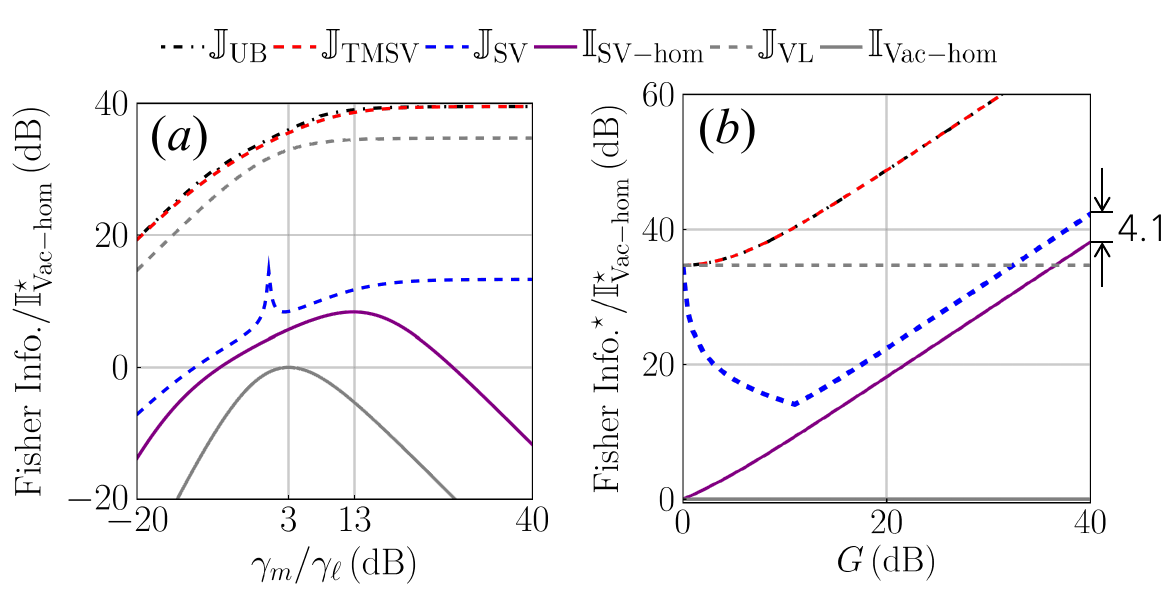}
	    \caption{ (a)Total Fisher information under various measurement coupling ratio over intrinsic loss $\gamma_m/\gamma_\ell$, squeezing strength $G=10$dB; (b) the optimized total Fisher information optimized with optimal $\gamma_m/\gamma_\ell$ under various squeezing strength $G$. Y axis normalized by $\mathbb I_{\rm Vac-hom}^\star$ as defined in Eq.~\eqref{eq:bbI_vachom_opt} \QZ{in appendix}, both axes plotted in decibel unit. \QZ{$\bbJ_{\rm TMSV}$, $\bbJ_{\rm SV}$ are achieved by nulling receivers as shown in Fig.~\ref{fig:FisherSpectrum_idealDM}.} The indicated gap \QZ{in (b)} is $10\log_{10}(3\sqrt{3}/2)\sim 4.1 $dB. Temperature $T=61mK$, cavity resonant frequency $\omega_c=2\pi\cdot 10$GHz, $\gamma_a/\gamma_\ell=10^{-12}$. }
    \label{fig:ScanRate_idealDMS}
\end{figure}

Now we proceed to analyze the scan rate from the total Fisher information. We begin with the performance enabled by homodyne detection, where closed-form solutions can be obtained. We put the lengthy expressions in Appendix~\ref{app:QFI_UB_total} and present our results for special cases or asymptotic analyses here.
For vacuum input of $N_{\rm S}=0$, we have
\be 
\mathbb{I}_{\rm Vac-hom}= 2\pi\gamma_\ell \tilde \gamma _a^2 
\frac{4 \tilde \gamma _m^2 }{\left(8 \tilde \gamma _m {n_T} +\left(1+\tilde \gamma _m\right)^2\right)^{3/2}}.
\label{I_total_vac_homo}
\ee 
When $n_T\ll1$, the optimum is achieved at $\tilde{\gamma}_m=2$ and
$ 
\mathbb{I}^\star_{ \rm Vac-hom}=2\pi\gamma_l \tilde{\gamma}_a^2 \times {16}/{27}.
$ 
As shown in Fig.~\ref{fig:ScanRate_idealDMS}(a) by the gray solid line, the optimal coupling ratio $\tilde \gamma _m\equiv \gamma_m/\gamma_l$ can be verified numerically, with peak value being zero dB due to normalization. For highly squeezed quantum source ($N_{\rm S}\gg1$) at low temperature ($n_T\ll1$)
\be 
{\mathbb I}_{\rm SV-hom}\simeq
2 \pi  \gamma_\ell\tilde\gamma _a^2 \frac{64 \tilde\gamma _m^2 N_{\rm S}^2}{\tilde\gamma _m^{3/2} \left[16 N_{\rm S} +\tilde\gamma _m\right]^{3/2}}\,.
\ee 
The optimum 
$ 
\mathbb I^\star_{\rm SV-hom}\simeq 2\pi \gamma_\ell \tilde \gamma _a^2 \times {8   N_{\rm S}}/{3^{3/2} }
$
is achieved at $\tilde{\gamma}_m\simeq 8N_{\rm S}\simeq  2G$.
Indeed in Fig.~\ref{fig:ScanRate_idealDMS}(a), we see the peak of the squeezing homodyne (purple solid) is when the coupling ratio is about $2G=13$ dB for the 10 dB squeezing. We see that squeezing homodyne provides an advantage of $\sim2.60 N_{\rm S}$ over vacuum homodyne, as we confirm in Fig.~\ref{fig:ScanRate_idealDMS}(b). Our analyses of homodyne-based strategy indeed recovers previous known results in Refs.~\cite{malnou2019,brady2022entangled}, even more precisely when we consider the extra thermal noise in the practical input source engineering case (see Appendix~\ref{app:practical_source_eng}). \QZ{We also note that the optimal scan rate increases with squeezing gain $G$ (equivalently $N_{\rm S}$) linearly. This is due to the larger effective bandwidth growing with $G$, while the peak sensitivity will saturate to a $G$ independent constant.}

Now we evaluate the performance limits. For the vacuum limit, the total Fisher information has a closed form solution,
\begin{align}
\mathbb{J}_{\rm VL}&=\int_{-\infty}^\infty {\rm d}\omega 
\calJ^{\rm VL}_{n_{\rm a}}
\\
&=2\pi \gamma_l\frac{\tilde{\gamma}_m \tilde{\gamma}_a^2}{n_T}\frac{1}{\sqrt{(1+\tilde{\gamma}_m)^2+4n_T\tilde{\gamma}_m }}
\\
&\simeq 2\pi \gamma_l \tilde{\gamma}_a^2\frac{1 }{n_T}\frac{\tilde{\gamma}_m}{1+\tilde{\gamma}_m},
\end{align}
where in the last step we considered the $n_T\ll1$ limit. 

For the single-mode squeezing performance limit, instead of presenting the lengthy closed-form result (see Appendix~\ref{app:QFI_UB_total}), we plot the results in Fig.~\ref{fig:ScanRate_idealDMS}. In subplot (a), a peak emerges for ${\mathbb{J}}_{ \rm SV}$ (blue dashed) at $\gamma_m/\gamma_\ell=1$ (0 dB), due to the peak in Fig.~\ref{fig:FisherSpectrum_idealDM} that emerges only at critical coupling. For strong squeezing $N_{\rm S}\gg 1$, the maximum total Fisher information 
\be 
\mathbb{J}^\star_{ \rm SV}=2\pi \gamma_l \tilde{\gamma}_a^2 \frac{(1+2N_{\rm S})^2}{N_{\rm S}+n_T+2N_{\rm S}n_T},
\label{eq:ScanRate_SV_overcouple}
\ee 
is achieved at the over coupling limit of $\tilde{\gamma}_m\to\infty$.
When $n_T\ll 1$, we can compare the performance of homodyne versus the limit enabled by squeezing, $\mathbb I^\star_{\rm SV-hom} / \mathbb{J}^\star_{ \rm SV} = 2/3\sqrt 3\simeq 0.385 \simeq -4.1$dB. This constant factor difference can be verified numerically in Fig.~\ref{fig:ScanRate_idealDMS}(b). At the same time, we note that it takes almost 40 dB of squeezing for the performance enabled by squeezed vacuum (purple solid and blue dashed) to reach the vacuum limit (gray dashed). This indicates the importance of a good photon counting detection in dark matter search. 

In Fig.~\ref{fig:ScanRate_idealDMS}(b), we also observe a minimum point of the total Fisher information with respect to the squeezing gain $G\equiv1+2N_{\rm S}+2\sqrt{(1+N_{\rm S})N_{\rm S}}$ for the squeezed vacuum source (blue dashed), in contrast to the monotonicity of all other sources. Indeed, the peak at $\tilde\gamma_m=1$ competes with the overcoupling limit at $\tilde\gamma_m=\infty$ as shown Appendix~\ref{app:QFI_UB_total}. After $G$ increases beyond a specific threshold, the overcoupling limit always dominates, which increases linearly with $G$. The linear growth at large $G$ verifies Eq.~\eqref{eq:ScanRate_SV_overcouple}, as $G\sim 4N_{\rm S}$ when $N_{\rm S}$ is large.

Finally, we address the total Fisher information enabled by the TMSV source,
\begin{align}
&\mathbb{J}_{\rm TMSV}=\int_{-\infty}^\infty  d\omega \calJ^{\rm TMSV}_{n_{\rm a}}(\omega)
\\
&=2\pi \gamma_l \frac{\tilde{\gamma}_m \tilde{\gamma}_a^2}{n_T} \frac{1+N_{\rm S}+n_T+2N_{\rm S} ~n_T}{(1+n_T)\sqrt{1+\tilde{\gamma}_m^2+2\tilde{\gamma}_m(1+2N_{\rm S})(1+2n_T)}}
\\
&\simeq 
2\pi \gamma_l \tilde{\gamma}_a^2 \frac{1 }{n_T} \frac{\left(1+N_{\rm S}\right)\tilde{\gamma}_m}{\sqrt{1+\tilde{\gamma}_m^2+2\tilde{\gamma}_m(1+2N_{\rm S})}},
\end{align}
where in the last step we take the low temperature limit of $n_T\ll1$.
In the over coupling limit of $\tilde{\gamma}_m\to\infty$, the maximum achieves the upper bound
\be 
\mathbb{J}^\star_{ \rm TMSV}\simeq \mathbb{J}^\star_{ \rm UB},
\label{TMSV_UB}
\ee 
as we can verify in Fig.~\ref{fig:ScanRate_idealDMS} by comparing $\mathbb{J}^\star_{ \rm TMSV}$ (red dashed) with the upper bound $\mathbb{J}^\star_{ \rm UB}$ (black dot-dash). We also note that in general TMSV performance overwhelms the single-mode squeezing by a large factor $1/n_T$. \QZ{As expected, the TMSV source yields an increasing advantage proportional to $G$ over the vacuum limit.}

\QZ{In the above, we have assumed the squeezed vacuum and two-mode squeezed vacuums can be prepared perfectsly. In practice, their preparations also have noise and we analyze this practical state engineering in Appendix~\ref{app:practical_source_eng}. In this regime, the TMSV QFI still achieves the large advantage over the vacuum limit, while it now falls below the upper bound with a constant gap of around 6.9dB.}

\section{Discussions}

In this work, we have shown that an entanglement-assisted strategy with two-mode squeezed vacuum as source and nulling receiver (anti-two-mode-squeezing+photon counting) as detector is optimal for noise sensing. In terms of dark matter search, such a strategy provides the optimal scan-rate and outperforms the single-mode squeezed vacuum and homodyne strategy by orders of magnitude. In this regard, developing a quantum-limited photon counting detector, such as that in Ref.~\cite{dixit2021} is crucial for the next generation microwave haloscopes. 

At the same time, our results reaffirm that other types of more exotic resources such as the Gottesman–Kitaev–Preskill (GKP) state~\cite{gottesman2001} are not necessary for microwave haloscopes~\cite{brady2022entangled}. For an energy constrained case, GKP states also obey the QFI upper bound, which is already achieved with squeezed vacuum states. Even when the energy constraint is relaxed, practical considerations also forbid GKP states to be worthwhile engineering in microwave haloscopes~\cite{brady2022entangled}.  

In our dark matter search model, we have not considered the case of a local array of microwave cavities~\cite{derevianko2018network,jeong2020prl,sikivie2020search,brady2022entangled}, where the dark matter induced noise at different sensors are correlated. However, as Ref.~\cite{brady2022entangled} showed, due to the correlation, the signals can be coherently combined and the problem can be reduced to a single sensor, especially for identical sensors (see Appendix~\ref{app:DQS}). The coherent combining between $M$ identical sensors will provide a $M^2$ boost to the scan rate, in addition to the quantum advantages considered here. One can adopt the protocols addressed in this work to sensor-networks via performing a passive linear network on the signal and send to all sensors, and then recombine with another passive linear network. Such sensor-network approach provides another approach of scan-rate boost without the need of a quantum-limited photon counting detector.



\begin{acknowledgements}
This material is based upon work supported by the U.S. Department of Energy, Office of Science, National Quantum Information Science Research Centers, Superconducting Quantum Materials and Systems Center (SQMS) under the contract No. DE-AC02-07CH11359.
QZ and HS acknowledge the support from National Science Foundation CAREER Award CCF-2142882 and Defense Advanced Research Projects Agency (DARPA) under Young Faculty Award (YFA)-Director’s Award, Grant No. N660012014029, National Science Foundation (NSF) Engineering Research Center for Quantum Networks Grant No. 1941583 to prove Theorem 1 and perform analyses of Section B-E.
\end{acknowledgements}



\begin{appendix}

\section{Details of formalism}
\label{app:methods}
\subsection{Bosonic Gaussian channel}

\QZ{
A phase-covariant bosonic Gaussian channel $\calN_{\kappa,n_B}$ is characterized by transmissivity/gain $\kappa$ and additive Gaussian noise $n_B$. Specifically, given a signal mode described by the annihilation operator $\hat{a}_S$, which satisfies canonical commutation relation $[\hat{a}_S^\dagger,\hat{a}_S]=1$, the annihilation operator of the return mode is given by the linear input-output relation
\be 
\hat a_R=\sqrt{\kappa}\hat a_S+\sqrt{1-\kappa}\hat a_E\,,
\ee
for $0\le \kappa< 1$, and 
\be 
\hat a_R=\sqrt{\kappa}\hat a_S+\sqrt{\kappa-1}\hat a_E^\dagger
\ee
for $\kappa>1$.
}

\QZ{
For $0\le \kappa< 1$, the channel mimics a beamsplitter, attenuates the mean of input signal mode $\hat a_S$ by $\sqrt{\kappa}$ and mixes in the environment mode $\hat a_E$ attenuated by $\sqrt{1-\kappa}$. The environment mode $\hat a_E$ is of mean thermal photon number $n_E$. Overall, the additive noise mixed into the return is $n_B=(1-\kappa)n_E$. Concretely, given a coherent-state input of mean $\alpha$, the output is in the displaced thermal state of mean $\sqrt{\kappa}\alpha$ and mean thermal photon number $n_B$.
}

\subsection{ Practical source engineering}

\QZ{
In an experimentally feasible scenario, the input is inevitably affected by thermal noise.  
To begin with, vacuum input is never perfect in an experiment. Practical vacuum input still has some weak thermal noise $n_{T}$. In this case, the output of Eq.~\eqref{eq:a_in_a_out} is a thermal state with mean photon number $\chi_{mm}^2n_{T}+(1-\chi_{mm}^2)n_{T}+\chi_{ma}^2 n_{\rm a}=n_{T}+\chi_{ma}^2 n_{\rm a}$. From Eq.~\eqref{eq:QFI_vl}, the vacuum limit for axion sensing is therefore  
\be 
\calJ^{\rm VL}_{n_{\rm a}}= \chi_{ma}^4(\omega)
\frac{1}{{n_T}({n_T}+1)}.
\label{J_na_VL}
\ee
From Eq.~\eqref{Fisher_vac_homo}, we have the performance of vacuum homodyne
\be 
\calI_{n_{\rm a}}^{\rm Vac-hom}(\omega)=\chi_{ma}^4(\omega)\frac{2}{(1+2{n_{T}})^2}.
\label{I_na_vac_homo}
\ee 
}

\QZ{
Similarly, the nonclassical sources is also affected by thermal noise. Instead of single or two-mode squeezing on vacuum, the squeezing operations are performed on thermal states with mean photon number $n_T$.
To characterize nonclassical sources, we use the squeezing strength $G$, for both the single-mode and two-mode squeezers. The input photon number $N_{\rm S}$ is contaminated by $n_T$ as $N_{\rm S}={\left[2 \left(G^2+1\right) {n_T}+(G-1)^2\right]}/{4 G}$. The upper bound Eq.~\eqref{eq:UB_all} with $N_{\rm S}$ as the mean photon number of the processed input still applies, however is much loser due to the inevitable initial noise. 
}

\QZ{
With the above input state adopting the thermal noise, the procedures for further analyses are the same as in the maintext: the squeezed sources are shined on the measurement port of the cavity, which is modelled by a phase-covariant BGC $\calN_{\chi_{mm}^2(\omega),n_{\rm B}(\omega)}$; finally the receiver measures the returned quantum states. 
}

\section{Proof of Theorem~1 of the main text} 
\label{app:proof}
\begin{proof}
Observe Eq.~\eqref{eq:QFI_fidelity} of the main text, if we can lower bound the fidelity $F\left[\hrho({n_{\rm B}}),\hrho({n_{\rm B}}^\prime) \right]$ then we will be able to upper bound the Fisher information. 
Indeed, consider the purifications $\ket{\psi({n_{\rm B}})}, \ket{\psi({n_{\rm B}}^\prime)}$ of $\hrho({n_{\rm B}}),\hrho({n_{\rm B}}^\prime)$, due to Uhlmann's theorem,
\begin{align} 
F\left[\hrho({n_{\rm B}}),\hrho({n_{\rm B}}^\prime) \right]
\ge |\expval{\psi({n_{\rm B}})|\psi({n_{\rm B}}^\prime)}|.
\label{eq:fidelity_pure}
\end{align}
Combining Eqs.~\eqref{eq:QFI_fidelity} of the main text and~\eqref{eq:fidelity_pure}, we derive an upper bound of QFI based on the overlap of any choice of purification
\be 
\calJ\le \calJ_{\rm UB}=-\frac{1}{M}4\partial^2_{{n_{\rm B}}^\prime} |\expval{\psi({n_{\rm B}})|\psi({n_{\rm B}}^\prime)}| \bigg|_{{n_{\rm B}}^\prime={n_{\rm B}}},
\label{F_UB_first}
\ee 
which is much easier to evaluate. 

To obtain the purifications, we adopt the Stinespring representation: as shown in Fig.~\ref{fig:channel}, the channel $\calN_{\kappa,n_{\rm B}}^{S\to R}$ is extended to a unitary transform $\calU_{\calN}^{SE_1E_2\to RE_1'E_2'}$ by further including two environment modes $E_1$, $E_2$ in vacuum state. By such means, the output state remains pure if the input is pure. To obtain a simple form of extension, we decompose a phase-covariant BGC $\calN_{\kappa,{n_{\rm B}}}$ to a concatenation of a quantum-limited loss and a quantum-limited amplifier as
\be 
\calN_{\kappa,{n_{\rm B}}}=\calA_{g({n_{\rm B}})}\circ \calL_{\eta({n_{\rm B}})}
\label{eq:channel_decomposition}
\ee
with $g({n_{\rm B}})=1+{n_{\rm B}}$, $\eta({n_{\rm B}})=\kappa/g({n_{\rm B}})=\kappa/(1+{n_{\rm B}})$, where $\calL_{\eta}=\calN_{\eta,0}$ and $\calA_{g}=\calN_{g,g-1}$ are the special cases of the general BGC. 
Therefore, the unitary extension also decomposes $\calU_{\calN}^{SE_1E_2\to RE_1'E_2'}=\calU_\calA^{S'E_2\to RE_2'}\circ\calU_\calL^{SE_1\to S'E_1'}$, as shown in Fig.~\ref{fig:channel}.

Now let the overall input state be $\ket{\psi_0}_{AS}\otimes \ket{\bm 0}_{E_1}\otimes \ket {\bm 0}_{E_2}$, where we have considered the $M$ channel uses, with environments in product of vacuum states. Using the decomposition of the unitary extension for each of the $M$ channel uses, the output can be expressed as
\begin{align}
&\ket{\psi({n_{\rm B}})}
\nonumber
\\
&=\sum_{\bm n,\bm \ell} \sum_{\bm k\le \bm n} \sqrt{p_{\bn}} A_{\bn-\bm k,\bm \ell} B_{\bn,\bm k}\ket{\chi_{\bn}, \bn-\bm k+\bm \ell, \bm k,\bm \ell}_{ARE_1E_2},
\label{psi_input}
\end{align} 
where the summation is over vectors with non-negative integer elements, $\bm k\le \bm n$ is element-wise and the coefficients are~\cite{ivan2011operator} 
\bal 
B_{\bn,\bm k}({n_{\rm B}})&= \prod_{j=1}^M\sqrt{ \left(1-\eta({n_{\rm B}}) \right)^{k_j} \binom{n_j}{k_j} \eta^{n_j-k_j}({n_{\rm B}}) }, \\
A_{\bn,\bm\ell}({n_{\rm B}})&=\prod_{j=1}^M \sqrt{\left(1-\frac{1}{g({n_{\rm B}})}\right)^{\ell_j} g^{-n_j-1}({n_{\rm B}}) \binom{\ell_j+n_j}{\ell_j}}\,.
\label{eq:Kraus}
\eal 
Here $\binom{a}{b}$ is the binomial coefficient $a$-choose-$b$.

\begin{figure}
    \centering
    \includegraphics[width=0.45\textwidth]{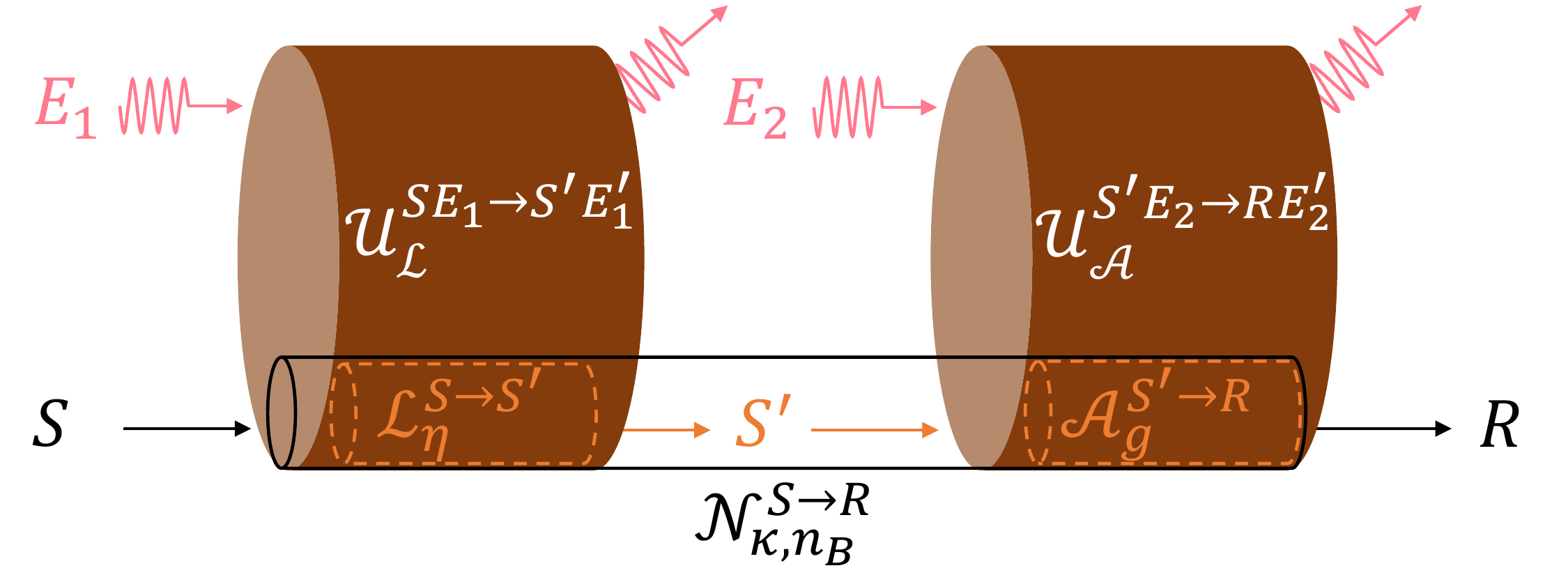}
    \caption{The Stinespring representation of channel $\calN_{\kappa,n_{\rm B}}^{S\to R}$.  In general, the unitary extension of a thermal bosonic Gaussian channel takes two environment modes $E_1,E_2$, due to the decomposition Eq.~\eqref{eq:channel_decomposition}.}
    \label{fig:channel}
\end{figure}

Combining Eq.~\eqref{psi_input} and Eq.~\eqref{eq:Kraus}, and utilizing the orthogonality of number bases, we obtain the fidelity
\be 
|\expval{\psi({n_{\rm B}})|\psi({n_{\rm B}}^\prime)}|
=\sum_{\bm n} p_{\bm n} \prod_{j=1}^M \zeta_1 \zeta_2^{ n_j}
=\sum_{n}  p_{ n} \zeta_1^M   \zeta_2^{n}
\label{fidelity_final}
\ee
here we define the distribution of total photon number $p_n=\sum_{\bm n: ||\bm n||_1=n} p_{\bm n}$ and the one-norm $||\bn||_1=\sum_j n_j$. We have also defined the parameters
\bal 
\zeta_1(n_{\rm B},n_{\rm B}')&=\left(\sqrt{\xi_{{n_{\rm B}},{n_{\rm B}}'}}-\sqrt{{n_{\rm B}} {n_{\rm B}}'}\right)^{-1} \,,\\
\zeta_2(n_{\rm B},n_{\rm B}')&=\frac{\sqrt{\nu  \nu '} \left(\sqrt{\xi_{{n_{\rm B}},{n_{\rm B}}'}}-\sqrt{{n_{\rm B}} {n_{\rm B}}'}\right)+\kappa }{\sqrt{\xi_{{n_{\rm B}},{n_{\rm B}}'}} \left(\sqrt{\xi_{{n_{\rm B}},{n_{\rm B}}'}}-\sqrt{{n_{\rm B}} {n_{\rm B}}'}\right) }\,,
\eal 
with $\nu={n_{\rm B}}-(\kappa-1)$, $\nu^\prime={n_{\rm B}}^\prime-(\kappa-1)$ and $\xi_{{n_{\rm B}},{n_{\rm B}}'}=\left({n_{\rm B}}+1\right) \left({n_{\rm B}}'+1\right)$.

With Ineq.~\eqref{F_UB_first} and Eq.~\eqref{fidelity_final}, from further simplification we have the upper bound of QFI $\calJ_{\rm UB, {UE}}$ in Eq.~\eqref{eq:QFI_UB} of the main text and the resulting additivity property. 

\paragraph*{Compound channel with heterogeneous structure.} Now we generalize the additivity to a compound channel $\otimes_{\ell=1}^K \calN_{\kappa_\ell,n_{{\rm B},\ell}(\theta) }$ with a heterogeneous structure of channel noises $\bm {n_{\rm B}}(\theta)$ that is determined by a single unknown parameter $\theta$. In this case, Eq.~\eqref{fidelity_final} stops at the first equality, because now $\zeta_1$ and $\zeta_2$ are non-identical across the channels and should also be defined accordingly as a vector $\bm \zeta_p\equiv [\zeta_p(n_{\rm B,1},n_{{\rm B},1}'), \ldots ,\zeta_p(n_{{\rm B},{K}},n_{{\rm B},{K}}') ]^T$ for $p=1,2$. However, we can show that the second derivative is linear to the photon numbers of each input mode:
\bal 
&\partial_{\theta'}^2\left[|\expval{\psi(\theta)|\psi(\theta^\prime)}|\right] |_{\theta=\theta'} 
\\
&=\sum_{\bm n} p_{\bm n} 
\partial_{\theta'}^2
\left[
\prod_{\ell=1}^K  \zeta_{1,{\ell}} \zeta_{2,{\ell}}^{ n_{\ell}}\right]|_{\theta'=\theta}\\
&=\sum_{\bm n}  p_{\bm  n}  \Bigg[ \left(\sum_{\ell=1}^K\partial_{\theta'}^2 \zeta_{1,{\ell}} 
\right)+\Bigg( \sum_{\ell=1}^K n_{\ell} \zeta_{2,{\ell}}^{n_{\ell}-1}\partial_{\theta'}^2\zeta_{2,{\ell}}  \Bigg)\Bigg] |_{\theta'=\theta}\\
\label{fidelity_derivative}
\eal 
Here we have used the fact that $\bm \zeta_1|_{\theta'=\theta}=\bm 1$, $\bm \zeta_2|_{\theta'=\theta}=\bm 1$, $\partial_{\theta'}\bm \zeta_1|_{\theta'=\theta}=\bm 0$, $\partial_{\theta'}\bm \zeta_2|_{\theta'=\theta}=\bm 0$ and therefore only the terms with second order derivatives of the same variable remains nonzero. Due to the photon-number linearity of the fidelity, intermodal correlation never increases the quantum Fisher information. Formally, let us define the marginal probability of $n_\ell$ as $p_{n_\ell}\equiv \left(\sum_{n_1, n_2,\ldots, n_{\ell-1},n_{\ell+1},\ldots, n_K} p_{\bm n} \right) $. Note that the upper bound for each channel 
\begin{align}
&\calJ_{n_{{\rm B},\ell}}^{\rm UB, {UE}}=
\nonumber
\\
&\!\!-4\left(\partial_{\theta} n_{{\rm B},\ell}\right)^{-2}  \sum_{n_\ell} p_{n_\ell} \Bigg[ \left(\partial_{\theta'}^2 \zeta_{1,{\ell}} 
\right)+\Bigg( n_{\ell} \zeta_{2,{\ell}}^{n_{\ell}-1}\partial_{\theta'}^2\zeta_{2,{\ell}}  \Bigg)\Bigg] |_{\theta'=\theta} \,,
\end{align} 
we obtain the additivity
\bal 
&\calJ^{\rm UB, {UE}}_\theta =-4\partial_{\theta'}^2|\expval{\psi(\theta)|\psi(\theta^\prime)}| |_{\theta=\theta'} \\
&=\!\!-4\sum_{\bm n}  p_{\bm n}  \Bigg[ \left(\sum_{\ell=1}^K\partial_{\theta'}^2 \zeta_{1,{\ell}} 
\right)+\Bigg( \sum_{\ell=1}^K n_{\ell} \zeta_{2,{\ell}}^{n_{\ell}-1}\partial_{\theta'}^2\zeta_{2,{\ell}}  \Bigg)\Bigg] |_{\theta'=\theta}\\
&=\sum_{\ell=1}^K  -4\sum_{n_\ell} p_{n_\ell} \Bigg[ \left(\partial_{\theta'}^2 \zeta_{1,{\ell}} 
\right)+\Bigg( n_{\ell} \zeta_{2,{\ell}}^{n_{\ell}-1}\partial_{\theta'}^2\zeta_{2,{\ell}}  \Bigg)\Bigg] |_{\theta'=\theta} \\
&=\sum_{\ell=1}^K \calJ_{n_{{\rm B},\ell}}^{\rm UB, {UE}}.
\eal
\end{proof}

\section{Gaussian-state evaluation} 
\label{app:Gaussian}

Here we derive formula of QFI for two well-studied Gaussian-state quantum probes: single-mode (without ancilla) squeezed vacuum state and two-mode squeezed vacuum (TMSV) state. To describe an $n$-mode state $\hat \rho$, we define a vector of annihilation operators $\hat{\bm a}=[\hat a_1,\hat a_1^\dagger, \ldots, \hat a_n,\hat a_n^\dagger]$, satisfying the commutation relation $[\hat a_i,\hat a_j]=\Omega_{ij}$, where the symplectic metric
$ 
\Omega=\oplus_{k=1}^n
i \mathbb{Y}
$ and $\mathbb{Y}$ is the Pauli-Y matrix.
We define its mean 
$
\bm d\equiv \expval{\hat{\bm a}}\,,
$
and covariance matrix 
$
\Sigma_{j\ell} \equiv \frac{1}{2}\expval{(\hat{a}_j-d_j)(\hat{a}_{\ell}- d_\ell)+(\hat{a}_\ell-d_\ell)(\hat{a}_{j}- d_j)} \,,
$
where $\expval{\hat A}\equiv \tr[\hat \rho \hat A]$ is the expectation value. A Gaussian state is entirely characterized by its mean and covariance matrix~\cite{weedbrook2012gaussian}.
We are interested in the zero-mean case $\bm d=\bm 0$ in our analyses. 

In a quantum sensing problem, the transmitter prepares a quantum state (Gaussian state in this work) and pass it through a bosonic Gaussian channel $\calN_{\kappa,n_{\rm B}}$ defined in the main text. The receiver performs measurement on the output state and estimate the unknown parameter upon the obtained quantum state.

In practice, the input source comes with a thermal noise ${n_T}$---squeezed thermal state and two-mode squeezed thermal state. When $n_T>0$, the input photon number
\be 
N_{\rm S}=\frac{2 \left(G^2+1\right) {n_T}+(G-1)^2}{4 G}.
\ee 
is contaminated by the thermal noise $n_T$, where $G$ is the single-mode or two-mode squeezing strength.
Thus we characterize nonclassical sources using the squeezing strength $G$, for both the single-mode and two-mode squeezers, on thermal inputs with mean photon number $n_T$. 

The covariance matrix of the channel output from a noisy squeezed state is
\be 
\Sigma_{\rm SV}=
\frac{1}{4G}\left(
\begin{array}{cc}
\left(G^2-1\right) \kappa  \nu & 
\nu\kappa
   \left(G^2+1\right)+\mu G 
\\
\nu\kappa \left(G^2+1\right)+\mu G 
& 
\left(G^2-1\right) \kappa  \nu
\end{array}
\right)
\,,
\label{output_cv_sv}
\ee
where we have defined 
\begin{subequations}
\begin{align} 
\mu&= 4 n_{\rm B}+2(1- \kappa)\,,
\\
\nu&=2 n_T  +1\,.
\end{align}
\label{mu_nu_def}
\end{subequations}
The covariance matrix of the channel output from a TMSV state is
\be 
\Sigma_{\rm TMSV}=\left(
\begin{array}{cc}
\frac{G \mu+\left(G^2+1\right) \kappa \nu}{4 G} {\bm X} 
&  
\frac{\left(G^2-1\right) \sqrt{\kappa } \nu}{4 G} {\bm I}
\\
\frac{\left(G^2-1\right) \sqrt{\kappa } \nu}{4 G} {\bm I}
& 
\frac{\left(G^2+1\right) \nu}{4 G} {\bm X}
\end{array}
\right)\,,
\label{output_cv_tmsv}
\ee 
where $\bm X, \bm I$ are the Pauli matrices.
Although we have not made the input state covariance matrix explicit in the above, it can be directly obtained by setting $\kappa=1,n_{\rm B}=0$ in Eq.~\eqref{output_cv_sv} and Eq.~\eqref{output_cv_tmsv}.

Based on the covariance matrices, the QFIs of these zero-mean Gaussian states are accessible via the formula \cite{gao2014bounds}.
\be 
\calJ=\frac{1}{2} {\mathscr R}^{-1}_{\alpha\beta,\gamma\delta} \partial_\theta\Sigma_{\alpha\beta}\partial_\theta\Sigma_{\gamma\delta}\,,
\label{eq:QFI_Gaussian}
\ee
where $\Sigma_\pm\equiv\Sigma\pm \Omega/2$, ${\mathscr R}\equiv\Sigma\otimes \Sigma+\Omega\otimes\Omega/4$. Here $\theta$ can be an arbitrary parameter, while we focus on the estimation of the additive Gaussian noise ${n_{\rm B}}$ in this paper. We present the results for ideal $n_T=0$ input sources in the maintext. We omit the results for general $n_T>0$ cases as they are too lengthy.

\section{Details on the measurements designs}
\label{app:measurement}
In this section, we will frequently use the quadrature covariance matrix, which completely characterizes a Gaussian state. Based on the annihilation operator $\hat a$, the position and momentum quadratures are defined as $\hat q=\hat a+\hat a^\dagger, \hat p=-i(\hat a-\hat a^\dagger)$. For $M$-mode Gaussian state, one can define quadrature vector $\bm {\hat x}=[\hat q_1,\hat p_1,\ldots,\hat q_M,\hat p_M]^T$. For zero-mean Gaussian states, the quadrature covariance matrix is defined as
$
V\equiv \expval{\bm {\hat x} \bm {\hat x}^T},
$
which is equivalent to the annihilation operator covariance matrix up to a bi-linear transform.

Following the formalism in Appendix~\ref{app:Gaussian}, we consider Gaussian input states to probe a bosonic Gaussian channel $\calN_{\kappa,n_{\rm B}}$ defined in the main text. Here we consider practical (noise present with mean photon number $n_T$) input source without loss of generality, as the ideal (noiseless) input source reduces to the $n_T=0$ practical source, while maintaining $n_{\rm B}$ the same.

\subsection{Single-mode squeezing }

After channel $\calN_{\kappa,{n_{\rm B}}}$, the squeezed state has the quadrature covariance matrix 
\be 
V=
\left(
\begin{array}{cc}
 \frac{1}{2} G\kappa  \nu+\frac{1}{4}\mu & 0 \\
 0 & \frac{1}{ 2G} \kappa\nu+\frac{1}{4}\mu
\end{array}
\right)\,.
\ee
As a reminder, $\mu,\nu$ are defined in Eq.~\eqref{mu_nu_def}.

\subsubsection{Homodyne measurement}
\label{app:SVhom}

Homodyne detection measures the squeezed quadrature of the output state, here the momentum quadrature $\hat p$. The readout is a zero-mean Gaussian random variable with variance $\sigma^2={\kappa} [2n_T - (G-1)]/ 2G+n_{\rm B}+1/2$.
It yields the Fisher information
\begin{align}
\calI_{\rm SV-hom}= \frac{2 G^2}{\left(-\left(G-1\right) \kappa+ G \left(2n_{\rm B}+1\right)+2 \kappa  n_T \right)^2}.
\label{eq:_app}
\end{align}
For ideal source, $G=1+2N_{\rm S}+2C_{\rm p}$, $\sigma^2=n_{\rm B}+\kappa  \left(N_{\rm S}-C_{\rm p}\right)+1/2$ then 
\begin{align}
&\calI_{\rm SV-hom}|_{n_T=0}= \left(\frac{\partial \sigma^2}{\partial {n_{\rm B}}}\right)^2 \cdot \frac{1}{2\sigma^4}|_{n_T=0}
\\
&=\frac{2}{\left(2 n_{\rm B}+2 \kappa  \left(N_{\rm S}-C_{\rm p}\right)+1\right)^2}\,,
\end{align}
where we have defined the notation $C_{\rm p}=\sqrt{N_{\rm S} \left(N_{\rm S}+1\right)}$.
Note that the above performance is invariant if any squeezing is further performed before the final homodyne detection.


\subsubsection{Nulling receiver}

In the nulling receiver, one squeezes the return mode $e^{-r^\star(\hat a_R^2-\hat a_R^{\dagger 2})/2}$ with 
\be 
r^\star=\frac{1}{2}\log[\frac{1}{G}].
\ee 
For an identity channel $\kappa=1,n_T=0,n_{\rm B}=0$, it nulls the return mode to vacuum. The derivation of covariance matrix of the general $n_T\neq 0$ case is trivial but the result is too lengthy, thus here we present the result for the ideal $n_T=0$ case. The covariance matrix is
\bal 
V^{\rm null}&=
\left(
\begin{array}{cc}
 \frac{\mu}{2G}+\kappa \nu & 0 \\
 0 & \frac{\mu  G}{2}+ \kappa\nu\\
\end{array}
\right)\,.
\eal 
The photon count distribution of such a covariance is in the Legendre function~\cite{marian1993squeezed}
\begin{align}
P(n)=&\left(A^2-B^2\right)^{n/2} \left(A \left(A+2\right)-B^2+1\right)^{-\left(n+1\right)/2}
\nonumber
\\
&\times P_n\left(\frac{A^2+A-B^2}{\sqrt{\left(A^2-B^2\right) \left(-B^2+A \left(A+2\right)+1\right)}}\right)\,,
\end{align}
where $P_n$ is the Legendre function of the first kind, $A= [ \frac{\mu}{2}(G+\frac{1}{G})+2\kappa \nu -2]/4,B=\frac{\mu}{2}(\frac{1}{G}-G)/4$. The Fisher information is evaluated as
\be 
\calI_{\rm SV-null}=\sum_{n}\left(\frac{\partial \log P(n)}{\partial {n_{\rm B}}}\right)^2 P (n)\,.
\label{eq:FI_Sqznull_app}
\ee

Remarkably, at the asymptotic identity-channel limit $n_B\to 0, \kappa\to 1$, we derive
\bal  
\calI_{\rm SV-null}= \frac{(1+2N_{\rm S})^2}{(1-\kappa)N_{\rm S}}+O(1)\,.
\eal 
Compared to the asymptotic value of Eq.~\eqref{eq:QFI_SQZ} of the main text, we find that the nulling receiver is optimum for SV probes at this limit. For $\kappa< 1$, the output state is highly involved and the performance of SV probes degrades rapidly, thus we omit the analysis in this paper.

\subsection{Entanglement-asssisted strategy}
After channel $\calN_{\kappa,{n_{\rm B}}}$, the TMSV state has the quadrature covariance matrix
\be 
V_{RA}=\left(
\begin{array}{cc}
\frac{G \mu+\left(G^2+1\right) \kappa \nu}{2 G} {\bm I} 
&  
\frac{\left(G^2-1\right) \sqrt{\kappa } \nu}{2 G} {\bm Z}
\\
\frac{\left(G^2-1\right) \sqrt{\kappa } \nu}{2 G} {\bm Z}
& 
\frac{\left(G^2+1\right) \nu}{2 G} {\bm I}
\end{array}
\right)\,,
\ee 
where $\mu,\nu$ are defined in Eq.~\eqref{mu_nu_def} and $\bm I,Z$ are the Pauli matrices.

\subsubsection{Bell measurement}
In a Bell measurement, one first passes the return mode and ancilla mode through a balanced beamsplitter, outputting a two-mode Gaussian state with covariance matrix 
\be 
V_{RA}^{\rm Bell}=
\left(
\begin{array}{cccc}
 a & 0 & c & 0 \\
 0 & b & 0 & c \\
 c & 0 & b & 0 \\
 0 & c & 0 & a \\
\end{array}
\right)
\label{eq:V_bell_app}
\ee 
where 
\begin{align}
a&=\frac{G \mu+G^2 \left(\sqrt{\kappa }+1\right)^2 \nu+\left(\sqrt{\kappa }-1\right)^2 \nu}{4 G},
\\
b&=\frac{G \mu+G^2 \left(\sqrt{\kappa }-1\right)^2 \nu+\left(\sqrt{\kappa }+1\right)^2 \nu}{4 G},
\\
c&=-\frac{1}{4} \left(\mu+\frac{\left(G^2+1\right) (\kappa -1) \nu}{G}\right).
\end{align} 
Then the quadrature measurements on $\hat p_R$ and $\hat q_A$ gives two i.i.d. Gaussian variables with variance 
\be 
\sigma^2=\frac{G \mu+G^2 \left(\sqrt{\kappa }-1\right)^2 \nu+\left(\sqrt{\kappa }+1\right)^2 \nu}{4 G} \,.
\ee
The joint 2-D Gaussian distribution yields the classical Fisher information 
\begin{align}
\calI_{\rm Bell}&=2\cdot \left(\frac{\partial \sigma^2}{\partial {n_{\rm B}}}\right)^2 \cdot \frac{1}{2\sigma^4}
\\
&=
\frac{16 G^2}{\left(G \mu+G^2 \left(\sqrt{\kappa }-1\right)^2 \nu+\left(\sqrt{\kappa }+1\right)^2 \nu\right)^2}\,.
\end{align} 
For the ideal $n_T=0$ input, $G=1+2N_{\rm S}+2C_{\rm p}$, we have
\begin{align}
\calI_{\rm Bell}|_{n_T=0}&=2\cdot \left(\frac{\partial \sigma^2}{\partial {n_{\rm B}}}\right)^2 \cdot \frac{1}{2\sigma^4}|_{n_T=0}
\\
&=\frac{1}{\left({n_{\rm B}}+\kappa  N_{\rm S}-2 \sqrt{\kappa}C_{\rm p}+N_{\rm S}+1\right)^2}\,.
\end{align} 

\subsubsection{Nulling receiver}
\label{app:nullRx}

In the nulling receiver, one squeezes the return mode and the ancilla mode via a two-mode squeezing process $e^{-r_2^\star(\hat a_R\hat a_A -\hat a_R^\dagger \hat a_A^\dagger)}$ with $r_2^\star=\log[\frac{2 (G+1)}{-(G-1) \sqrt{\kappa }+G+1}-1]/2$. For $n_T=0, n_{\rm B}=0$, it nulls the returned signal mode to vacuum. The derivation of the covariance matrix of the general $n_T\neq 0$ case is straightforward but the result is too lengthy to display, thus here we present the result for the ideal $n_T=0$ case. The resulting covariance matrix is
\be 
V^{\rm null}_{RA}=\bp 
 e {\bm I} & c {\bm Z} \\
 c {\bm Z} & s {\bm I}\,,
 \ep 
\ee 
where
\begin{align}
e&=\frac{(1-\kappa)  N_{\rm S}+2 {n_{\rm B}} \left(N_{\rm S}+1\right)+1}{(1-\kappa)  N_{\rm S}+1},
\\
s&=\frac{-2 (\kappa -1)^2 N_{\rm S}^2+\left(\kappa  \left(3-2 {n_{\rm B}}\right)-3\right) N_{\rm S}-1}{(\kappa -1) N_{\rm S}-1} ,
\\
c&=-\frac{2 {n_{\rm B}} \sqrt{\kappa}C_{\rm p}}{(1-\kappa)  N_{\rm S}+1}.
\end{align}
The photon count distribution of such a two-mode state is subject to the hypergeometric distribution
\begin{align}
P(n_R,n_A)&=-4 y^{-n_A-1} x^{-n_R-1} 
\nonumber
\\
&\times \left(c^2-e s+e+s-1\right)^{n_A+n_R+1} \,
\nonumber
\\
&\times_2F_1\left(n_A+1,n_R+1;1;\frac{4 c^2}{x y}\right)\,,
\end{align} 
where $_2F_1$ is the regularized hypergeometric function, $x=c^2-(e+1) s+e+1$, $y=c^2-(e-1) (s+1)$. The Fisher information is evaluated as
\be 
\calI_{\rm TMSV-null}=\sum_{n_R,n_A}\left(\frac{\partial \log P(n_R,n_A)}{\partial {n_{\rm B}}}\right)^2 P (n_R,n_A)\,.
\label{eq:FI_TMSVnull_app}
\ee

\begin{figure}
    \centering
    \includegraphics[width=0.25\textwidth]{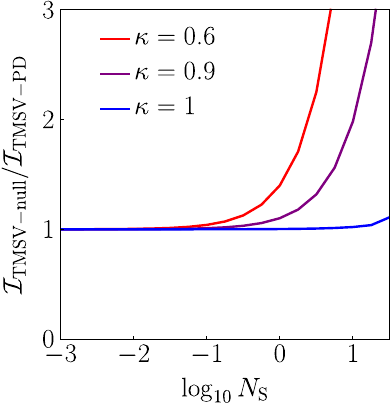}
    \caption{The ratio of Fisher information of the nulling receiver over the direct photon detection, both using TMSV probes. $N_{\rm B}=10^{-3}.$}
    \label{fig:nullingbetter}
\end{figure}

Note that the nulling step is indispensable. As shown in Fig.~\ref{fig:nullingbetter}, the Fisher information of a nulling receiver is strictly larger than direct photon detection. The advantage expands significantly when $N_{\rm S}$ increases. For $\kappa\to 1$ the advantage is negligible for small $N_{\rm S}$, nevertheless we still see an increasing trend with $N_{\rm S}$.

Remarkably, at the asymptotically low-noise limit $n_B\to 0$, we derive
\bal  
\calI_{\rm TMSV-null}= \frac{1+N_{\rm S}}{[1+N_{\rm S}(1-\kappa)]n_{\rm B}}+O(1)\,.
\eal 
Compared to the asymptotic value of Eq.~\eqref{eq:QFI_TMSV} of the main text, we find that the nulling receiver is optimum for TMSV probes at this limit.

One may consider the strategy of measuring only one of the two ports. No matter which port one measures, the reduced state is always a thermal state. A thermal state with mean photon number $\overline N$ has photon count $n$ satisfying the distribution
\be 
P_{\rm th}(n)=\frac{\overline N^n}{(1+\overline N)^{n+1}}.
\ee
Consider $\overline N=\overline N(n_B)$, one can immediately obtain the Fisher information about $n_B$ as
\bal 
\calI &=\left(\frac{\partial \overline N}{\partial {n_{\rm B}}}\right)^2 \sum_{n}\left(\frac{\partial \log P_{\rm th}(n)}{\partial {\overline N}}\right)^2 P (n)\\
&=\left(\frac{\partial \overline N}{\partial {n_{\rm B}}}\right)^2\frac{1}{{\overline N} ({\overline N}+1)}.
\eal

If one only measures the ancilla, the mean photon number is
\be 
\overline N_A=\frac{n_B \left(N_S+1\right)}{(1-\kappa)  N_S+1}.
\ee
If one only measures the returned signal, the mean photon number is
\be 
\overline N_R=\frac{N_S \left(\kappa  \left(n_B+(\kappa -2) N_S-1\right)+N_S+1\right)}{(1-\kappa ) N_S+1}.
\ee

Indeed, only measuring the returned signal also achieves the Fisher information scaling of measuring both when the noise $n_B\ll 1$:
\bal  
\calI_{\rm TMSV-null,signal}
&=\frac{1+N_{\rm S}}{[1+N_{\rm S}(1-\kappa)]n_{\rm B}}+O(1)\,.
\eal 

However, if one only measures the idler,
\bal 
&\calI_{\rm TMSV-null,idler}=\frac{\kappa ^2 N_S}{(\kappa -1) \left((\kappa -1) N_S-1\right){}^3}+O(n_B)
\eal 

\begin{figure}[htbp]
    \centering
    \includegraphics[width=0.5\textwidth]{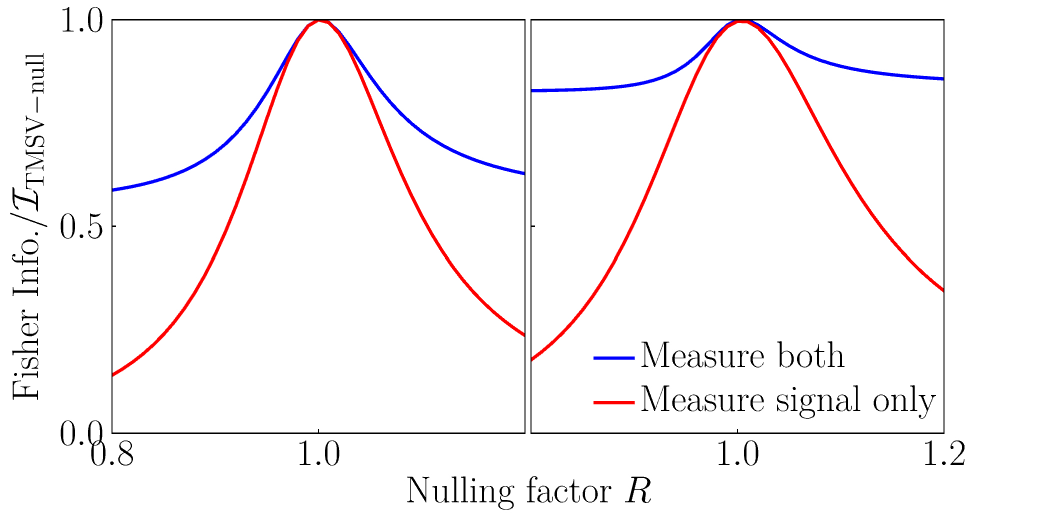}
    \caption{The ratio of Fisher information of the nulling receiver with nulling (deviation) factor $R$ over the ideal $\calI_{\rm TMSV-null}$ with $R=1$, using TMSV probes. (a) $\kappa=0.6$; (b) $\kappa=0.9$. The measurement on both the returned signal and ancilla modes (blue) is compared with that on the returned signal mode only (red). $N_{\rm B}=10^{-3}$, $G=10$dB ($N_S=2.025$).}
    \label{fig:nullingrobust}
\end{figure}

In general, we can consider an arbitrary nulling parameter as $r_2=R\cdot r_2^\star$, where $R$ is the nulling (deviation) factor that characterizes its deviation from our proposal $r_2^\star$, ideally $R=1$. We find that measuring both signal and ancilla modes significantly improves the robustness against such deviation, as shown in Fig.~\ref{fig:nullingrobust}. When the nulling factor $R$ deviates from $1$, the Fisher information degrades from $\calI_{\rm TMSV-null}$ achieved by $r_2^\star$, while the measure-both strategy (blue) decays much slower than the measure-signal-only strategy (red).

\section{Fisher information in dark matter search}
\label{app:QFI_UB_total}

Before presenting the formulas, we clarify our treatment of the thermal background of input source in derivations. For the upper bound of the total quantum Fisher information, we consider the energy constraint on the input state to the channel, rather than the output state of the channel. Such a treatment greatly simplifies the optimization over source states and yields analytical formulas. In contrast, for all the other Fisher information quantities, in consistence with Appendix~\ref{app:Gaussian}, we regard the input state as a possibly noisy quantum state contaminated by the thermal noise.

For the upper bound, we present the full formula for Eq.~\eqref{J_total_UB_asym} of the main text and the derivation has been described in the main text.
In the ideal source case, the spectra of Fisher information quantities can be derived by simply substituting Eqs.~\eqref{eq:QFI_vl}~\eqref{Fisher_vac_homo}~\eqref{eq:QFI_SQZ}~\eqref{Fisher_SV_homo}~\eqref{eq:QFI_TMSV} into Eq.~\eqref{eq:QFI_freq_perfect} of the main text. We derive the total Fisher information upper bound and total Fisher information for various quantum sources and measurements as follows.
\begin{widetext}
\begin{align} 
&\bbJ_{\rm UB}=
\frac{2 \pi  \gamma _a^2 \gamma _m }{{{n_T}
   \left({n_T}+1\right) \left(\gamma _a+\gamma _l\right) \left[4 \gamma _m {n_T}
   \left(\gamma _a+\gamma _l\right)+\left(\gamma _a+\gamma _l+\gamma
   _m\right)^2\right]^{3/2}}} \nonumber \\
& \quad \quad \quad \times \Big\{2 \gamma _l \big[\gamma _a \left(2 N_{\rm S}
   {n_T}+N_{\rm S}+{n_T}+1\right)+\gamma _m \left(2 {n_T}+1\right) \left(N_{\rm S}
   {n_T}+{n_T}+1\right)\big]+2 \gamma _a \gamma _m \left(2 {n_T}+1\right) \left(N_{\rm S}
   {n_T}+{n_T}+1\right) \nonumber\\
& \quad \quad \quad +\gamma _a^2 \left(2 N_{\rm S} {n_T}+N_{\rm S}+{n_T}+1\right)+\gamma _l^2 \left(2
   N_{\rm S} {n_T}+N_{\rm S}+{n_T}+1\right)+\gamma _m^2 \left(2 N_{\rm S} {n_T}+N_{\rm S}+{n_T}+1\right)\Big\}\,,
\\
&\bbJ_{\rm VL}=\frac{2 \pi  \gamma _a^2 \gamma _m}{n_T \left(\gamma _a+\gamma _l\right) \sqrt{4 \gamma _m n_T \left(\gamma _a+\gamma _l\right)+\left(\gamma _a+\gamma _l+\gamma _m\right)^2}}\,,
\\
&\mathbb I_{\rm Vac-hom}=\frac{8 \pi  \gamma _a^2 \gamma _m^2}{\left(8 \gamma _m n_T \left(\gamma _a+\gamma _l\right)+\left(\gamma _a+\gamma _l+\gamma _m\right)^2\right)^{3/2}}\,,
\\
 &\mathbb I_{\rm SV-hom}=\frac{8 \pi  \gamma _a^2 \gamma _m^2 \left(2 N_{\rm S}-2 C_{\rm p}+1\right)^{3/2} \left[8 N_{\rm S} \left(N_{\rm S}+C_{\rm p}+1\right)+4 C_{\rm p}+1\right]}{\left(2 \gamma _l \gamma _m \left(4 n_T-2 N_{\rm S}+2 C_{\rm p}+1\right)+\gamma _l^2 \left(2 N_{\rm S}-2 C_{\rm p}+1\right)+\gamma _m^2 \left(2 N_{\rm S}-2 C_{\rm p}+1\right)\right)^{3/2}}\,,
\\
&\bbJ_{\rm TMSV}=\frac{2 \pi  \gamma _a^2 \gamma _m \left(2 n_T N_{\rm S}+n_T+N_{\rm S}+1\right)}{n_T \left(n_T+1\right) \left(\gamma _a+\gamma _l\right) \sqrt{2 \gamma _l \left[\gamma _a+\gamma _m \nu \left(2 N_{\rm S}+1\right)\right]+2 \gamma _a \gamma _m \nu \left(2 N_{\rm S}+1\right)+\gamma _a^2+\gamma _l^2+\gamma _m^2}}\,.
\end{align}
\end{widetext}

\begin{figure}
    \centering
    \includegraphics[width=0.3\textwidth]{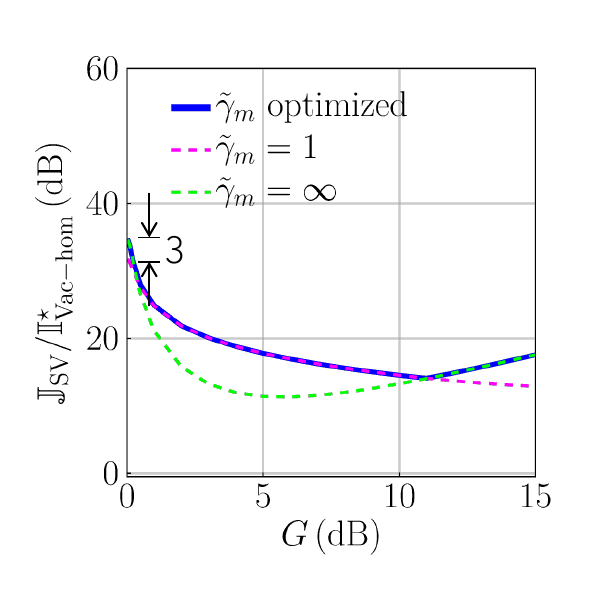}
    \caption{The piece-wise optima of $\bbJ_{\rm SV}$. Parameters are chosen identical with Fig.~\ref{fig:ScanRate_idealDMS} of the main text. For small and large gain $G$, the optimum is at $\tilde\gamma=1$; for intermediate $G$, the optimum is at $\tilde \gamma= \infty$.  }
    \label{fig:piecewise_ScanRateSVgm}
\end{figure}

Here we omit the formula for $\bbJ_{\rm SV}$ as it is too lengthy to display. At the same time, we provide an additional Fig.~\ref{fig:piecewise_ScanRateSVgm} to explain the optimization of $\bbJ_{\rm SV}$ over $\tilde \gamma$ mentioned in the main text. We see that when the squeezing $G$ is very small, the optimal coupling ratio is $\tilde\gamma=\infty$ (green dashed), the same as the vacuum limit; while when squeezing $G$ is in a certain range, the optimal coupling ratio $\tilde\gamma=1$ (magenta dashed); in the large squeezing region above a threshold, the optimal coupling ratio is again $\tilde\gamma=\infty$ (green dashed). The $\tilde\gamma_m=1$ peak always decays as $G$ increases, because in this case the on-resonance peak, which suffers severe loss for $\kappa(\omega)\simeq 0$, contributes most of the total Fisher information (See Fig.~\ref{fig:FisherSpectrum_idealDM} of the main text) and after such a severe loss the squeezing on the source degenerates to harmful noises. Similar phenomenon occurs to the overcoupling limit for small $N_{\rm S}$, as Eq.~\eqref{eq:ScanRate_SV_overcouple} of the main text shows.

\section{Practical input engineering}
\label{app:practical_source_eng}

As discussed in Appendix~\ref{app:methods}, in an experimentally feasible scenario, the input will also be affected by thermal noise. Here we give detailed analysis of such effect.  


The squeezed-vacuum homodyne performance can be obtained as (see Appendix~\ref{app:FI_practical} for derivations)
\be 
\calI^{\rm SV-hom}_{n_{\rm a}}(\omega)=
\frac{2 \gamma _m^2 \gamma_a^2}{\left(2
   n_T+1\right)^2\left(\frac{
   \left(\gamma /2\right)^2+\omega ^2-\gamma _l
   \gamma _m}{G}+ \gamma
   _m\gamma _l \right)^2},
   \label{eq:FisherSpectrum_SVhomo}
\ee
where $\gamma\equiv\gamma_m+\gamma_\ell+\gamma_a$ is the total coupling strength. From Eq.~\eqref{eq:FisherSpectrum_SVhomo}, we can see that the continuous-spectrum Fisher information is identical to $\alpha^2(\omega)/2n_a^2$, i.e., only a constant factor different from the square of visibility $\alpha(\omega)$ defined in Eq.~(1) of Ref.~\cite{malnou2019}. Therefore, our results provide the Fisher information interpretation of the visibility for DM signal.

\begin{figure}[t]
    \centering
    \includegraphics[width=0.4\textwidth]{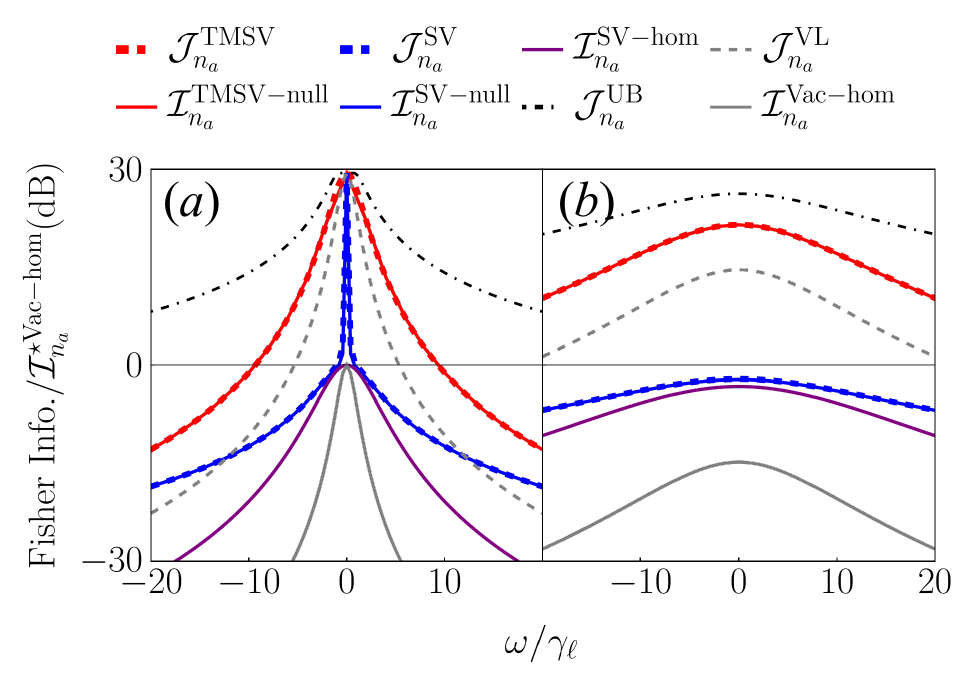}
    \caption{Frequency spectrum of the Fisher information with respect to the axion occupation number $N_{\rm S}$, normalized by the optimized peak value of vacuum-homodyne Fisher ${\calI^\star}_{n_{\rm a}}^{ \rm Vac-hom} $and plotted in decibel unit. (a) $\gamma_m/\gamma_\ell=1$; (b) $\gamma_m/\gamma_\ell=2G$. Temperature $T=61mK$, cavity resonant frequency $\omega_c=2\pi\cdot 10$GHz, squeezing strength $G=10$dB, $\gamma_a/\gamma_\ell=10^{-12}$.}
    \label{fig:FisherSpectrum_DM}
\end{figure}

In the practical input case, the closed-form formulas for $\calJ^{\rm TMSV}_{n_{\rm a}}$ and $\calJ^{\rm SV}_{n_{\rm a}}$ are too lengthy to display (See Appendix~\ref{app:FI_practical} for derivations); instead, we directly plot the results for a comparison. As shown in Fig.~\ref{fig:FisherSpectrum_DM}, due to the additional thermal noise, the performance of different sources and measurements is overall worse than the performance in the ideal case. In this practical case, we cannot show the optimality of the noisy TMSV, as there is a gap between the upper bound (black dot-dash) and the TMSV performance (red dashed). The homodyne detection performance (purple and gray solid) is still worse than the vacuum limit (gray dashed). While the same nulling receivers are still optimal given the noisy TMSV and noisy single-mode squeezed vacuum sources (red solid and blue solid).


\begin{figure}[t]
    \centering
    \includegraphics[width=0.5\textwidth]{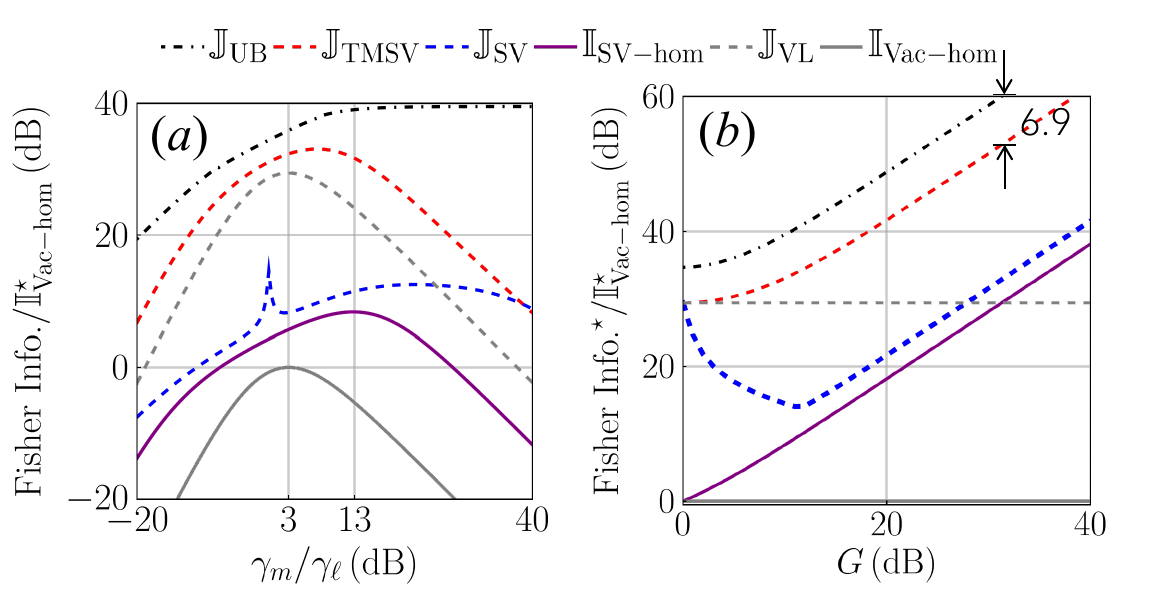}
    \caption{ (a)Total Fisher information under various measurement coupling ratio over intrinsic loss $\gamma_m/\gamma_\ell$, squeezing strength $G=10$dB; (b) the optimized total Fisher information optimized with optimal $\gamma_m/\gamma_\ell$ under various squeezing strength $G$. Y axis normalized by $\mathbb I_{\rm Vac-hom}^\star$ as defined in Eq.~\eqref{eq:bbI_vachom_opt}, both axes plotted in decibel unit. 
    Temperature $T=61mK$, cavity resonant frequency $\omega_c=2\pi\cdot 10$GHz, $\gamma_a/\gamma_\ell=10^{-12}$.}
    \label{fig:ScanRate_DMS}
\end{figure}

Now we consider the total Fisher information to obtain insights into the scan rate.
Consider homodyne detection on vacuum input (up to weak thermal noise), combining Eq.~\eqref{I_na_vac_homo} and Eq.~\eqref{I_total_define} of the main text we have the total Fisher information
\begin{align}
\mathbb{I}_{\rm Vac-hom}&=\frac{2}{(1+2 n_{T})^2}\int_{-\infty}^\infty {\rm d}\omega\, \chi_{ma}^4
\\
&=4\pi\gamma_\ell\frac{2 \tilde{\gamma}_s^2}{(2 n_{T}+1)^2}\frac{\tilde{\gamma}_m^2}{(\tilde{\gamma}_m+1)^3}.
\end{align}
The result agrees with the ideal case Eq.~\eqref{I_total_vac_homo} in the low tempetaure limit of $n_T=0$, as expected.
We can see that the total Fisher information is indeed proportional to the scan rate~\cite{brady2022entangled,backes2021}. Similarly, the maximal total Fisher information 
\be 
\mathbb{I}_{\rm Vac-hom}^\star=4\pi\gamma_\ell \frac{8\tilde{\gamma}_s^2}{27(1+2 n_{T})^2}
\label{eq:bbI_vachom_opt}
\ee 
is achieved at $\tilde{\gamma}_m=2$, as we numerically verify in Fig.~\ref{fig:ScanRate_DMS}(a) with the gray solid plot. When there is squeezing, from Eq.~\eqref{eq:FisherSpectrum_SVhomo} and Eq.~\eqref{I_total_define} of the main text, we can obtain the corresponding total Fisher information
\begin{align}
\mathbb I_{\rm SV-hom}=\frac{8 \pi  G^2 \gamma _a^2 \gamma _m^2}{(1+2n_T)^2\left[2 \left(2 G-1\right) \gamma _l \gamma _m+\gamma _l^2+\gamma _m^2\right]^{3/2}},
\end{align}
which is again proportional to the scan rate in Refs.~\cite{malnou2019,brady2022entangled}.
For a highly squeezed quantum source with $G\gg 1$, the maximum total Fisher information is again achieved at $\tilde \gamma_m=2G$ (same as the ideal case), leading to 
\be 
\mathbb{I}_{\rm SV-hom}^\star= 2 \pi \gamma_\ell \tilde\gamma_a^2 \frac{8  G }{3^3  \left(2 n_T+1\right)^2}+O\left(\sqrt{\frac{1}{G}}\right)\,.
\ee
Similarly, the optimal coupling rate is verified in Fig.~\ref{fig:ScanRate_DMS}(a) by the purple solid line.

Now we consider the quantum limits in the practical source case. For the vacuum source (up to weak thermal noise), from Eq.~\eqref{J_na_VL} and Eq.~\eqref{J_total_define} of the main text we have
\begin{align}
\mathbb{J}_{\rm VL}&=4\pi \gamma_\ell\frac{ \tilde{\gamma}_s^2}{n_{T}(n_{T}+1)}\frac{\tilde{\gamma}_m^2}{(\tilde{\gamma}_m+1)^3},
\end{align}
which is achieved by photon counting. When the noise $n_{T}$ is small at low temperature, we see that $\mathbb{J}_{\rm VL}$ will have an advantage scaling with $1/n_{T}$ compared with homodyne detection. More specifically, at $\tilde{\gamma}_m=2$, the optimal vacuum limit total Fisher information $\mathbb{J}_{\rm VL}^\star$ satisfies the relation 
\be 
\frac{\mathbb{J}_{\rm VL}^\star}{\mathbb{I}_{\rm Vac-hom}^\star}=\frac{(1+2 n_{T})^2}{2 n_{T}(1+n_{T})}.
\ee 
The optimal coupling ratio is verified numerically in Fig.~\ref{fig:ScanRate_DMS}(a) as the gray dashed curve.
We see that the vacuum limit achieved by photon counting has a huge advantage over homodyne when noise $n_{T}$ is low. Indeed, Ref.~\cite{dixit2021} proposed an approach of photon counting in microwave to explore such an advantage, which is made rigorous in our work.

For the quantum limit of single-mode squeezed vacuum, the formula is too long to display here and we plot the results in Fig.~\ref{fig:ScanRate_DMS}. For single-mode squeezing, there is still gap between the QFI performance limit (blue dashed) and the performance enabled by homodyne detection (purple solid). In subplot (b), similar to the ideal case, the gap is large at low squeezing, while being a constant factor at large squeezing. Overall, it takes almost $30$ dB of squeezing for the squeezing homodyne performance (purple solid) to overcome the vacuum limit (gray dashed).

For the quantum limit of TMSV, at the limit $G\gg 1, n_T\ll 1$ we find that it is optimized at $\tilde\gamma=G/2$ to 
\be 
\bbJ^\star_{\rm TMSV}\simeq 2\pi\gamma_\ell  \tilde\gamma_a^2\frac{  G}{12 \sqrt{3} n_T}
\ee
Plugging in $G\simeq 4 N_{\rm S}$, we see a constant gap between the above equation and the upper bound Eq.~\eqref{eq:ScanRateUB_opt},
\be 
\bbJ^\star_{\rm TMSV}/\bbJ^\star_{\rm UB}\simeq \frac{1}{3\sqrt{3}}\simeq 0.19\,.
\ee
This is verified in Fig.~\ref{fig:ScanRate_DMS}(b), by a constant gap of 6.9dB between $\bbJ^\star_{\rm UB}$ (black dotdashed) and $\bbJ^\star_{\rm TMSV}$ (red dashed). Here the gap is almost equal to $3\sqrt{3}=7.1$dB. The tiny difference is due to utilizing the combination of two bounds in Eq.~\eqref{eq:UB_all} of the main text in the numerical evaluation, while the asymptotic result is only from the unitary extension upper bound of Eq.~\eqref{eq:QFI_UB} of the main text. We see that in terms of the total Fisher information, TMSV is optimal up to a constant factor of $\sim0.2$. Moreover, we expect this constant factor off is due to our upper bound being not tight in the practical source engineering case, where all operations are contaminated by the thermal noise.



\subsection{Fisher information for practical input engineering}
\label{app:FI_practical}

In the practical source case, with respect to the overall background $n_{\rm B}$, the spectra of the Fisher information can be obtained following the procedure in Appendix~\ref{app:Gaussian} for QFIs and following Appendix~\ref{app:measurement} for measurement. Then the Fisher information quantities with respect to the axion occupation number $n_a$ are immediately obtained by multiplying $\left(\partial n_{\rm B}/\partial n_a\right)^2$. The Fisher spectra of homodyne measurements are
\be 
\calI_{n_{\rm a}}^{\rm Vac-hom}(\omega)=\frac{32 \gamma _m^2 \gamma_a^2}{\left(2 n_T+1\right)^2\left(\left(\gamma _l+\gamma _m\right)^2+4 \omega ^2\right)^2}\,,
\ee
\be 
\calI_{n_{\rm a}}^{\rm SV-hom}(\omega)=\frac{32  \gamma _m^2 \gamma_a^2}{\left(2 n_T+1\right)^2\left(4 \gamma _l \gamma _m+\frac{\left(\gamma _l-\gamma _m\right)^2+4 \omega ^2}{G}\right)^2}\,.
\ee
The exact formulas for the QFIs are too lengthy and the derivation is straightforward, here we present asymptotic results.
At the limit $G\gg 1$, we have the QFIs
\be 
\calJ_{n_{\rm a}}^{\rm SV}(\omega)\simeq \frac{2\tilde \gamma_a^2}{(1+2n_T)^2}
\ee
\be 
\calJ_{n_{\rm a}}^{\rm TMSV}(\omega)\simeq \frac{\tilde \gamma_a^2}{n_T(1+n_T)}
\ee
for squeezed-vacuum probes and TMSV probes.
We see that the spectra are extended to infinity with the assistance of non-classical sources, thus the infinite squeezing strength provides a diverging quantum advantage in the total Fisher information. 
\begin{figure}
    \centering
    \includegraphics[width=0.45\textwidth]{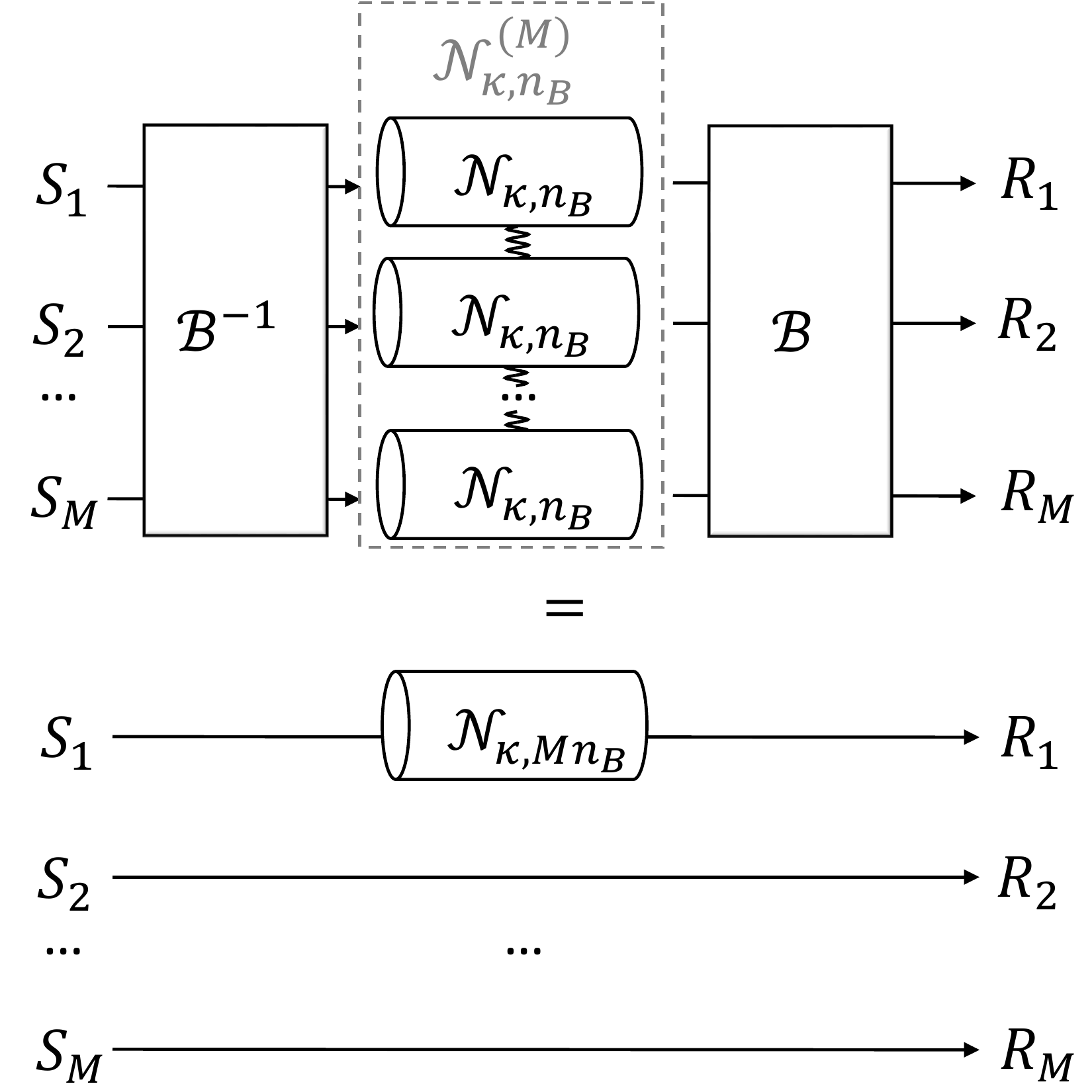}
    \caption{Reduction of distributed noise sensing by applying beamsplitters. The fully-correlated thermal noise channel $\calN_{\kappa,n_B}^{(M)}$ consists of $M$ fully-correlated, entanglement-free subchannels $\calN_{\kappa,n_B}$. Sandwiched by beamsplitters $\calB^{-1}$, $\calB$, $\calN_{\kappa,n_B}^{(M)}$ reduces to a single-mode thermal noise channel $\calN_{\kappa,Mn_B}$.}
    \label{fig:equivalence_distributed}
\end{figure}
With the spectra in hand, we derive the formulas of total Fisher information about $n_a$ as follows.
\begin{align}
&\bbJ_{\rm VL}=\frac{4 \pi  \gamma _a^2 \gamma _m^2}{n_T \left(n_T+1\right) \left(\gamma _a+\gamma _l+\gamma _m\right)^3}\,,
\\
&\mathbb I_{\rm Vac-hom}=\frac{8 \pi  \gamma _a^2 \gamma _m^2}{\nu^2\left(\gamma _a+\gamma _l+\gamma _m\right)^3}\,,
\\ 
&\mathbb I_{\rm SV-hom}=\frac{8 \pi  G^2 \gamma _a^2 \gamma _m^2}{\nu^2\left[2 \left(2 G-1\right) \gamma _l \gamma _m+\gamma _l^2+\gamma _m^2\right]^{3/2}}\,,
\end{align}
\begin{widetext}
 \bal 
\bbJ_{\rm TMSV}=
&-\frac{\pi  \gamma _a^2 \gamma _m }{2 (G-1)^2 G \gamma _l n_T^2 \left(n_T+1\right)^2\nu} \times \\
&\Bigg[n_T^{5/2} \left(G^2 \nu-2 G+2 n_T+1\right)^2\sqrt{\frac{G}{\gamma _l \gamma _m \left((G^2+1) \nu-2 G \left(n_T+1\right)\right)+G (\gamma _l^2+\gamma_m^2) n_T}}\\
&-\left(n_T+1\right)^{5/2} \left((G^2+1) \nu+2 G\right)^2\sqrt{\frac{G}{\gamma _l \gamma _m \left((G^2+1) \nu-2 G n_T\right)+G (\gamma _l^2+\gamma_m^2) \left(n_T+1\right)}}\\
&+(G+1)^2 \nu \left(\nu^2(G^2+1)+2 G\right) \cdot \\
&\quad \sqrt{\frac{G \left(2 n_T^2+2 n_T+1\right)}{\gamma _l \gamma _m \left(\nu^2(G^2+1)-4 G n_T \left(n_T+1\right)\right)+G (\gamma _l^2+\gamma_m^2) \left(2 n_T^2+2 n_T+1\right)}}\Bigg]\,.
\eal
\end{widetext}
Here $\nu$ is defined in Eq.~\eqref{mu_nu_def}. The formula for $\bbJ_{\rm SV}$ is too long to display.

\section{Memory channels and distributed sensing} 
\label{app:DQS}

So far we have been focusing on the memoryless channels, where the $M$-mode probe travel through $M$-product of independent identical channels. Nevertheless, a highlighted scenario is to estimate the fully-correlated thermal noise, which is modelled by an $M$-mode memory channel $\calN_{\kappa,{n_{\rm B}}}^{(M)}$. Indeed, as illustrated in Fig.~\ref{fig:equivalence_distributed}, based on the Gaussian-state theory~\cite{weedbrook2012gaussian} $\calN_{\kappa,{n_{\rm B}}}^{(M)}$ can be reduced to a single-mode channel $\calN_{\kappa,M{n_{\rm B}}}$ after being sandwiched by two beamsplitters as
\be 
\calN_{\kappa,M{n_{\rm B}}}\otimes \calI^{\otimes (M-1)}=\calB\circ\calN_{\kappa,{n_{\rm B}}}^{(M)}\circ\calB^{-1}\,.
\ee
Here $\calB=\calB^{-1}$ is a beamsplitter transform with uniform interference ratios. It is straightforward to show that our QFI upper bound $\calJ_{\rm UB}$ of $\calN_{\kappa,M{n_{\rm B}}}$ also upper bounds $\calN_{\kappa,{n_{\rm B}}}^{(M)}$:
\bal 
\calJ_{\rm UB} &\ge \max_{\hat \sigma} \calJ(\{\calN_{\kappa,M{n_{\rm B}}}\otimes \calI^{\otimes (M-1)}[\hat \sigma]\})\\
&=  \max_{\hat \sigma} \calJ(\{\calB\circ\calN_{\kappa,{n_{\rm B}}}^{(M)}\circ\calB^{-1}[\hat \sigma]\})\\
&=\max_{\hat \sigma} \calJ(\{\calN_{\kappa,{n_{\rm B}}}^{(M)}\circ\calB^{-1}[\hat \sigma]\})\\
&=\max_{\hat \sigma'} \calJ(\{\calN_{\kappa,{n_{\rm B}}}^{(M)}[\hat \sigma']\})
\eal 
where $\calJ$ is the QFI defined by \eqref{eq:QFI_fidelity} of the main text. Here the second equality is because that $\calB$ is a unitary transform, and the third equality is because that for any optimum state $\sigma$ that maximizes the QFI for channel $\calN_{\kappa,{n_{\rm B}}}^{(M)}\circ\calB^{-1}$, one can always achieve it by letting $\hat\sigma'=\calB^{-1}\hat\sigma^\prime$ for channel $\calN_{\kappa,{n_{\rm B}}}^{(M)}$.



\begin{thebibliography}{46}%
\makeatletter
\providecommand \@ifxundefined [1]{%
 \@ifx{#1\undefined}
}%
\providecommand \@ifnum [1]{%
 \ifnum #1\expandafter \@firstoftwo
 \else \expandafter \@secondoftwo
 \fi
}%
\providecommand \@ifx [1]{%
 \ifx #1\expandafter \@firstoftwo
 \else \expandafter \@secondoftwo
 \fi
}%
\providecommand \natexlab [1]{#1}%
\providecommand \enquote  [1]{``#1''}%
\providecommand \bibnamefont  [1]{#1}%
\providecommand \bibfnamefont [1]{#1}%
\providecommand \citenamefont [1]{#1}%
\providecommand \href@noop [0]{\@secondoftwo}%
\providecommand \href [0]{\begingroup \@sanitize@url \@href}%
\providecommand \@href[1]{\@@startlink{#1}\@@href}%
\providecommand \@@href[1]{\endgroup#1\@@endlink}%
\providecommand \@sanitize@url [0]{\catcode `\\12\catcode `\$12\catcode
  `\&12\catcode `\#12\catcode `\^12\catcode `\_12\catcode `\%12\relax}%
\providecommand \@@startlink[1]{}%
\providecommand \@@endlink[0]{}%
\providecommand \url  [0]{\begingroup\@sanitize@url \@url }%
\providecommand \@url [1]{\endgroup\@href {#1}{\urlprefix }}%
\providecommand \urlprefix  [0]{URL }%
\providecommand \Eprint [0]{\href }%
\providecommand \doibase [0]{https://doi.org/}%
\providecommand \selectlanguage [0]{\@gobble}%
\providecommand \bibinfo  [0]{\@secondoftwo}%
\providecommand \bibfield  [0]{\@secondoftwo}%
\providecommand \translation [1]{[#1]}%
\providecommand \BibitemOpen [0]{}%
\providecommand \bibitemStop [0]{}%
\providecommand \bibitemNoStop [0]{.\EOS\space}%
\providecommand \EOS [0]{\spacefactor3000\relax}%
\providecommand \BibitemShut  [1]{\csname bibitem#1\endcsname}%
\let\auto@bib@innerbib\@empty


\bibitem [{\citenamefont {Aghanim}\ \emph {et~al.}(2020)\citenamefont
  {Aghanim}, \citenamefont {Akrami}, \citenamefont {Ashdown}, \citenamefont
  {Aumont}, \citenamefont {Baccigalupi}, \citenamefont {Ballardini},
  \citenamefont {Banday}, \citenamefont {Barreiro}, \citenamefont {Bartolo},
  \citenamefont {Basak} \emph {et~al.}}]{aghanim2020planck}%
  \BibitemOpen
  \bibfield  {author} {\bibinfo {author} {\bibfnamefont {N.}~\bibnamefont
  {Aghanim}}, \bibinfo {author} {\bibfnamefont {Y.}~\bibnamefont {Akrami}},
  \bibinfo {author} {\bibfnamefont {M.}~\bibnamefont {Ashdown}}, \bibinfo
  {author} {\bibfnamefont {J.}~\bibnamefont {Aumont}}, \bibinfo {author}
  {\bibfnamefont {C.}~\bibnamefont {Baccigalupi}}, \bibinfo {author}
  {\bibfnamefont {M.}~\bibnamefont {Ballardini}}, \bibinfo {author}
  {\bibfnamefont {A.}~\bibnamefont {Banday}}, \bibinfo {author} {\bibfnamefont
  {R.}~\bibnamefont {Barreiro}}, \bibinfo {author} {\bibfnamefont
  {N.}~\bibnamefont {Bartolo}}, \bibinfo {author} {\bibfnamefont
  {S.}~\bibnamefont {Basak}}, \emph {et~al.},\ }\bibfield  {title} {\bibinfo
  {title} {Planck 2018 results-vi. cosmological parameters},\ }\href@noop {}
  {\bibfield  {journal} {\bibinfo  {journal} {Astron. Astrophys.}\ }\textbf
  {\bibinfo {volume} {641}},\ \bibinfo {pages} {A6} (\bibinfo {year}
  {2020})}\BibitemShut {NoStop}%
\bibitem [{\citenamefont {Massey}\ \emph {et~al.}(2010)\citenamefont {Massey},
  \citenamefont {Kitching},\ and\ \citenamefont {Richard}}]{massey2010dark}%
  \BibitemOpen
  \bibfield  {author} {\bibinfo {author} {\bibfnamefont {R.}~\bibnamefont
  {Massey}}, \bibinfo {author} {\bibfnamefont {T.}~\bibnamefont {Kitching}},\
  and\ \bibinfo {author} {\bibfnamefont {J.}~\bibnamefont {Richard}},\
  }\bibfield  {title} {\bibinfo {title} {The dark matter of gravitational
  lensing},\ }\href@noop {} {\bibfield  {journal} {\bibinfo  {journal} {Rep.
  Prog. Phys.}\ }\textbf {\bibinfo {volume} {73}},\ \bibinfo {pages} {086901}
  (\bibinfo {year} {2010})}\BibitemShut {NoStop}%
\bibitem [{\citenamefont {Sofue}\ and\ \citenamefont
  {Rubin}(2001)}]{sofue2001rotation}%
  \BibitemOpen
  \bibfield  {author} {\bibinfo {author} {\bibfnamefont {Y.}~\bibnamefont
  {Sofue}}\ and\ \bibinfo {author} {\bibfnamefont {V.}~\bibnamefont {Rubin}},\
  }\bibfield  {title} {\bibinfo {title} {Rotation curves of spiral galaxies},\
  }\href@noop {} {\bibfield  {journal} {\bibinfo  {journal} {Annu. Rev. Astron.
  Astr.}\ }\textbf {\bibinfo {volume} {39}},\ \bibinfo {pages} {137} (\bibinfo
  {year} {2001})}\BibitemShut {NoStop}%
\bibitem [{\citenamefont {Carney}\ \emph {et~al.}(2020)\citenamefont {Carney},
  \citenamefont {Ghosh}, \citenamefont {Krnjaic},\ and\ \citenamefont
  {Taylor}}]{carney2020PRD}%
  \BibitemOpen
  \bibfield  {author} {\bibinfo {author} {\bibfnamefont {D.}~\bibnamefont
  {Carney}}, \bibinfo {author} {\bibfnamefont {S.}~\bibnamefont {Ghosh}},
  \bibinfo {author} {\bibfnamefont {G.}~\bibnamefont {Krnjaic}},\ and\ \bibinfo
  {author} {\bibfnamefont {J.~M.}\ \bibnamefont {Taylor}},\ }\bibfield  {title}
  {\bibinfo {title} {Proposal for gravitational direct detection of dark
  matter},\ }\href {https://doi.org/10.1103/PhysRevD.102.072003} {\bibfield
  {journal} {\bibinfo  {journal} {Phys. Rev. D}\ }\textbf {\bibinfo {volume}
  {102}},\ \bibinfo {pages} {072003} (\bibinfo {year} {2020})}\BibitemShut
  {NoStop}%
\bibitem [{\citenamefont {Manley}\ \emph {et~al.}(2020)\citenamefont {Manley},
  \citenamefont {Wilson}, \citenamefont {Stump}, \citenamefont {Grin},\ and\
  \citenamefont {Singh}}]{dal2020resonators}%
  \BibitemOpen
  \bibfield  {author} {\bibinfo {author} {\bibfnamefont {J.}~\bibnamefont
  {Manley}}, \bibinfo {author} {\bibfnamefont {D.~J.}\ \bibnamefont {Wilson}},
  \bibinfo {author} {\bibfnamefont {R.}~\bibnamefont {Stump}}, \bibinfo
  {author} {\bibfnamefont {D.}~\bibnamefont {Grin}},\ and\ \bibinfo {author}
  {\bibfnamefont {S.}~\bibnamefont {Singh}},\ }\bibfield  {title} {\bibinfo
  {title} {Searching for scalar dark matter with compact mechanical
  resonators},\ }\href {https://doi.org/10.1103/PhysRevLett.124.151301}
  {\bibfield  {journal} {\bibinfo  {journal} {Phys. Rev. Lett.}\ }\textbf
  {\bibinfo {volume} {124}},\ \bibinfo {pages} {151301} (\bibinfo {year}
  {2020})}\BibitemShut {NoStop}%
\bibitem [{\citenamefont {Manley}\ \emph {et~al.}(2021)\citenamefont {Manley},
  \citenamefont {Chowdhury}, \citenamefont {Grin}, \citenamefont {Singh},\ and\
  \citenamefont {Wilson}}]{dal2021VDM}%
  \BibitemOpen
  \bibfield  {author} {\bibinfo {author} {\bibfnamefont {J.}~\bibnamefont
  {Manley}}, \bibinfo {author} {\bibfnamefont {M.~D.}\ \bibnamefont
  {Chowdhury}}, \bibinfo {author} {\bibfnamefont {D.}~\bibnamefont {Grin}},
  \bibinfo {author} {\bibfnamefont {S.}~\bibnamefont {Singh}},\ and\ \bibinfo
  {author} {\bibfnamefont {D.~J.}\ \bibnamefont {Wilson}},\ }\bibfield  {title}
  {\bibinfo {title} {Searching for vector dark matter with an optomechanical
  accelerometer},\ }\href {https://doi.org/10.1103/PhysRevLett.126.061301}
  {\bibfield  {journal} {\bibinfo  {journal} {Phys. Rev. Lett.}\ }\textbf
  {\bibinfo {volume} {126}},\ \bibinfo {pages} {061301} (\bibinfo {year}
  {2021})}\BibitemShut {NoStop}%
\bibitem [{\citenamefont {Carney}\ \emph
  {et~al.}(2021{\natexlab{a}})\citenamefont {Carney}, \citenamefont {Hook},
  \citenamefont {Liu}, \citenamefont {Taylor},\ and\ \citenamefont
  {Zhao}}]{carney2021DM}%
  \BibitemOpen
  \bibfield  {author} {\bibinfo {author} {\bibfnamefont {D.}~\bibnamefont
  {Carney}}, \bibinfo {author} {\bibfnamefont {A.}~\bibnamefont {Hook}},
  \bibinfo {author} {\bibfnamefont {Z.}~\bibnamefont {Liu}}, \bibinfo {author}
  {\bibfnamefont {J.~M.}\ \bibnamefont {Taylor}},\ and\ \bibinfo {author}
  {\bibfnamefont {Y.}~\bibnamefont {Zhao}},\ }\bibfield  {title} {\bibinfo
  {title} {Ultralight dark matter detection with mechanical quantum sensors},\
  }\href {https://doi.org/10.1088/1367-2630/abd9e7} {\bibfield  {journal}
  {\bibinfo  {journal} {New J. Phys.}\ }\textbf {\bibinfo {volume} {23}},\
  \bibinfo {pages} {023041} (\bibinfo {year} {2021}{\natexlab{a}})}\BibitemShut
  {NoStop}%
\bibitem [{\citenamefont {Carney}\ \emph
  {et~al.}(2021{\natexlab{b}})\citenamefont {Carney}, \citenamefont {Krnjaic},
  \citenamefont {Moore}, \citenamefont {Regal}, \citenamefont {Afek},
  \citenamefont {Bhave}, \citenamefont {Brubaker}, \citenamefont {Corbitt},
  \citenamefont {Cripe}, \citenamefont {Crisosto} \emph
  {et~al.}}]{carney2021white_paper}%
  \BibitemOpen
  \bibfield  {author} {\bibinfo {author} {\bibfnamefont {D.}~\bibnamefont
  {Carney}}, \bibinfo {author} {\bibfnamefont {G.}~\bibnamefont {Krnjaic}},
  \bibinfo {author} {\bibfnamefont {D.~C.}\ \bibnamefont {Moore}}, \bibinfo
  {author} {\bibfnamefont {C.~A.}\ \bibnamefont {Regal}}, \bibinfo {author}
  {\bibfnamefont {G.}~\bibnamefont {Afek}}, \bibinfo {author} {\bibfnamefont
  {S.}~\bibnamefont {Bhave}}, \bibinfo {author} {\bibfnamefont
  {B.}~\bibnamefont {Brubaker}}, \bibinfo {author} {\bibfnamefont
  {T.}~\bibnamefont {Corbitt}}, \bibinfo {author} {\bibfnamefont
  {J.}~\bibnamefont {Cripe}}, \bibinfo {author} {\bibfnamefont
  {N.}~\bibnamefont {Crisosto}}, \emph {et~al.},\ }\bibfield  {title} {\bibinfo
  {title} {Mechanical quantum sensing in the search for dark matter},\ }\href
  {https://doi.org/10.1088/2058-9565/abcfcd} {\bibfield  {journal} {\bibinfo
  {journal} {Quantum Sci. Technol.}\ }\textbf {\bibinfo {volume} {6}},\
  \bibinfo {pages} {024002} (\bibinfo {year} {2021}{\natexlab{b}})}\BibitemShut
  {NoStop}%
\bibitem [{\citenamefont {Monteiro}\ \emph {et~al.}(2020)\citenamefont
  {Monteiro}, \citenamefont {Afek}, \citenamefont {Carney}, \citenamefont
  {Krnjaic}, \citenamefont {Wang},\ and\ \citenamefont
  {Moore}}]{monteiro2020PRL}%
  \BibitemOpen
  \bibfield  {author} {\bibinfo {author} {\bibfnamefont {F.}~\bibnamefont
  {Monteiro}}, \bibinfo {author} {\bibfnamefont {G.}~\bibnamefont {Afek}},
  \bibinfo {author} {\bibfnamefont {D.}~\bibnamefont {Carney}}, \bibinfo
  {author} {\bibfnamefont {G.}~\bibnamefont {Krnjaic}}, \bibinfo {author}
  {\bibfnamefont {J.}~\bibnamefont {Wang}},\ and\ \bibinfo {author}
  {\bibfnamefont {D.~C.}\ \bibnamefont {Moore}},\ }\bibfield  {title} {\bibinfo
  {title} {Search for composite dark matter with optically levitated sensors},\
  }\href {https://doi.org/10.1103/PhysRevLett.125.181102} {\bibfield  {journal}
  {\bibinfo  {journal} {Phys. Rev. Lett.}\ }\textbf {\bibinfo {volume} {125}},\
  \bibinfo {pages} {181102} (\bibinfo {year} {2020})}\BibitemShut {NoStop}%
\bibitem [{\citenamefont {Moore}\ and\ \citenamefont
  {Geraci}(2021)}]{moore2021levitated}%
  \BibitemOpen
  \bibfield  {author} {\bibinfo {author} {\bibfnamefont {D.~C.}\ \bibnamefont
  {Moore}}\ and\ \bibinfo {author} {\bibfnamefont {A.~A.}\ \bibnamefont
  {Geraci}},\ }\bibfield  {title} {\bibinfo {title} {Searching for new physics
  using optically levitated sensors},\ }\href
  {https://doi.org/10.1088/2058-9565/abcf8a} {\bibfield  {journal} {\bibinfo
  {journal} {Quantum Sci. Technol.}\ }\textbf {\bibinfo {volume} {6}},\
  \bibinfo {pages} {014008} (\bibinfo {year} {2021})}\BibitemShut {NoStop}%
\bibitem [{\citenamefont {Afek}\ \emph {et~al.}(2022)\citenamefont {Afek},
  \citenamefont {Carney},\ and\ \citenamefont
  {Moore}}]{afek2022trapped_sensors}%
  \BibitemOpen
  \bibfield  {author} {\bibinfo {author} {\bibfnamefont {G.}~\bibnamefont
  {Afek}}, \bibinfo {author} {\bibfnamefont {D.}~\bibnamefont {Carney}},\ and\
  \bibinfo {author} {\bibfnamefont {D.~C.}\ \bibnamefont {Moore}},\ }\bibfield
  {title} {\bibinfo {title} {Coherent scattering of low mass dark matter from
  optically trapped sensors},\ }\href
  {https://doi.org/10.1103/PhysRevLett.128.101301} {\bibfield  {journal}
  {\bibinfo  {journal} {Phys. Rev. Lett.}\ }\textbf {\bibinfo {volume} {128}},\
  \bibinfo {pages} {101301} (\bibinfo {year} {2022})}\BibitemShut {NoStop}%
\bibitem [{\citenamefont {Yin}\ \emph {et~al.}(2022)\citenamefont {Yin},
  \citenamefont {Li}, \citenamefont {Yin}, \citenamefont {Xu}, \citenamefont
  {Bian}, \citenamefont {Xie}, \citenamefont {Duan}, \citenamefont {Huang},
  \citenamefont {He},\ and\ \citenamefont {Du}}]{yin2022}%
  \BibitemOpen
  \bibfield  {author} {\bibinfo {author} {\bibfnamefont {P.}~\bibnamefont
  {Yin}}, \bibinfo {author} {\bibfnamefont {R.}~\bibnamefont {Li}}, \bibinfo
  {author} {\bibfnamefont {C.}~\bibnamefont {Yin}}, \bibinfo {author}
  {\bibfnamefont {X.}~\bibnamefont {Xu}}, \bibinfo {author} {\bibfnamefont
  {X.}~\bibnamefont {Bian}}, \bibinfo {author} {\bibfnamefont {H.}~\bibnamefont
  {Xie}}, \bibinfo {author} {\bibfnamefont {C.-K.}\ \bibnamefont {Duan}},
  \bibinfo {author} {\bibfnamefont {P.}~\bibnamefont {Huang}}, \bibinfo
  {author} {\bibfnamefont {J.-h.}\ \bibnamefont {He}},\ and\ \bibinfo {author}
  {\bibfnamefont {J.}~\bibnamefont {Du}},\ }\bibfield  {title} {\bibinfo
  {title} {Experiments with levitated force sensor challenge theories of dark
  energy},\ }\href@noop {} {\bibfield  {journal} {\bibinfo  {journal} {Nat.
  Phys.}\ } (\bibinfo {year} {2022})}\BibitemShut {NoStop}%
\bibitem [{\citenamefont {Sikivie}(1983)}]{Sikivie:1983ip}%
  \BibitemOpen
  \bibfield  {author} {\bibinfo {author} {\bibfnamefont {P.}~\bibnamefont
  {Sikivie}},\ }\bibfield  {title} {\bibinfo {title} {Experimental tests of the
  invisible axion},\ }\href {https://doi.org/10.1103/PhysRevLett.51.1415}
  {\bibfield  {journal} {\bibinfo  {journal} {Phys. Rev. Lett.}\ }\textbf
  {\bibinfo {volume} {51}},\ \bibinfo {pages} {1415} (\bibinfo {year}
  {1983})},\ \bibinfo {note} {[Erratum: Phys.Rev.Lett. 52, 695
  (1984)]}\BibitemShut {NoStop}%
\bibitem [{\citenamefont {Zheng}\ \emph
  {et~al.}(2016{\natexlab{a}})\citenamefont {Zheng}, \citenamefont {Silveri},
  \citenamefont {Brierley}, \citenamefont {Girvin},\ and\ \citenamefont
  {Lehnert}}]{girvin2016axdm}%
  \BibitemOpen
  \bibfield  {author} {\bibinfo {author} {\bibfnamefont {H.}~\bibnamefont
  {Zheng}}, \bibinfo {author} {\bibfnamefont {M.}~\bibnamefont {Silveri}},
  \bibinfo {author} {\bibfnamefont {R.~T.}\ \bibnamefont {Brierley}}, \bibinfo
  {author} {\bibfnamefont {S.~M.}\ \bibnamefont {Girvin}},\ and\ \bibinfo
  {author} {\bibfnamefont {K.~W.}\ \bibnamefont {Lehnert}},\ }\href@noop {}
  {\bibinfo {title} {Accelerating dark-matter axion searches with quantum
  measurement technology}} (\bibinfo {year} {2016}{\natexlab{a}}),\ \Eprint
  {https://arxiv.org/abs/1607.02529} {arXiv:1607.02529 [hep-ph]} \BibitemShut
  {NoStop}%
\bibitem [{\citenamefont {Malnou}\ \emph {et~al.}(2019)\citenamefont {Malnou},
  \citenamefont {Palken}, \citenamefont {Brubaker}, \citenamefont {Vale},
  \citenamefont {Hilton},\ and\ \citenamefont {Lehnert}}]{malnou2019}%
  \BibitemOpen
  \bibfield  {author} {\bibinfo {author} {\bibfnamefont {M.}~\bibnamefont
  {Malnou}}, \bibinfo {author} {\bibfnamefont {D.}~\bibnamefont {Palken}},
  \bibinfo {author} {\bibfnamefont {B.}~\bibnamefont {Brubaker}}, \bibinfo
  {author} {\bibfnamefont {L.~R.}\ \bibnamefont {Vale}}, \bibinfo {author}
  {\bibfnamefont {G.~C.}\ \bibnamefont {Hilton}},\ and\ \bibinfo {author}
  {\bibfnamefont {K.}~\bibnamefont {Lehnert}},\ }\bibfield  {title} {\bibinfo
  {title} {Squeezed vacuum used to accelerate the search for a weak classical
  signal},\ }\href {https://doi.org/10.1103/PhysRevX.9.021023} {\bibfield
  {journal} {\bibinfo  {journal} {Phys. Rev. X}\ }\textbf {\bibinfo {volume}
  {9}},\ \bibinfo {pages} {021023} (\bibinfo {year} {2019})}\BibitemShut
  {NoStop}%
\bibitem [{\citenamefont {Dixit}\ \emph {et~al.}(2021)\citenamefont {Dixit},
  \citenamefont {Chakram}, \citenamefont {He}, \citenamefont {Agrawal},
  \citenamefont {Naik}, \citenamefont {Schuster},\ and\ \citenamefont
  {Chou}}]{dixit2021}%
  \BibitemOpen
  \bibfield  {author} {\bibinfo {author} {\bibfnamefont {A.~V.}\ \bibnamefont
  {Dixit}}, \bibinfo {author} {\bibfnamefont {S.}~\bibnamefont {Chakram}},
  \bibinfo {author} {\bibfnamefont {K.}~\bibnamefont {He}}, \bibinfo {author}
  {\bibfnamefont {A.}~\bibnamefont {Agrawal}}, \bibinfo {author} {\bibfnamefont
  {R.~K.}\ \bibnamefont {Naik}}, \bibinfo {author} {\bibfnamefont {D.~I.}\
  \bibnamefont {Schuster}},\ and\ \bibinfo {author} {\bibfnamefont
  {A.}~\bibnamefont {Chou}},\ }\bibfield  {title} {\bibinfo {title} {Searching
  for dark matter with a superconducting qubit},\ }\href
  {https://doi.org/10.1103/PhysRevLett.126.141302} {\bibfield  {journal}
  {\bibinfo  {journal} {Phys. Rev. Lett.}\ }\textbf {\bibinfo {volume} {126}},\
  \bibinfo {pages} {141302} (\bibinfo {year} {2021})}\BibitemShut {NoStop}%
\bibitem [{\citenamefont {Backes}\ \emph {et~al.}(2021)\citenamefont {Backes},
  \citenamefont {Palken}, \citenamefont {Al~Kenany}, \citenamefont {Brubaker},
  \citenamefont {Cahn}, \citenamefont {Droster}, \citenamefont {Hilton},
  \citenamefont {Ghosh}, \citenamefont {Jackson}, \citenamefont {Lamoreaux}
  \emph {et~al.}}]{backes2021}%
  \BibitemOpen
  \bibfield  {author} {\bibinfo {author} {\bibfnamefont {K.}~\bibnamefont
  {Backes}}, \bibinfo {author} {\bibfnamefont {D.}~\bibnamefont {Palken}},
  \bibinfo {author} {\bibfnamefont {S.}~\bibnamefont {Al~Kenany}}, \bibinfo
  {author} {\bibfnamefont {B.}~\bibnamefont {Brubaker}}, \bibinfo {author}
  {\bibfnamefont {S.}~\bibnamefont {Cahn}}, \bibinfo {author} {\bibfnamefont
  {A.}~\bibnamefont {Droster}}, \bibinfo {author} {\bibfnamefont {G.~C.}\
  \bibnamefont {Hilton}}, \bibinfo {author} {\bibfnamefont {S.}~\bibnamefont
  {Ghosh}}, \bibinfo {author} {\bibfnamefont {H.}~\bibnamefont {Jackson}},
  \bibinfo {author} {\bibfnamefont {S.}~\bibnamefont {Lamoreaux}}, \emph
  {et~al.},\ }\bibfield  {title} {\bibinfo {title} {A quantum enhanced search
  for dark matter axions},\ }\href {https://doi.org/10.1038/s41586-021-03226-7}
  {\bibfield  {journal} {\bibinfo  {journal} {Nature}\ }\textbf {\bibinfo
  {volume} {590}},\ \bibinfo {pages} {238} (\bibinfo {year}
  {2021})}\BibitemShut {NoStop}%
\bibitem [{\citenamefont {Berlin}\ \emph {et~al.}(2022)\citenamefont {Berlin},
  \citenamefont {Belomestnykh}, \citenamefont {Blas}, \citenamefont {Frolov},
  \citenamefont {Brady}, \citenamefont {Braggio}, \citenamefont {Carena},
  \citenamefont {Cervantes}, \citenamefont {Checchin}, \citenamefont
  {Contreras-Martinez} \emph {et~al.}}]{berlin2022searches}%
  \BibitemOpen
  \bibfield  {author} {\bibinfo {author} {\bibfnamefont {A.}~\bibnamefont
  {Berlin}}, \bibinfo {author} {\bibfnamefont {S.}~\bibnamefont
  {Belomestnykh}}, \bibinfo {author} {\bibfnamefont {D.}~\bibnamefont {Blas}},
  \bibinfo {author} {\bibfnamefont {D.}~\bibnamefont {Frolov}}, \bibinfo
  {author} {\bibfnamefont {A.~J.}\ \bibnamefont {Brady}}, \bibinfo {author}
  {\bibfnamefont {C.}~\bibnamefont {Braggio}}, \bibinfo {author} {\bibfnamefont
  {M.}~\bibnamefont {Carena}}, \bibinfo {author} {\bibfnamefont
  {R.}~\bibnamefont {Cervantes}}, \bibinfo {author} {\bibfnamefont
  {M.}~\bibnamefont {Checchin}}, \bibinfo {author} {\bibfnamefont
  {C.}~\bibnamefont {Contreras-Martinez}}, \emph {et~al.},\ }\bibfield  {title}
  {\bibinfo {title} {Searches for new particles, dark matter, and gravitational
  waves with srf cavities},\ }\href@noop {} {\bibfield  {journal} {\bibinfo
  {journal} {arXiv:2203.12714}\ } (\bibinfo {year} {2022})}\BibitemShut
  {NoStop}%
\bibitem [{\citenamefont {Brady}\ \emph {et~al.}(2022)\citenamefont {Brady},
  \citenamefont {Gao}, \citenamefont {Harnik}, \citenamefont {Liu},
  \citenamefont {Zhang},\ and\ \citenamefont {Zhuang}}]{brady2022entangled}%
  \BibitemOpen
  \bibfield  {author} {\bibinfo {author} {\bibfnamefont {A.~J.}\ \bibnamefont
  {Brady}}, \bibinfo {author} {\bibfnamefont {C.}~\bibnamefont {Gao}}, \bibinfo
  {author} {\bibfnamefont {R.}~\bibnamefont {Harnik}}, \bibinfo {author}
  {\bibfnamefont {Z.}~\bibnamefont {Liu}}, \bibinfo {author} {\bibfnamefont
  {Z.}~\bibnamefont {Zhang}},\ and\ \bibinfo {author} {\bibfnamefont
  {Q.}~\bibnamefont {Zhuang}},\ }\bibfield  {title} {\bibinfo {title}
  {Entangled sensor-networks for dark-matter searches},\ }\href
  {https://doi.org/10.1103/PRXQuantum.3.030333} {\bibfield  {journal} {\bibinfo
   {journal} {PRX Quantum}\ }\textbf {\bibinfo {volume} {3}},\ \bibinfo {pages}
  {030333} (\bibinfo {year} {2022})}\BibitemShut {NoStop}%
\bibitem [{\citenamefont {Holevo}(2007)}]{holevo2007one}%
  \BibitemOpen
  \bibfield  {author} {\bibinfo {author} {\bibfnamefont {A.~S.}\ \bibnamefont
  {Holevo}},\ }\bibfield  {title} {\bibinfo {title} {One-mode quantum gaussian
  channels: Structure and quantum capacity},\ }\href@noop {} {\bibfield
  {journal} {\bibinfo  {journal} {Probl. Inform. Transm.}\ }\textbf {\bibinfo
  {volume} {43}},\ \bibinfo {pages} {1} (\bibinfo {year} {2007})}\BibitemShut
  {NoStop}%
\bibitem [{\citenamefont {Weedbrook}\ \emph {et~al.}(2012)\citenamefont
  {Weedbrook}, \citenamefont {Pirandola}, \citenamefont {García-Patrón},
  \citenamefont {Cerf}, \citenamefont {Ralph}, \citenamefont {Shapiro},\ and\
  \citenamefont {Lloyd}}]{weedbrook2012gaussian}%
  \BibitemOpen
  \bibfield  {author} {\bibinfo {author} {\bibfnamefont {C.}~\bibnamefont
  {Weedbrook}}, \bibinfo {author} {\bibfnamefont {S.}~\bibnamefont
  {Pirandola}}, \bibinfo {author} {\bibfnamefont {R.}~\bibnamefont
  {Garc\'{i}a-Patr\'{o}n}}, \bibinfo {author} {\bibfnamefont {N.~J.}\ \bibnamefont
  {Cerf}}, \bibinfo {author} {\bibfnamefont {T.~C.}\ \bibnamefont {Ralph}},
  \bibinfo {author} {\bibfnamefont {J.~H.}\ \bibnamefont {Shapiro}},\ and\
  \bibinfo {author} {\bibfnamefont {S.}~\bibnamefont {Lloyd}},\ }\bibfield
  {title} {\bibinfo {title} {Gaussian quantum information},\ }\href@noop {}
  {\bibfield  {journal} {\bibinfo  {journal} {Rev. Mod. Phys.}\ }\textbf
  {\bibinfo {volume} {84}},\ \bibinfo {pages} {621} (\bibinfo {year}
  {2012})}\BibitemShut {NoStop}%
\bibitem [{\citenamefont {Escher}\ \emph {et~al.}(2011)\citenamefont {Escher},
  \citenamefont {de~Matos~Filho},\ and\ \citenamefont
  {Davidovich}}]{escher2011general}%
  \BibitemOpen
  \bibfield  {author} {\bibinfo {author} {\bibfnamefont {B.}~\bibnamefont
  {Escher}}, \bibinfo {author} {\bibfnamefont {R.}~\bibnamefont
  {de~Matos~Filho}},\ and\ \bibinfo {author} {\bibfnamefont {L.}~\bibnamefont
  {Davidovich}},\ }\bibfield  {title} {\bibinfo {title} {General framework for
  estimating the ultimate precision limit in noisy quantum-enhanced
  metrology},\ }\href@noop {} {\bibfield  {journal} {\bibinfo  {journal} {Nat.
  Phys.}\ }\textbf {\bibinfo {volume} {7}},\ \bibinfo {pages} {406} (\bibinfo
  {year} {2011})}\BibitemShut {NoStop}%
\bibitem [{\citenamefont {Zhuang}\ \emph {et~al.}(2018)\citenamefont {Zhuang},
  \citenamefont {Zhang},\ and\ \citenamefont {Shapiro}}]{zhuang2018DQSCV}%
  \BibitemOpen
  \bibfield  {author} {\bibinfo {author} {\bibfnamefont {Q.}~\bibnamefont
  {Zhuang}}, \bibinfo {author} {\bibfnamefont {Z.}~\bibnamefont {Zhang}},\ and\
  \bibinfo {author} {\bibfnamefont {J.~H.}\ \bibnamefont {Shapiro}},\
  }\bibfield  {title} {\bibinfo {title} {Distributed quantum sensing using
  continuous-variable multipartite entanglement},\ }\href
  {https://doi.org/10.1103/PhysRevA.97.032329} {\bibfield  {journal} {\bibinfo
  {journal} {Phys. Rev. A}\ }\textbf {\bibinfo {volume} {97}},\ \bibinfo
  {pages} {032329} (\bibinfo {year} {2018})}\BibitemShut {NoStop}%
\bibitem [{\citenamefont {Xia}\ \emph {et~al.}(2020)\citenamefont {Xia},
  \citenamefont {Li}, \citenamefont {Clark}, \citenamefont {Hart},
  \citenamefont {Zhuang},\ and\ \citenamefont {Zhang}}]{xia2020demonstration}%
  \BibitemOpen
  \bibfield  {author} {\bibinfo {author} {\bibfnamefont {Y.}~\bibnamefont
  {Xia}}, \bibinfo {author} {\bibfnamefont {W.}~\bibnamefont {Li}}, \bibinfo
  {author} {\bibfnamefont {W.}~\bibnamefont {Clark}}, \bibinfo {author}
  {\bibfnamefont {D.}~\bibnamefont {Hart}}, \bibinfo {author} {\bibfnamefont
  {Q.}~\bibnamefont {Zhuang}},\ and\ \bibinfo {author} {\bibfnamefont
  {Z.}~\bibnamefont {Zhang}},\ }\bibfield  {title} {\bibinfo {title}
  {Demonstration of a reconfigurable entangled radio-frequency photonic sensor
  network},\ }\href {https://doi.org/10.1103/PhysRevLett.124.150502} {\bibfield
   {journal} {\bibinfo  {journal} {Phys. Rev. Lett.}\ }\textbf {\bibinfo
  {volume} {124}},\ \bibinfo {pages} {150502} (\bibinfo {year}
  {2020})}\BibitemShut {NoStop}%
\bibitem [{\citenamefont {Monras}\ and\ \citenamefont
  {Paris}(2007)}]{monras2007optimal}%
  \BibitemOpen
  \bibfield  {author} {\bibinfo {author} {\bibfnamefont {A.}~\bibnamefont
  {Monras}}\ and\ \bibinfo {author} {\bibfnamefont {M.~G.}\ \bibnamefont
  {Paris}},\ }\bibfield  {title} {\bibinfo {title} {Optimal quantum estimation
  of loss in bosonic channels},\ }\href@noop {} {\bibfield  {journal} {\bibinfo
   {journal} {Phys. Rev. Lett.}\ }\textbf {\bibinfo {volume} {98}},\ \bibinfo
  {pages} {160401} (\bibinfo {year} {2007})}\BibitemShut {NoStop}%
\bibitem [{\citenamefont {Nair}(2018)}]{nair2018quantum}%
  \BibitemOpen
  \bibfield  {author} {\bibinfo {author} {\bibfnamefont {R.}~\bibnamefont
  {Nair}},\ }\bibfield  {title} {\bibinfo {title} {Quantum-limited loss
  sensing: Multiparameter estimation and bures distance between loss
  channels},\ }\href {https://doi.org/10.1103/PhysRevLett.121.230801}
  {\bibfield  {journal} {\bibinfo  {journal} {Phys. Rev. Lett.}\ }\textbf
  {\bibinfo {volume} {121}},\ \bibinfo {pages} {230801} (\bibinfo {year}
  {2018})}\BibitemShut {NoStop}%
\bibitem [{\citenamefont {Nair}\ \emph {et~al.}(2022)\citenamefont {Nair},
  \citenamefont {Tham},\ and\ \citenamefont {Gu}}]{nair2022optimal}%
  \BibitemOpen
  \bibfield  {author} {\bibinfo {author} {\bibfnamefont {R.}~\bibnamefont
  {Nair}}, \bibinfo {author} {\bibfnamefont {G.~Y.}\ \bibnamefont {Tham}},\
  and\ \bibinfo {author} {\bibfnamefont {M.}~\bibnamefont {Gu}},\ }\bibfield
  {title} {\bibinfo {title} {Optimal gain sensing of quantum-limited
  phase-insensitive amplifiers},\ }\href@noop {} {\bibfield  {journal}
  {\bibinfo  {journal} {Phys. Rev. Lett.}\ }\textbf {\bibinfo {volume} {128}},\
  \bibinfo {pages} {180506} (\bibinfo {year} {2022})}\BibitemShut {NoStop}%
\bibitem [{\citenamefont {Pirandola}\ and\ \citenamefont
  {Lupo}(2017)}]{pirandola2017ultimate}%
  \BibitemOpen
  \bibfield  {author} {\bibinfo {author} {\bibfnamefont {S.}~\bibnamefont
  {Pirandola}}\ and\ \bibinfo {author} {\bibfnamefont {C.}~\bibnamefont
  {Lupo}},\ }\bibfield  {title} {\bibinfo {title} {Ultimate precision of
  adaptive noise estimation},\ }\href@noop {} {\bibfield  {journal} {\bibinfo
  {journal} {Phys. Rev. Lett.}\ }\textbf {\bibinfo {volume} {118}},\ \bibinfo
  {pages} {100502} (\bibinfo {year} {2017})}\BibitemShut {NoStop}%
\bibitem [{\citenamefont {Zheng}\ \emph
  {et~al.}(2016{\natexlab{b}})\citenamefont {Zheng}, \citenamefont {Silveri},
  \citenamefont {Brierley}, \citenamefont {Girvin},\ and\ \citenamefont
  {Lehnert}}]{zheng2016accelerating}%
  \BibitemOpen
  \bibfield  {author} {\bibinfo {author} {\bibfnamefont {H.}~\bibnamefont
  {Zheng}}, \bibinfo {author} {\bibfnamefont {M.}~\bibnamefont {Silveri}},
  \bibinfo {author} {\bibfnamefont {R.}~\bibnamefont {Brierley}}, \bibinfo
  {author} {\bibfnamefont {S.}~\bibnamefont {Girvin}},\ and\ \bibinfo {author}
  {\bibfnamefont {K.}~\bibnamefont {Lehnert}},\ }\bibfield  {title} {\bibinfo
  {title} {Accelerating dark-matter axion searches with quantum measurement
  technology},\ }\href@noop {} {\bibfield  {journal} {\bibinfo  {journal}
  {arXiv:1607.02529}\ } (\bibinfo {year} {2016}{\natexlab{b}})}\BibitemShut
  {NoStop}%
\bibitem [{\citenamefont {Gottesman}\ \emph {et~al.}(2001)\citenamefont
  {Gottesman}, \citenamefont {Kitaev},\ and\ \citenamefont
  {Preskill}}]{gottesman2001}%
  \BibitemOpen
  \bibfield  {author} {\bibinfo {author} {\bibfnamefont {D.}~\bibnamefont
  {Gottesman}}, \bibinfo {author} {\bibfnamefont {A.}~\bibnamefont {Kitaev}},\
  and\ \bibinfo {author} {\bibfnamefont {J.}~\bibnamefont {Preskill}},\
  }\bibfield  {title} {\bibinfo {title} {Encoding a qubit in an oscillator},\
  }\href {https://doi.org/10.1103/PhysRevA.64.012310} {\bibfield  {journal}
  {\bibinfo  {journal} {Phys. Rev. A}\ }\textbf {\bibinfo {volume} {64}},\
  \bibinfo {pages} {012310} (\bibinfo {year} {2001})}\BibitemShut {NoStop}%
\bibitem [{\citenamefont {Brady~et al.}(2022)}]{brady2022entanglement}%
  \BibitemOpen
  \bibfield  {author} {\bibinfo {author} {\bibfnamefont {A.~J.}\ \bibnamefont
  {Brady~et al.}},\ }\bibfield  {title} {\bibinfo {title}
  {Entanglement-enhanced optomechanical sensor array for dark matter
  searches},\ }\href@noop {} {\bibfield  {journal} {\bibinfo  {journal} {arXiv preprint arXiv:2210.07291}\ } (\bibinfo {year} {2022})}\BibitemShut {NoStop}%
\bibitem [{\citenamefont {Tan}\ \emph {et~al.}(2008)\citenamefont {Tan},
  \citenamefont {Erkmen}, \citenamefont {Giovannetti}, \citenamefont {Guha},
  \citenamefont {Lloyd}, \citenamefont {Maccone}, \citenamefont {Pirandola},\
  and\ \citenamefont {Shapiro}}]{tan2008quantum}%
  \BibitemOpen
  \bibfield  {author} {\bibinfo {author} {\bibfnamefont {S.-H.}\ \bibnamefont
  {Tan}}, \bibinfo {author} {\bibfnamefont {B.~I.}\ \bibnamefont {Erkmen}},
  \bibinfo {author} {\bibfnamefont {V.}~\bibnamefont {Giovannetti}}, \bibinfo
  {author} {\bibfnamefont {S.}~\bibnamefont {Guha}}, \bibinfo {author}
  {\bibfnamefont {S.}~\bibnamefont {Lloyd}}, \bibinfo {author} {\bibfnamefont
  {L.}~\bibnamefont {Maccone}}, \bibinfo {author} {\bibfnamefont
  {S.}~\bibnamefont {Pirandola}},\ and\ \bibinfo {author} {\bibfnamefont
  {J.~H.}\ \bibnamefont {Shapiro}},\ }\bibfield  {title} {\bibinfo {title}
  {Quantum illumination with gaussian states},\ }\href@noop {} {\bibfield
  {journal} {\bibinfo  {journal} {Phys. Rev. Lett.}\ }\textbf {\bibinfo
  {volume} {101}},\ \bibinfo {pages} {253601} (\bibinfo {year}
  {2008})}\BibitemShut {NoStop}%
\bibitem [{\citenamefont {Shi}\ \emph {et~al.}(2020)\citenamefont {Shi},
  \citenamefont {Zhang}, \citenamefont {Pirandola},\ and\ \citenamefont
  {Zhuang}}]{shi2020entanglement}%
  \BibitemOpen
  \bibfield  {author} {\bibinfo {author} {\bibfnamefont {H.}~\bibnamefont
  {Shi}}, \bibinfo {author} {\bibfnamefont {Z.}~\bibnamefont {Zhang}}, \bibinfo
  {author} {\bibfnamefont {S.}~\bibnamefont {Pirandola}},\ and\ \bibinfo
  {author} {\bibfnamefont {Q.}~\bibnamefont {Zhuang}},\ }\bibfield  {title}
  {\bibinfo {title} {Entanglement-assisted absorption spectroscopy},\ }\href
  {https://doi.org/10.1103/PhysRevLett.125.180502} {\bibfield  {journal}
  {\bibinfo  {journal} {Phys. Rev. Lett.}\ }\textbf {\bibinfo {volume} {125}},\
  \bibinfo {pages} {180502} (\bibinfo {year} {2020})}\BibitemShut {NoStop}%
\bibitem [{\citenamefont {Zhuang}\ and\ \citenamefont
  {Shapiro}(2022)}]{zhuang2022}%
  \BibitemOpen
  \bibfield  {author} {\bibinfo {author} {\bibfnamefont {Q.}~\bibnamefont
  {Zhuang}}\ and\ \bibinfo {author} {\bibfnamefont {J.~H.}\ \bibnamefont
  {Shapiro}},\ }\bibfield  {title} {\bibinfo {title} {Ultimate accuracy limit
  of quantum pulse-compression ranging},\ }\href
  {https://doi.org/10.1103/PhysRevLett.128.010501} {\bibfield  {journal}
  {\bibinfo  {journal} {Phys. Rev. Lett.}\ }\textbf {\bibinfo {volume} {128}},\
  \bibinfo {pages} {010501} (\bibinfo {year} {2022})}\BibitemShut {NoStop}%
\bibitem [{\citenamefont {Helstrom}(1967)}]{Helstrom_1967}%
  \BibitemOpen
  \bibfield  {author} {\bibinfo {author} {\bibfnamefont {C.}~\bibnamefont
  {Helstrom}},\ }\bibfield  {title} {\bibinfo {title} {Minimum mean-squared
  error of estimates in quantum statistics},\ }\href@noop {} {\bibfield
  {journal} {\bibinfo  {journal} {Phys. Lett. A}\ }\textbf {\bibinfo {volume}
  {25}},\ \bibinfo {pages} {101} (\bibinfo {year} {1967})}\BibitemShut
  {NoStop}%
\bibitem [{\citenamefont {Yuen}\ and\ \citenamefont {Lax}(1973)}]{Yuen_1973}%
  \BibitemOpen
  \bibfield  {author} {\bibinfo {author} {\bibfnamefont {H.}~\bibnamefont
  {Yuen}}\ and\ \bibinfo {author} {\bibfnamefont {M.}~\bibnamefont {Lax}},\
  }\bibfield  {title} {\bibinfo {title} {Multiple-parameter quantum estimation
  and measurement of nonselfadjoint observables},\ }\href@noop {} {\bibfield
  {journal} {\bibinfo  {journal} {IEEE Trans. Inf. Theory}\ }\textbf {\bibinfo
  {volume} {19}},\ \bibinfo {pages} {740} (\bibinfo {year} {1973})}\BibitemShut
  {NoStop}%
\bibitem [{\citenamefont {Holevo}(2011)}]{holevo2011probabilistic}%
  \BibitemOpen
  \bibfield  {author} {\bibinfo {author} {\bibfnamefont {A.~S.}\ \bibnamefont
  {Holevo}},\ }\href@noop {} {\emph {\bibinfo {title} {Probabilistic and
  statistical aspects of quantum theory}}},\ Vol.~\bibinfo {volume} {1}\
  (\bibinfo  {publisher} {Springer Science \& Business Media},\ \bibinfo {year}
  {2011})\BibitemShut {NoStop}%
\bibitem [{\citenamefont {Braunstein}\ and\ \citenamefont
  {Caves}(1994)}]{braunstein1994statistical}%
  \BibitemOpen
  \bibfield  {author} {\bibinfo {author} {\bibfnamefont {S.~L.}\ \bibnamefont
  {Braunstein}}\ and\ \bibinfo {author} {\bibfnamefont {C.~M.}\ \bibnamefont
  {Caves}},\ }\bibfield  {title} {\bibinfo {title} {Statistical distance and
  the geometry of quantum states},\ }\href@noop {} {\bibfield  {journal}
  {\bibinfo  {journal} {Phys. Rev. Lett.}\ }\textbf {\bibinfo {volume} {72}},\
  \bibinfo {pages} {3439} (\bibinfo {year} {1994})}\BibitemShut {NoStop}%
\bibitem [{\citenamefont {Górecki}\ \emph {et~al.}(2022)\citenamefont
  {G\'{o}recki}, \citenamefont {Riccardi},\ and\ \citenamefont
  {Maccone}}]{gorecki2022}%
  \BibitemOpen
  \bibfield  {author} {\bibinfo {author} {\bibfnamefont {W.}~\bibnamefont
  {G\'{o}recki}}, \bibinfo {author} {\bibfnamefont {A.}~\bibnamefont {Riccardi}},\
  and\ \bibinfo {author} {\bibfnamefont {L.}~\bibnamefont {Maccone}},\
  }\bibfield  {title} {\bibinfo {title} {Quantum metrology of noisy spreading
  channels},\ }\href@noop {} {\bibfield  {journal} {\bibinfo  {journal}
  {arXiv:2208.09386}\ } (\bibinfo {year} {2022})}\BibitemShut {NoStop}%
\bibitem [{\citenamefont {Polloreno}\ \emph {et~al.}(2022)\citenamefont
  {Polloreno}, \citenamefont {Beckey}, \citenamefont {Levin}, \citenamefont
  {Shlosberg}, \citenamefont {Thompson}, \citenamefont {Foss-Feig},
  \citenamefont {Hayes},\ and\ \citenamefont
  {Smith}}]{polloreno2022opportunities}%
  \BibitemOpen
  \bibfield  {author} {\bibinfo {author} {\bibfnamefont {A.~M.}\ \bibnamefont
  {Polloreno}}, \bibinfo {author} {\bibfnamefont {J.~L.}\ \bibnamefont
  {Beckey}}, \bibinfo {author} {\bibfnamefont {J.}~\bibnamefont {Levin}},
  \bibinfo {author} {\bibfnamefont {A.}~\bibnamefont {Shlosberg}}, \bibinfo
  {author} {\bibfnamefont {J.~K.}\ \bibnamefont {Thompson}}, \bibinfo {author}
  {\bibfnamefont {M.}~\bibnamefont {Foss-Feig}}, \bibinfo {author}
  {\bibfnamefont {D.}~\bibnamefont {Hayes}},\ and\ \bibinfo {author}
  {\bibfnamefont {G.}~\bibnamefont {Smith}},\ }\bibfield  {title} {\bibinfo
  {title} {Opportunities and limitations in broadband sensing},\ }\href@noop {}
  {\bibfield  {journal} {\bibinfo  {journal} {arXiv:2203.05520}\ } (\bibinfo
  {year} {2022})}\BibitemShut {NoStop}%
\bibitem [{\citenamefont {Derevianko}(2018)}]{derevianko2018network}%
  \BibitemOpen
  \bibfield  {author} {\bibinfo {author} {\bibfnamefont {A.}~\bibnamefont
  {Derevianko}},\ }\bibfield  {title} {\bibinfo {title} {Detecting dark-matter
  waves with a network of precision-measurement tools},\ }\href
  {https://doi.org/10.1103/PhysRevA.97.042506} {\bibfield  {journal} {\bibinfo
  {journal} {Phys. Rev. A}\ }\textbf {\bibinfo {volume} {97}},\ \bibinfo
  {pages} {042506} (\bibinfo {year} {2018})}\BibitemShut {NoStop}%
\bibitem [{\citenamefont {Jeong}\ \emph {et~al.}(2020)\citenamefont {Jeong},
  \citenamefont {Youn}, \citenamefont {Bae}, \citenamefont {Kim}, \citenamefont
  {Seong}, \citenamefont {Kim},\ and\ \citenamefont
  {Semertzidis}}]{jeong2020prl}%
  \BibitemOpen
  \bibfield  {author} {\bibinfo {author} {\bibfnamefont {J.}~\bibnamefont
  {Jeong}}, \bibinfo {author} {\bibfnamefont {S.}~\bibnamefont {Youn}},
  \bibinfo {author} {\bibfnamefont {S.}~\bibnamefont {Bae}}, \bibinfo {author}
  {\bibfnamefont {J.}~\bibnamefont {Kim}}, \bibinfo {author} {\bibfnamefont
  {T.}~\bibnamefont {Seong}}, \bibinfo {author} {\bibfnamefont {J.~E.}\
  \bibnamefont {Kim}},\ and\ \bibinfo {author} {\bibfnamefont {Y.~K.}\
  \bibnamefont {Semertzidis}},\ }\bibfield  {title} {\bibinfo {title} {Search
  for invisible axion dark matter with a multiple-cell haloscope},\ }\href
  {https://doi.org/10.1103/PhysRevLett.125.221302} {\bibfield  {journal}
  {\bibinfo  {journal} {Phys. Rev. Lett.}\ }\textbf {\bibinfo {volume} {125}},\
  \bibinfo {pages} {221302} (\bibinfo {year} {2020})}\BibitemShut {NoStop}%
\bibitem [{\citenamefont {Yang}\ \emph {et~al.}(2020)\citenamefont {Yang},
  \citenamefont {Gleason}, \citenamefont {Jois}, \citenamefont {Stern},
  \citenamefont {Sikivie}, \citenamefont {Sullivan},\ and\ \citenamefont
  {Tanner}}]{sikivie2020search}%
  \BibitemOpen
  \bibfield  {author} {\bibinfo {author} {\bibfnamefont {J.}~\bibnamefont
  {Yang}}, \bibinfo {author} {\bibfnamefont {J.~R.}\ \bibnamefont {Gleason}},
  \bibinfo {author} {\bibfnamefont {S.}~\bibnamefont {Jois}}, \bibinfo {author}
  {\bibfnamefont {I.}~\bibnamefont {Stern}}, \bibinfo {author} {\bibfnamefont
  {P.}~\bibnamefont {Sikivie}}, \bibinfo {author} {\bibfnamefont {N.~S.}\
  \bibnamefont {Sullivan}},\ and\ \bibinfo {author} {\bibfnamefont {D.~B.}\
  \bibnamefont {Tanner}},\ }\bibfield  {title} {\bibinfo {title} {{Search for
  5--9 $\mu$eV Axions with ADMX Four-Cavity Array}},\ }in\ \href
  {https://doi.org/10.1007/978-3-030-43761-9_7} {\emph {\bibinfo {booktitle}
  {Microwave Cavities and Detectors for Axion Research}}}\ (\bibinfo
  {publisher} {Springer},\ \bibinfo {year} {2020})\ pp.\ \bibinfo {pages}
  {53--62}\BibitemShut {NoStop}%
  
\bibitem [{\citenamefont {Ivan}\ \emph {et~al.}(2011)\citenamefont {Ivan},
  \citenamefont {Sabapathy},\ and\ \citenamefont {Simon}}]{ivan2011operator}%
  \BibitemOpen
  \bibfield  {author} {\bibinfo {author} {\bibfnamefont {J.~S.}\ \bibnamefont
  {Ivan}}, \bibinfo {author} {\bibfnamefont {K.~K.}\ \bibnamefont
  {Sabapathy}},\ and\ \bibinfo {author} {\bibfnamefont {R.}~\bibnamefont
  {Simon}},\ }\bibfield  {title} {\bibinfo {title} {Operator-sum representation
  for bosonic gaussian channels},\ }\href@noop {} {\bibfield  {journal}
  {\bibinfo  {journal} {Phys. Rev. A}\ }\textbf {\bibinfo {volume} {84}},\
  \bibinfo {pages} {042311} (\bibinfo {year} {2011})}\BibitemShut {NoStop}%
\bibitem [{\citenamefont {Gao}\ and\ \citenamefont
  {Lee}(2014)}]{gao2014bounds}%
  \BibitemOpen
  \bibfield  {author} {\bibinfo {author} {\bibfnamefont {Y.}~\bibnamefont
  {Gao}}\ and\ \bibinfo {author} {\bibfnamefont {H.}~\bibnamefont {Lee}},\
  }\bibfield  {title} {\bibinfo {title} {Bounds on quantum multiple-parameter
  estimation with gaussian state},\ }\href@noop {} {\bibfield  {journal}
  {\bibinfo  {journal} {Eur. Phys. J. D}\ }\textbf {\bibinfo {volume} {68}},\
  \bibinfo {pages} {1} (\bibinfo {year} {2014})}\BibitemShut {NoStop}%
\bibitem [{\citenamefont {Marian}\ and\ \citenamefont
  {Marian}(1993)}]{marian1993squeezed}%
  \BibitemOpen
  \bibfield  {author} {\bibinfo {author} {\bibfnamefont {P.}~\bibnamefont
  {Marian}}\ and\ \bibinfo {author} {\bibfnamefont {T.~A.}\ \bibnamefont
  {Marian}},\ }\bibfield  {title} {\bibinfo {title} {Squeezed states with
  thermal noise. i. photon-number statistics},\ }\href@noop {} {\bibfield
  {journal} {\bibinfo  {journal} {Phys. Rev. A}\ }\textbf {\bibinfo {volume}
  {47}},\ \bibinfo {pages} {4474} (\bibinfo {year} {1993})}\BibitemShut
  {NoStop}%
\end{thebibliography}

%

\end{appendix}

\end{document}